\documentclass[12pt,a4paper]{article}
\usepackage{amsmath}
\usepackage{amssymb}
\usepackage{amsfonts}
\usepackage{amsthm}
\usepackage{mathrsfs}
\usepackage{graphicx}
\usepackage{subfig} 
\usepackage{xcolor}
\usepackage{multirow}
\usepackage{changes}
\usepackage{comment}

\usepackage{float}

\newtheorem{definition}{Definition}[section]

\theoremstyle{definition}
\newtheorem{remark}{Remark}
\DeclareMathOperator{\spn}{\mathrm{Span}}

\newcommand*{\diff}{\mathop{}\!\mathrm{d}}

\title{Functional architecture of M1 cells encoding movement direction}
\date{}\author{C. Mazzetti, A. Sarti, G. Citti}
\begin{document}
\maketitle
\tableofcontents
\section*{Abstract}
In this paper we propose a neurogeometrical model of the behaviour of cells of the arm area of the primary motor cortex (M1). We will mathematically express as a fiber bundle the hypercolumnar organization of this cortical area, first modelled by Georgopoulos in \cite{georgopoulos1982relations, georgopoulos2015columnar}.  
On this structure, we will consider the selective tuning of M1 neurons of kinematic variables of positions and directions of movement. We will then extend this model to encode the notion of fragments introduced by Hatsopoulos \cite{Encoding} which describes the selectivity of neurons to movement direction varying in time. This leads to consider a higher dimensional geometrical structure where fragments are represented as integral curves. 
A comparison with the curves obtained through numerical simulations and experimental data will be presented. Moreover, neural activity shows coherent behaviours represented in terms of movement trajectories pointing to a specific pattern of movement decomposition \cite{kadmon2019movement}. Here, we will recover this pattern through a spectral clustering algorithm in the subriemannian structure we introduced, and compare our results with the neurophysiological one of \cite{kadmon2019movement}.

\section*{Introduction}
A fundamental problem regarding the study of motor cortex deals with the information conveyed by the discharge pattern of motor cortical cells. This is a quite difficult topic if we compare it to sensory areas, indeed, for any specific sensory stimulus, there are many inputs captured by sensory receptors and one output signal processed in the cortex. Inputs the motor system come from basal ganglia, cerebellum and fronto-parietal cortex and there are as many output signals directed to interneurons and motorneurons of the spinal cord (\cite{WISE200110137}, \cite{carpenter2012neurophysiology}, \cite{economo2018distinct}). In the motor system the existence of a notion of receptive profiles, or maybe in this case, of ``actuator profiles", is not established nor well understood. Nevertheless, it is recognized that the cortex, including the motor area itself, has a modular structure (see \cite{hubel1962receptive}, \cite{evarts1967representation}, \cite{mountcastle1997columnar},  \cite{georgopoulos2015columnar}) and whose constituent modules, linked together simultaneously or in series in time, are considered to be responsible for the broad domain of voluntary movements (\cite{flash2005motor}, \cite{mussa2004neural}). With regard to the arm area of primates motor cortex, several studies reveal how neuronal activity involves the processing of the spatial-motor information (see for example \cite{schwartz1988primate}, \cite{georgopoulos1988primate}, \cite{kettner1988primate}, \cite{caminiti1990making}). Starting from 1978, a pioneering work has been developed by A. Georgopoulos, whose experiments allow to recognize at least two main features of the arm area functional architecture. First, he discovered that cells of this area are sensible to the position and direction of the hand movement (see \cite{georgopoulos1984static, kettner1988primate} and \cite{georgopoulos1982relations, schwartz1988primate}): cells response is maximal when hand position and direction coincide with a position and direction, characteristic of the cell. Secondly, the columnar structure, which organizes motor cortical cells in columns corresponding to movement directions (see \cite{georgopoulos2007mapping}, \cite{amirikian2003modular}, \cite{georgopoulos2015columnar}). After the work of Georgopoulos, other experiments proved that activity of neurons in the primary motor cortex correlates with a
broader variety of movement-related variables, including endpoint position, velocity, acceleration  (see for example \cite{kettner1988primate}, \cite{moran1999motor}, \cite{schwartz2007useful}), as well as joint angles (see \cite{ajemian2001model}, \cite{teka2017motor}), endpoint force \cite{georgopoulos1992motor}, muscle tensions (\cite{evarts1967representation},  \cite{todorov2000direct}, \cite{holdefer2002primary}). It is also proved that the tuning for movement parameters is not static, but varies with time (\cite{ashe1994movement}, \cite{moran1999motor}, \cite{churchland2007temporal}, \cite{paninski2004spatiotemporal}) and 
for this reason Hatsopoulos (\cite{Encoding, reimer2009problem}) argues that individual motor cortical cells rather encode ``movement fragments", i.e. movement trajectories.
This feature can be considered in a more general perspective developed by M.S.A. Graziano who proposed that the motor cortex is organized into action maps (see \cite{graziano2002cortical},  \cite{graziano2007mapping}, \cite{aflalo2006possible}). This means that the motor cortex organization reflects the complexity of a movement related to a specific task. His results allowed an extension of previous neural models; in fact, he proved that cellular tuning to a limited set of movement variables could stably account for neural activity with respect to a restricted set of movements \cite{aflalo2006partial}. The reverse is also true, meaning that motor cortex neurons become tuned to movement parameters that are relevant to the task being performed. For example, in the case of center-out movements, directional tuning still plays a central role in the variation of neuronal activity, as it is strongly supported from Georgopoulos results (\cite{georgopoulos1982relations, georgopoulos1986neuronal}).
 
The aim of this paper is to propose a mathematical model inspired by the functional architecture of the arm area of motor cortex for movements related to reaching tasks. 
We will model the selective behaviour of each neuron, which changes in time, through integral curves in a suitable space of kinematic variables. We will start with a static model expressing the A. Georgopoulos' data of directional selectivity and columnar organization, and gradually we will add new aspects to get to describe the evidence of time dependence provided by Hatsopoulos \cite{Encoding} and Churchland \cite{churchland2007temporal}. As a starting point, we propose a first fiber bundle structure compatible with the hypercolumnar organization of directionally tuned cells. We assume that to every point on the cortical surface coding for hand's position in the plane $\left(x,y\right)\in\mathbb{R}^2$ is associated a full fiber of possible movement directions. More specifically, we suggest that a motor neuron can be represented by a point $\left(x,y,\theta\right)\in\mathbb{R}^2\times S^1$, where $\left(x,y\right)$  denotes the hand's position in a two dimensional plane and $\theta$ denotes a movement direction at position $\left(x,y\right)$. We then present a comparison with models of the primary visual cortex V1 (\cite{petitot1999vers,bressloff2003functional, citti2006cortical}). 
Finally we combine the two previous model in a more complex one which takes into account the hand’s position $\left(x,y\right)$ at time $t$, hand's movement direction described by an angle $\theta\in S^1$ and hand’s speed and acceleration along $\theta$, denoted by $\left(v, a\right)$. The resulting space is then $\mathcal{M}= \mathbb{R}^{3}_{\left(x,y,t\right)} \times S^1_{\theta} \times \mathbb{R}^{2}_{\left(v,a\right)}$.
In this model we describe the dependence on time of the preferred direction of cells as a curve of preferred direction, evolving in time. In section \ref{result}, we show that it is possible to choose suitable parameters in such a way to recover the experimental data of \cite{Encoding} and \cite{churchland2007temporal}. 
A comparison with Cocci's \cite{cocci2014spatio} model for movement in the visual areas is presented.

We then validate the model as follows. We define a connectivity kernel   
\begin{equation}\label{nucleo}
\omega_{\mathcal{M}}\left(\eta_i,\eta_j\right)= e^{-d_{\mathcal{M}}\left(\eta_i,\eta_j\right)^2},
\end{equation}
where $\left(\eta_i,\eta_j\right)\in\mathcal{M}$ and $d_{\mathcal{M}}$ is the sub-Riemannian metric of the space.
Equation \eqref{nucleo} is an estimate of the heat kernel, and we propose it as a model of the local connectivity between the cortical tuning points $\eta_i$ and $\eta_j$. We use this kernel, expressed in terms of kinematic variables, and a spectral clustering algorithm to detect a set of hand trajectories. These resulting paths are well in accordance with the ones obtained by Hatsopoulos et al. \cite{Encoding} and by Kadmon Harpaz et al. \cite{kadmon2019movement} with a clustering algorithm applied directly on cortical variables. 

The structure of the paper is the following. In section \ref{nback} we provide a short description of the neurophysiological structure of the motor cortex. Section \ref{math model} is devoted to the setting of our neurogeometrical model. In section \ref{result} we discuss the structure of the model by fitting its parameters with the experimental data of \cite{churchland2007temporal} and we compare the present model with the one of \cite{cocci2014spatio}, describing visual areas devoted to movement coding.
In section \ref{data_analysis} we validate the model by comparing the pattern of movement decomposition with the curves found with a grouping algorithm, driven by the proposed distance. Finally, section \ref{concl} provides a conclusion and future development of this work. In Appendices \ref{sub} and \ref{app_B} some essential definitions and properties of fiber bundles and sub-Riemannian distances are recalled.

\section{Basic neurophysiology of the primate arm area of motor cortex}\label{nback}
\subsection{Role of the movement direction information}\label{georg}
One of the key functions of the motor cortex consists on the control of the direction of movement trajectory (see \cite{georgopoulos1982relations, schwartz1988primate}). 
Each cell is sensible to a specific direction of movement in the sense that its discharge rate is highest before and during the execution movements in a specific direction, called preferred direction of the cell (PD). This single cell behaviour is modelled in \cite{georgopoulos1982relations, schwartz1988primate} through a sinusoidal function of the movement direction:
\begin{equation}\label{prima_tuning_curve}
f\left(\theta\right)= b + k\cos\left(\theta- \theta_{\text{PD}}\right),
\end{equation}
where $\theta_{\text{PD}}$ represents the preferred direction of the cell, and the coefficient $k$ denotes the increase in discharge over the overall mean $b$ at the preferred direction $\theta_{\text{PD}}$. Equation \eqref{prima_tuning_curve} is called directional tuning curve.
PDs differ from different cells and a large proportion of arm-related cells are active during reaching movements, hence in \cite{georgopoulos1988spatial}, \cite{georgopoulos1988primate} the authors proposed to estimate the direction of the movement via a weighted vectorial sum of cells PDs

\begin{equation}\label{NPV_sum}
P\left(\theta\right)= \sum_{i=1}^{N} \theta_{\text{PD}}^i\, w_i\left(\theta\right), 
\end{equation}
where $N$ is the number of cells in the population and $w_i$ is a symmetric function with respect to the preferred direction $\theta_{\text{PD}}^i$ of the $i$-th cell. 

The sum \eqref{NPV_sum} is called the neuronal population vector \cite{georgopoulos1986neuronal} and represents the direction of hand movement not only during the execution of the movement, but even before its starting.
Best predictions for the upcoming direction of movement (see Table 2 of \cite{georgopoulos1988primate}) were made with weights of the form 
\begin{equation}\label{w_g}
w_{\theta_{\text{PD}}}\left(\theta\right)= \left(f\left(\theta\right)- b\right)/k= \cos\left(\theta-\theta_{\text{PD}}\right).
\end{equation} 
 
Motor cortical cells activity is also related with the position at which the hand is actively maintained in space (see in particular \cite{georgopoulos1984static} and \cite{kettner1988primate}). In \cite{georgopoulos1984static} a positional gradient tuning curve is presented as a function which locally acts as
\begin{equation}\label{positional_gradient_tc}
g\left(x,y\right)= b+ \alpha x + \beta y,
\end{equation}
where the value of $g\left(x,y\right)$ denotes cell's discharge rate at position $\left(x,y\right)$ with respect to an origin located at the central button of the center-out reaching apparatus; the quantity $b$ is the discharge rate in the origin and $\alpha, \beta$ are expressions of the slopes of cell discharge per unit length along the $x$ and $y$ axes of the plane.  
\subsubsection{Columnar organization}
\begin{figure}[htbp]
\centering
\subfloat[]{\includegraphics[scale=0.3]{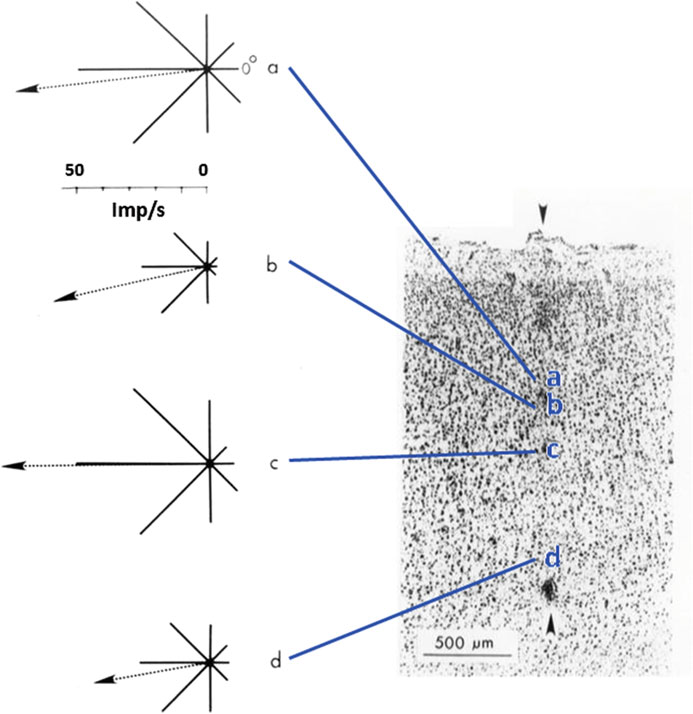}}
\qquad \qquad \qquad \qquad
\subfloat[]{\includegraphics[scale=0.095]{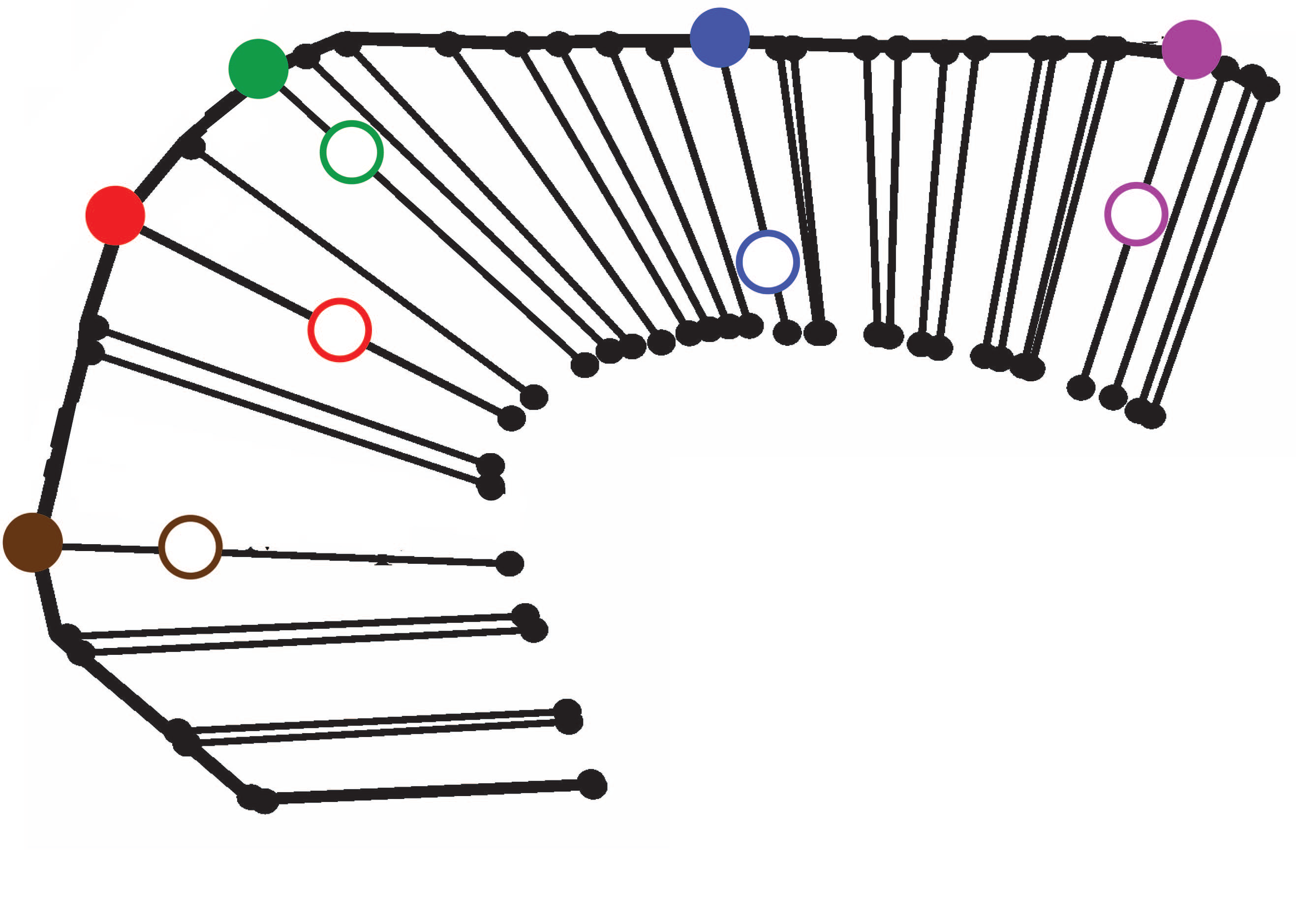}}
\caption{Columnar organization of the motor cortex. (a) Movement direction selectivity of four neurons recorded along the histologically identified penetration. The similarity of preferred directions for a penetration parallel to the cortical columns is shown. Source: \cite{georgopoulos2015columnar}. (b) Schematic illustration of the projection of recording sites onto the cortical surface along anatomical columns. Open and slashed circles denote recording and projected sites, respectively. Source: \cite{georgopoulos2007mapping}.}   
\label{colonne}
\end{figure}

\begin{figure}[htbp]
    \centering      
        \subfloat[]{\includegraphics[scale=0.046]{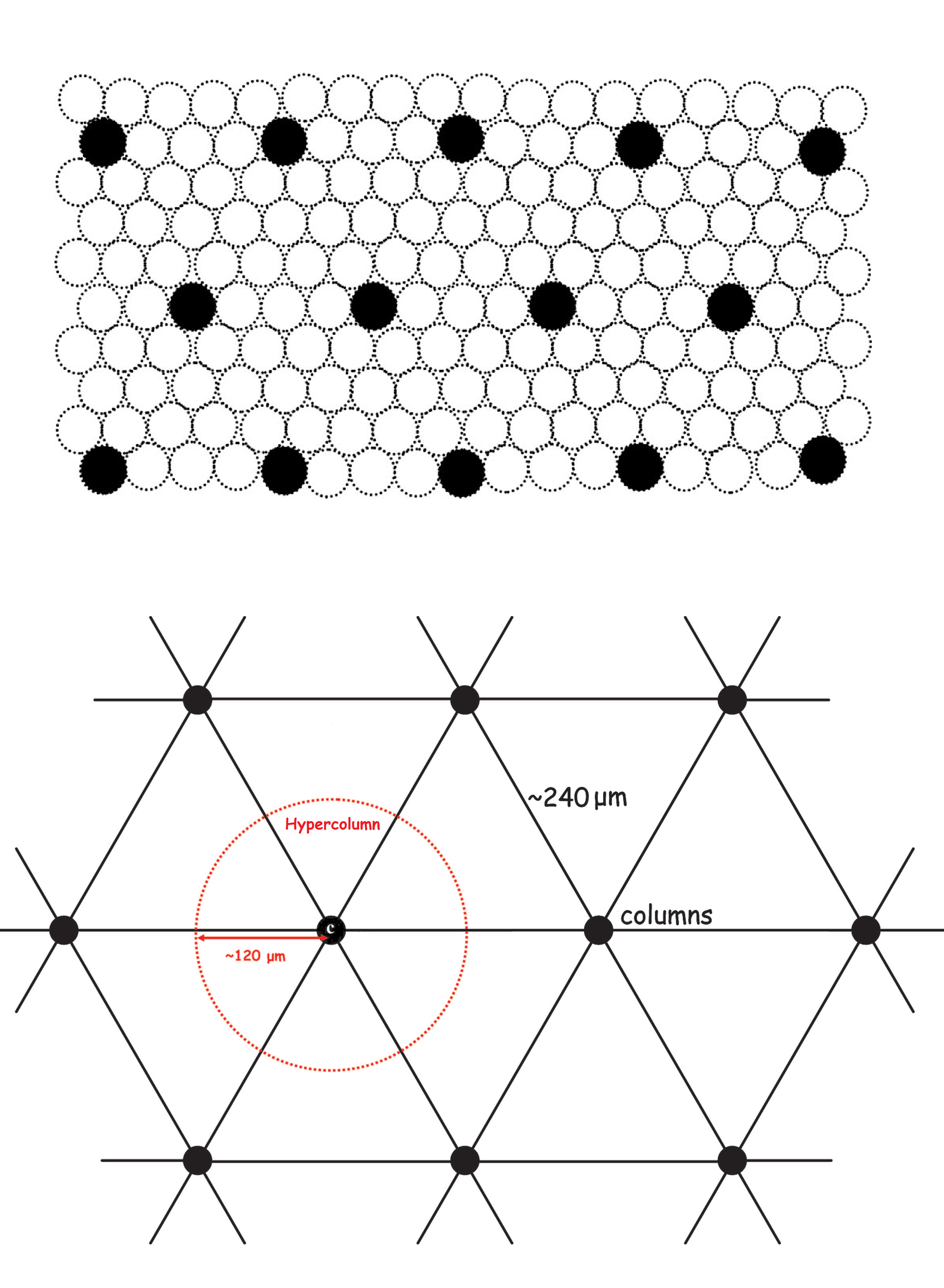}}
        \quad\quad
\subfloat[]{\includegraphics[scale=0.55]{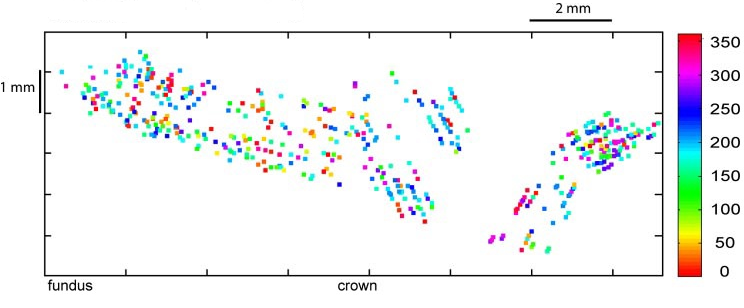}}
\caption{(a) Schematic lattice model of the repeated, regular mapping of the preferred directions in motor cortex. Adapted from \cite{georgopoulos2007mapping}. (b) Motor cortical map of preferred directions: each color denotes a cell's PD within the unit circle. Image adapted from \cite{naselaris2006large}.}\label{pinwheel_motori}     
\end{figure}

A crucial problem is the cortical representation of the preferred direction over the cortical surface. A. Georgopoulos \cite{georgopoulous1984representation} noted the location of cells with specific preferred movement direction along histologically identified penetrations and observed a change in PD in penetrations at an angle with anatomical cortical columns (see Figure \ref{colonne} as an illustrative explanation).
In \cite{amirikian2003modular} and subsequently in \cite{naselaris2006large} and \cite{georgopoulos2007mapping},
the spatial organization of PDs was examined and a columnar organization was discovered (see Figure \ref{pinwheel_motori}). 
A continuum of 500 $\mu$m in depth with cells of similar preferred directions and a repeating columnar pattern of similar PDs with a width of 50 to 100 $\mu$m together with a repetition distance of almost 200 $\mu$m were measured (see Figure \ref{pinwheel_motori}a for a schematic representation). 
A fundamental aspect of the above organization was that within each hypercolumn (i.e. assemblage of columns) of radius 120 $\mu$m, PDs were organized in such a way to represent any given direction of reach (\cite{naselaris2006large, georgopoulos2007mapping}). 

\subsection{Complexity of motor coding and temporal behaviour}\label{motor_coding}
Subsequent studies revealed that the arm area of the motor cortex is related to a more complex and heterogeneous set of movement variables (see \cite{kalaska2009intention}, \cite{scott20056} for a general review). A fundamental class is formed by ``extrinsic" or  ``hand-centered" parameters which typically describe cortical activity with respect to hand's movement. These variables mainly refer to endpoint position, velocity and acceleration of the hand both in two-dimensional and three-dimensional space (see \cite{ashe1994movement}, \cite{georgopoulos1984static}, \cite{kettner1988primate},\cite{moran1999motor}, \cite{stark2009motor}), in addition to the movement direction variable. This class of parameters has been broadly used for the characterization of the spatio-temporal form of movement (see for example the work of Flash and Hogan \cite{FH} and \cite{flash1992timing}, \cite{hogan1984organizing}). 
It is important to note that the sensibility of each neuron to these variables can depend on time (\cite{ashe1994movement}, \cite{moran1999motor}, \cite{paninski2004spatiotemporal}, \cite{churchland2007temporal},\cite{ reimer2009problem, Encoding}).
In particular, Hatsopolous \cite{Encoding} (see also \cite{reimer2009problem}) highlighted that tuning to movement parameters varies with time and proposed to describe the activity of neurons through a trajectory encoding model. In his model, the probability of spiking of a neuron is expressed as the exponential of the inner product between the ``preferred velocity trajectory" $\vec{k}$ and the normalized velocity trajectory of the hand $\vec{v}\,^{t_0}$ of duration 400 ms, as follows 
\begin{equation}\label{encoding_model}
p\left(\text{spike}\left(t_0\right)|\vec{v}\,^{t_0}\right)= \exp\left(\vec{k}^{t_0}\cdot \vec{v}\,^{t_0} + \gamma^{t_0}\right).
\end{equation}
The vector $\vec{k}$ is named as ``preferred velocity" since it maximizes the spike probability when it is aligned to $\vec{v}\,^{t_0}$, whereas the parameter $\gamma$ is an offset parameter of the model. Note how for fixed instant of time $t_0$ equation \eqref{encoding_model} reduces to equation \eqref{prima_tuning_curve} of Georgopolous model. Indeed if  $k = \left|k\right| \cos\theta_{\text{PD}}$, $v = \left|v\right|\cos\theta$, then 
$$k \cdot v + \gamma  = \left|k\right| \left|v\right| \cos(\theta - \theta_{\text{PD}}) + \gamma.$$

Consequently the argument of the exponential in equation \eqref{encoding_model} is exactly the function $f$ in \eqref{prima_tuning_curve}. The main difference is that now the same expression is considered at different instants of time $t_0$. 
In addition, equation \eqref{encoding_model} evaluates the output of a single cell in response to a trajectory fragment as the probability of spiking a neuron.  
The preferred path of the neuron is then obtained by integrating $\vec{k}$ over a time window which precedes and follows the spike time $t_0$. In the same work, it is also provided an extension to \eqref{encoding_model} by including the average speed $\bar{v}^{t_{0}}$, and average position $\left(\bar{x}^{t_{0}}, \bar{y}^{t_{0}}\right)$ of the hand trajectory:
\begin{equation}\label{encoding_model_esteso}
p\left(\text{spike}\left(t_0\right)|\vec{v}\,^{t_0}, \bar{v}^{t_{0}}, \bar{x}^{t_{0}}, \bar{y}^{t_{0}}\right)= \exp\left(\vec{k}^{t_0}\cdot\vec{v}\,^{t_0}+ a\bar{v}^{t_{0}}+ b \bar{x}^{t_{0}}+ c \bar{y}^{t_{0}} + \gamma^{t_0}\right).
\end{equation}

\begin{figure}[htbp]
\centering
\subfloat[]{\includegraphics[scale=0.25]{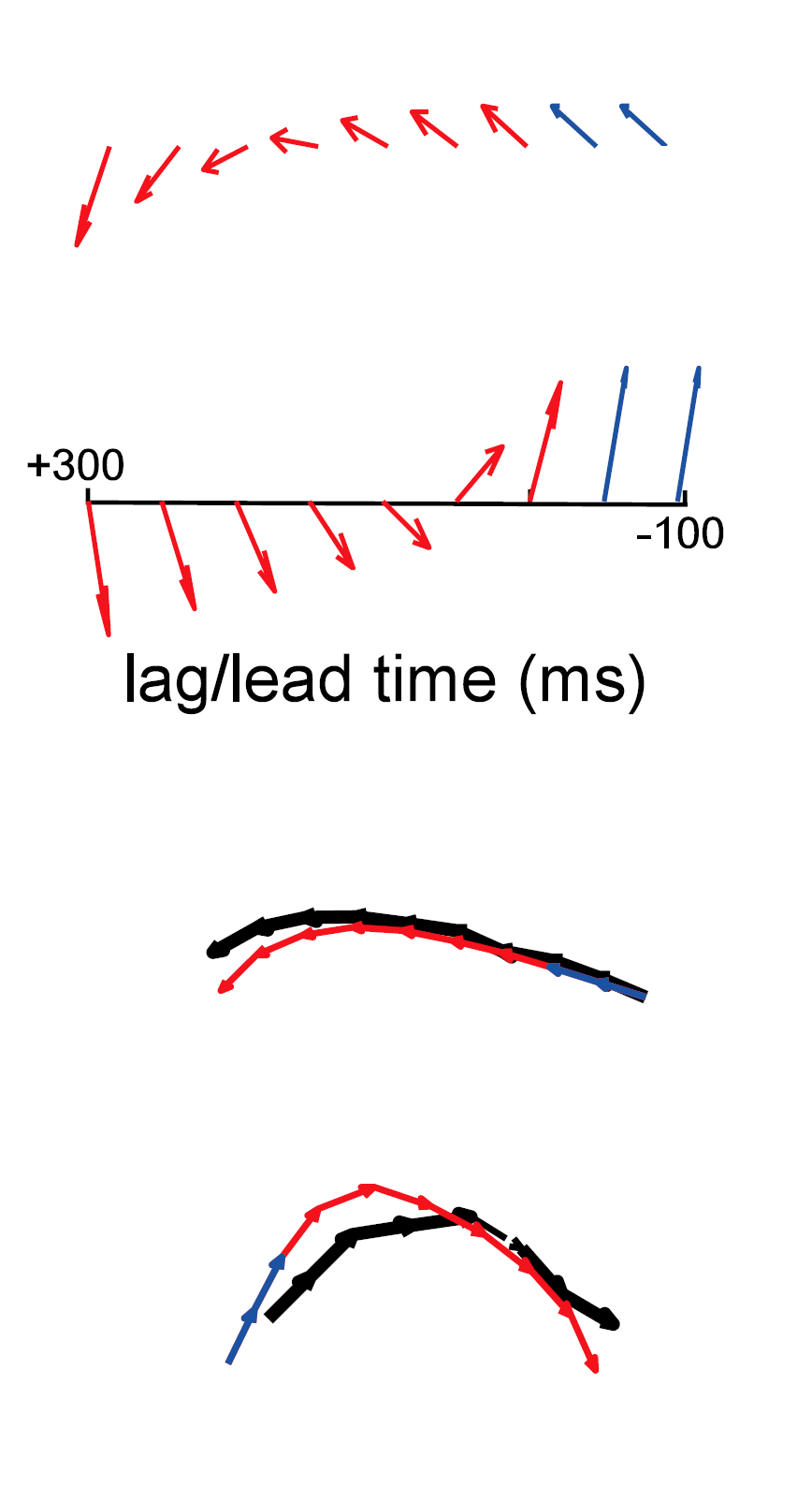}}
\qquad\qquad\qquad
\subfloat[]{\includegraphics[scale=0.7]{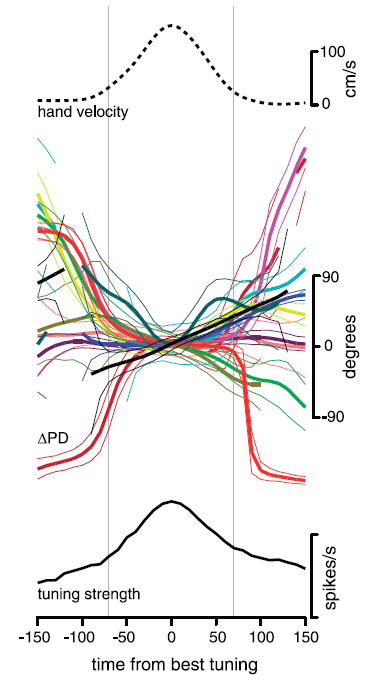}}
\caption{Temporal evolution of cells PDs through different representations. (a) Red and blue arrows show the encoded PDs for movements before and after the measured neuron firing rate, respectively. Below, the vectors of preferred directions added together give rise to the preferred trajectories. The black directional paths show the similarity of the encoded trajectories computed during a different task. Image adapted from \cite{Encoding}. (b) In the middle, change in PD expressed as continuous graphs for twelve M1 neurons. Below is shown the mean strength of direction tuning, where time 0 is assumed to be the strongest tuning instant. Above is represented the mean hand velocity profile whose peak is aligned at time $t=0$. Image adapted from \cite{churchland2007temporal}.}
\label{im_temporal_beh}
\end{figure}

Overall, Hatsopoulos \cite{Encoding} argues that M1 neurons are selective to a preferred ``movement fragment": a short trajectory describing a combination of parameters evolving in time (see also \cite{reimer2009problem} and \cite{omrani2017perspectives}).
Figure \ref{im_temporal_beh}a (from \cite{Encoding}) shows the temporal evolution of preferred directions for two neurons where each direction of movement (being an unit vector in $\mathbb{R}^2$) is represented by an angle in polar coordinates. Hence, in \cite{Encoding}, the temporal behaviour of directionally tuned cells is represented by a function  
\begin{equation}
t\mapsto\left(\cos(\theta(t)), \sin(\theta(t))\right)\in\mathbb{R}^2.
\end{equation}
At the bottom of Figure \ref{im_temporal_beh}a, vectors of preferred directions are added together giving rise to the movement fragment. 
 
Churchland and Shenoy \cite{churchland2007temporal} proposed an analogous model which describes the temporal properties of motor cortical responses.
Figure \ref{im_temporal_beh}b displays the temporal variation of the preferred directions of twelve M1 neurons during an instructed center-out reaching task. In this article, each direction of movement is expressed as a graph over a temporal interval, as follows 
\begin{equation}
t\mapsto \theta\left(t\right)\in\mathbb{R}.
\end{equation}
Through different representations of movement direction tuning, Hatsopoulos \cite{Encoding} and Churchland and Shenoy \cite{churchland2007temporal} stressed the importance of time dependence on cell sensitivity. In addition, both groups exploit a principal component analysis on a space of trajectory templates allowing a finite dimensional basis over the heterogeneity of neuronal patterns. In other words, even though neurons in M1 encode elementary movements, thank to the intrinsic connectivity of the cortex, they can generate the rich variety of complex motor behaviours. 
We present here a model able to describe the temporal dependence of the selective tuning of motor cortical cells with a finite number of kinematic variables. The differential constraints which relate these variables will be fundamental to give the structure of the space. 

\subsection{Neural states and movement output}\label{neural_states_movement_output}
\begin{figure}[htbp]
\centering
\includegraphics[scale= 0.4]{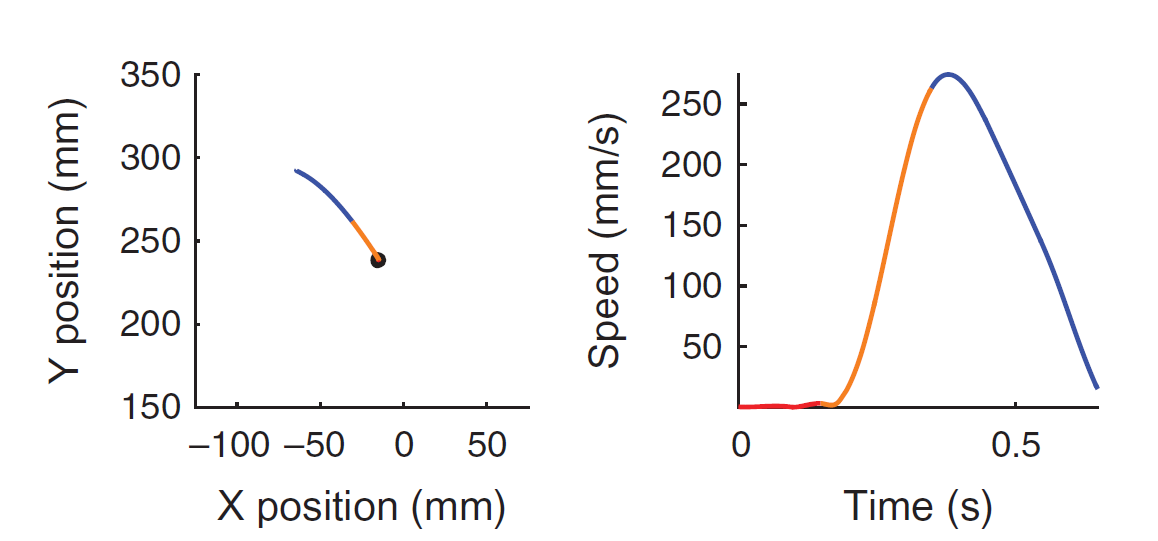}
\caption{Examples of a center-out task, with position (left) and speed profile (right) colored according to the identified neural states. Black dot represents the starting position. Each color represents a single state \cite{kadmon2019movement}.}
\label{fig_ret}
\end{figure}

\begin{figure}[htbp]
\centering
\includegraphics[scale= 0.45]{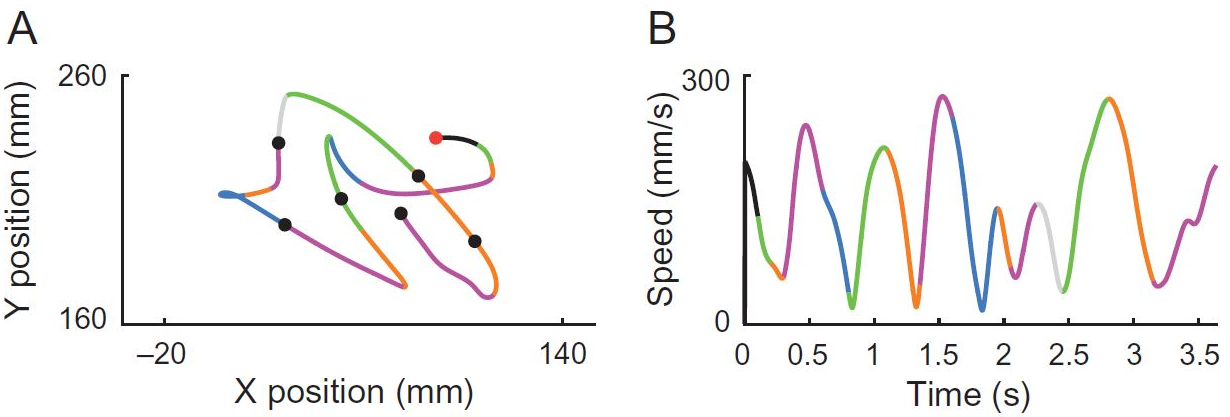}
\caption{(A) Position data of a random target pursuit task segmented and colored according to the decoded neural states. Filled circles represent the target locations, first target is colored in red.  (B) Corresponding speed profiles, colored as in A. Note the invariance of neural states to movement speed. Image taken from \cite{kadmon2019movement}.}
\label{fig_4}
\end{figure}

Another evidence of a trajectory encoding model comes from a recent study of the dynamics of neural population provided by the work of Kadmon Harpaz et al. \cite{kadmon2019movement} who studied the cell population in M1 with a 100-electrode array, thus studying the area not at the level of a single neuron, but at the local population level. The authors processed neural activity by identifying sequences of coherent behaviours, called neural states, by means of a Hidden Markov model (HMM). A HMM is used to describe  events (called hidden states)  which are not directly observable,  but are causal factors of other observed events. In their article, the recorded spike trains are interpreted as the observed states of the system, and are used to estimate global changes in cortical activity, considered as the hidden neural states of the model. The neural states (color-coded in Figures \ref{fig_ret}, \ref{fig_4}) associated with motion trajectories and identified by the model, coincide with acceleration and deceleration phases with directional selectivity of the entire reaching movement (see Figures \ref{fig_ret}, \ref{fig_4}).  Transitions between neural states systematically coincide with minima and maxima points of the tangential velocity of the end-effector, decomposing the movement into accelerating and decelerating phases. Nevertheless, the neural states do not show selectivity to movement speed and amplitude (see Figure \ref{fig_4}). A consistent decomposition of the bell-shaped speed profiles at both minima and maxima of the tangential velocity is still present in straight reaching movements (see Figure \ref{fig_ret}).

We stress that the observed segmentations in M1 were not directly predicted by previously proposed models of kinematic parameters, such as for example the one of Georgopoulos (see eq. \eqref{prima_tuning_curve}).  
The data analysis was performed at the neural level, and the authors posed the problem to recover the same decomposition by using only kinematic variables.


In this paper, we will provide a kinematic features space endowed with a metric which is in agreement with both neural models incorporating kinematic variables and both the above movement decomposition.

\section{A mathematical model}\label{math model}
The goal is to develop a mathematical model inspired by the functional architecture of the arm area of the primary motor cortex, specifically taking into account the organization of motor cortical cells and temporal behaviour. We start by showing a fiber bundle structure emerging from Georgopoulos neural models in terms of hand's position and movement direction in the plane (see section \ref{georg}).
Then we extend the model in order to include kinematic variables to which motor cortical cells are selective. 
In subsection \ref{cortical_features_structure}, we integrate in a unified framework the preceding model by considering a 6D space which codes time, position, direction of movement, speed and acceleration of the hand in the plane.

\subsection{Fiber bundle of positions and movement directions (static model)}\label{fiber_pos_dir}
As briefly exposed in section \ref{georg}, Georgopoulos  neurophysiological studies \cite{georgopoulos1982relations, georgopoulos1984static} experimentally verify that the basic functional properties of cellular activity in the arm area of M1 involve directional and positional tuning. We therefore consider that a motor cortical neuron can be represented by a point $\left(x,y,\theta\right)\in\mathbb{R}^2\times S^1$, where $\left(x,y\right)$  denotes cell's coding for hand's position in a two dimensional space and $\theta$ represents cell's preferred direction at position $\left(x,y\right)$. 
 Hence, in this first model, we propose 
to describe directionally tuned cells organization 
as a fiber bundle $\left(E, M, F, \pi\right)$ (see Definition \ref{fiber bundle}), where 

\begin{figure}[htbp]
\centering
\includegraphics[scale=0.105]{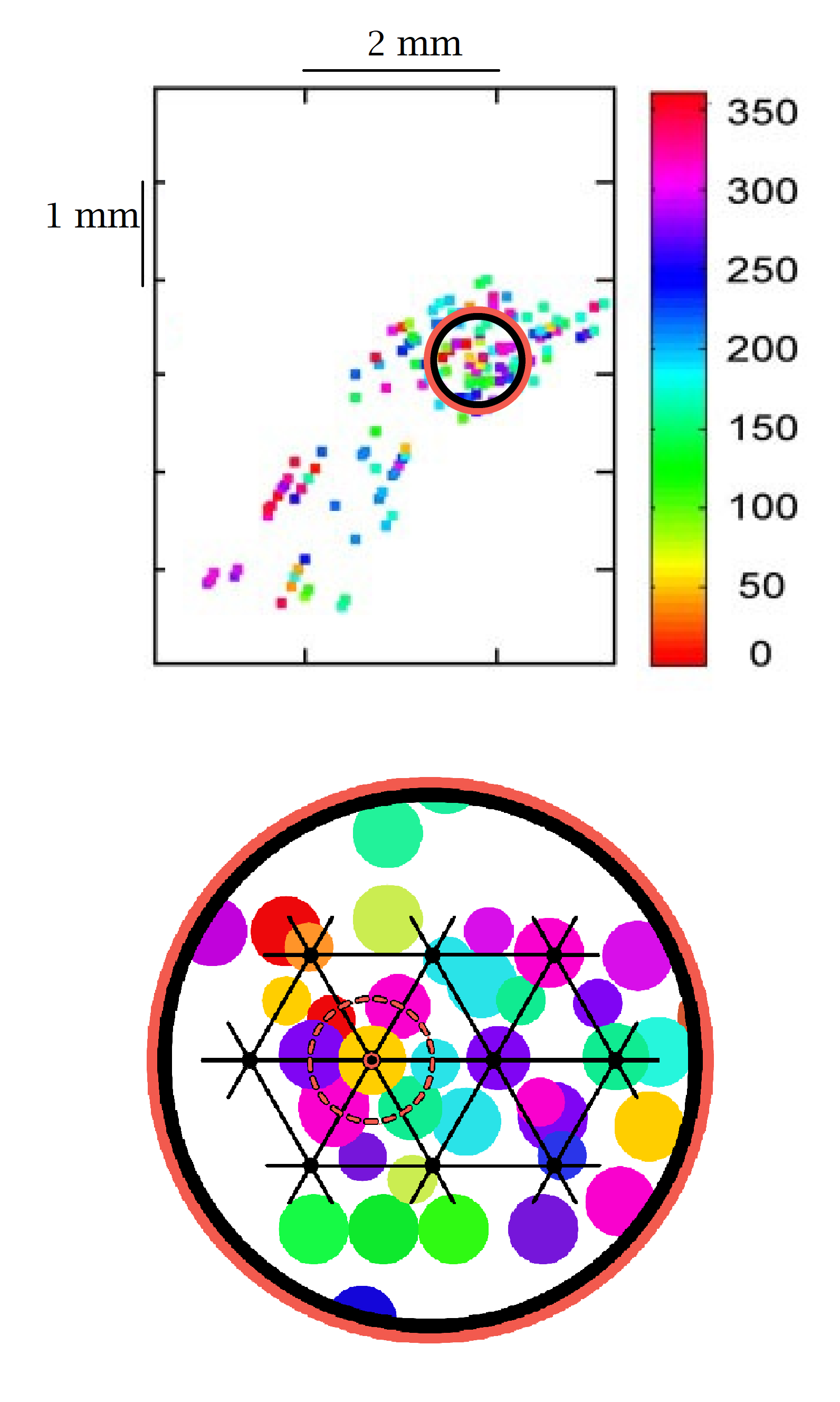}
\quad\quad\quad
\includegraphics[scale=0.45]{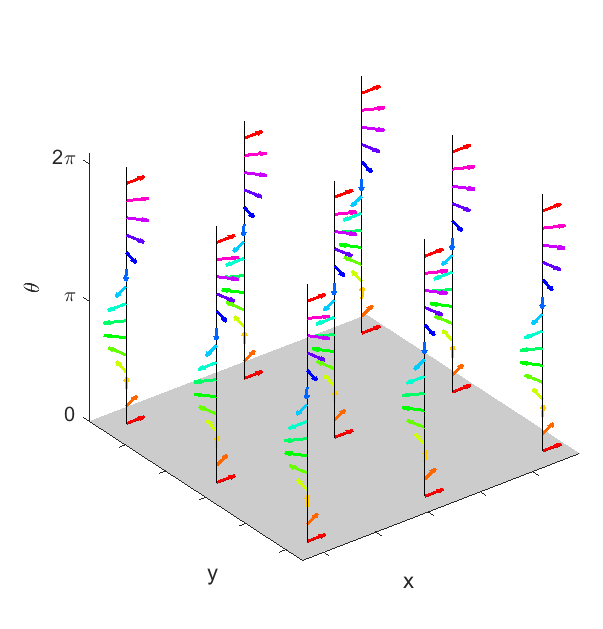}
\caption{(Left) In the up, motor cortical map of preferred directions referred to movements on a two-dimensional space (adapted from \cite{naselaris2006large}). Colors denote preferred directions within the interval $\left[0,2\pi\right]$. A conventional zoom and a superimposition of the lattice model (see Figure \ref{pinwheel_motori}) have been made in order to visualize the directional map referred to the size of the hypercolumns. (Right) Arm area of M1 modelled as a set of hypercolumns. Here, the angle $\theta$ lies in the interval $\left[0, 2\pi\right]$ and it is represented as an arrow.}
\label{model_M1}    
\end{figure}

\begin{itemize}
\item Due to the columnar representation, we identify the fiber $F$ with the set $S^{1}$ of the preferred directions of the cells in the plane. Moreover, as it is represented in Figure \ref{colonne}, cells with similar preferred directions are organized in columns perpendicular to the cortical surface. Directional columns are in turn grouped into hypercolumns (see Figure \ref{pinwheel_motori}), each of them coding for the full range of reaching directions.
\item Here we choose as a basis of the fiber bundle the cortical tuning of the position of the plane. Hence $M\subset \mathbb{R}^{2}$.
There is wide neural literature supporting that M1 neurons encode hand positions (see e.g. \cite{georgopoulos1984static}, \cite{kettner1988primate}, \cite{schwartz2007useful}, as well as sections \ref{georg} and \ref{motor_coding}), but the way that these positions are mapped on the cortical plane is not well understood. Possibly the position of the hand will be indirectly coded through the command to the specific group of muscles which will implement the movement \cite{schwartz1988primate, georgopoulos1988neural}. In particular, according to \cite{graziano2007mapping} (see also \cite{graziano2002cortical}, \cite{aflalo2006possible}, \cite{graziano2008intelligent}) the topographic organization in motor cortex emerge from a competition among three mappings: somatotopic map of the body; a map of hand location in space; a map of movements organization. Since these maps preserve a principle of local similarity, and we are considering here very simple hand movements, a fiber bundle structure in the position-directions is not inconsistent with these data. On the other side, from a functional point of view, it is clear that at every point of the 2D space the hand can move in any direction, and this aspect is captured by a fiber bundle of direction on a 2D spatial bundle. 
\item $E$ is the total tuning space to which motor cortical cells are selective and it is locally described by the product $\mathbb{R}^{2}\times S^1$;
\item $\pi: E\rightarrow M$ is a projection on the $\left(x,y\right)$ variables which acts as $\pi\left(x,y,\theta\right)= \left(x,y\right)$.
\end{itemize}
A section $\sigma: M\rightarrow E$ represents the selection of a point on a fiber of possible movement directions 
at position $\left(x,y\right)\in M$, namely, it associates the point $\left(x,y\right)$ to a point $\left(x,y,\theta\right)= \sigma\left(x,y\right)$. A fiber $E_{\left(x,y\right)}= \pi^{-1}{\left(x,y\right)}\simeq S^{1}$ corresponds to an entire hypercolumn. A schematic representation of the fiber bundle structure is shown in the right side of Figure \ref{model_M1}. 

We recall that formula \eqref{prima_tuning_curve}, which is equivalent of \eqref{encoding_model} for every fixed instant of time, selects the maximum of the scalar product in the direction of the trajectory of movement. 
This is equivalent to say that the spike probability is maximized if the scalar product in the direction orthogonal to that of motion vanishes. For our model it will be essential to consider this and to do so we will make use of the following definition. 
We call 1-form a function $\omega = a_1 dx + a_2 dy$ which acts on a vector $v$ as a scalar product: 
\begin{equation}\label{select}
\omega (v) = \left\langle a, v \right\rangle.
\end{equation}
In our case, to be compatible with equation \eqref{encoding_model}, we will choose $a= v^\perp$, where $v= \left(\cos\theta, \sin\theta\right)$ denotes cell's preferred movement direction. 
As a result $a=(-\sin(\theta), \cos(\theta))$ and 
\begin{equation}\label{origine_di_tutto}
\omega= -\sin\theta\diff x + \cos\theta\diff y.
\end{equation}

\subsubsection{Neuronal population vector and distance}\label{npv_distance}
As we clarified, we are interested in the set of vectors on which the 1-form \eqref{origine_di_tutto} vanishes. This set identifies a two-dimensional subset of the tangent space at every point, called horizontal distribution (see Definition \ref{def_distribuzione}). It can be represented as  
\begin{equation}\label{D_xytheta}
D_{\left(x,y,\theta\right)}= \{\alpha_1 \vec{X}_1 + \alpha_2 \vec{X}_2: \alpha_1, \alpha_2\in\mathbb{R}\},
\end{equation}
where the generators are
\begin{equation}\label{m1_vision_vec}
\vec{X}_1= \left(\cos\theta, \sin\theta, 0\right)\quad , \quad \vec{X}_2= \left(0, 0, 1\right).
\end{equation}
In terms of vector fields, they will be denoted respectively 
\begin{equation}\label{m1_vision}
X_1= \cos\theta\frac{\partial}{\partial x} + \sin\theta\frac{\partial}{\partial y}\quad , \quad X_2= \frac{\partial}{\partial \theta}.
\end{equation}

According to Chow's Theorem (see \cite{Montgomery} and \cite{agrachev2019comprehensive} for a detailed analysis), vector fields \eqref{m1_vision} induce on the space $\mathbb{R}^2\times S^1$ a distance $d$ in terms of Definition \ref{cc_distance_def}.

We saw in section \ref{georg} (see also \cite{georgopoulos1986neuronal, georgopoulos1988spatial,georgopoulos1988primate}) that one estimate concerning the output of a population of M1 cells is given by the neuronal population vector \eqref{NPV_sum}. 
Its formula describes an expectation value weighted by the w-functions with respect to all possible cells preferred directions $\theta'\in S^{1}$. Basically each cell assigns a contribution to the output given by its own preferred direction modulated by the distance between the actual direction of movement and cell's preferred direction itself. As reported in \cite{naselaris2006large}, within each hypercolumn the neuronal population vector ensures a good estimate of a reaching direction. These results suggest that within each hypercolumn of M1 there is a local and isotropic activity pattern characterized by the weight functions. We observe that the weight \eqref{w_g} can locally be approximated through Taylor expansion by
\begin{equation*}
\cos(\theta - \theta') \simeq  1 - \frac{|\theta - \theta'|^2}{2} \simeq e^{-\frac{|\theta - \theta'|^2}{2}}.
\end{equation*} 
Since for small values of $\theta$ the distance in the circumference is $\left|\theta- \theta'\right|$, this suggests approximating the discrete formula \eqref{NPV_sum} with the continuous correspective in which the weight \eqref{w_g} is replaced with the exponential 
\begin{equation}
P\left(\theta\right)= \int_0^{2\pi} e^{i\theta'} e^{-\frac{|\theta - \theta'|^2}{2}} d \theta'.
\end{equation}

This formula can also be exploited in our case. Indeed, if we denote by $d$ the distance induced by vector fields $X_1$ and $X_2$ (see Definition \ref{cc_distance_def}), 
we can provide an estimate of the collective behaviour of cells tuning with respect to a selective point $\left(x,y,\theta\right)$ within a hypercolumn of positions and directions of movement:
\begin{equation}\label{npv_esteso_r2_s1}
P\left(x,y,\theta\right):= \int_{D}\int_{0}^{2\pi} g_{x,y, \theta}\left(x',y', \theta'\right)\omega\left(\left(x,y,\theta\right), \left(x',y',\theta'\right)\right) dx'dy'd\theta',
\end{equation}
where $D\subset\mathbb{R}^2$ is a subset of a cortical module 
and the function $\left(x',y', \theta'\right)\mapsto g_{x,y,\theta}\left(x',y', \theta'\right)$ represents the single cell's spike probability density in response to $\left(x,y,\theta\right)$:
\begin{equation}\label{prob_xytheta}
g_{x,y,\theta}\left(x',y', \theta'\right)= e^{\langle \left(x,y,\theta\right), \left(x',y', \theta'\right) \rangle}.
\end{equation}
The above equation is the analogue of Hatsopoulos model \eqref{encoding_model_esteso} only in relation to the variables of position and direction of movement. 
The new weighting function 
\begin{equation}\label{w_loc_m1_static}
\omega\left(\left(x,y,\theta\right), \left(x',y',\theta'\right)\right)= e^{-\frac{d^2\left(\left(x,y,\theta\right),\left(x',y',\theta'\right)\right)}{2}}
\end{equation}
measures the closeness between the cellular selectivity of the points $\left(x,y,\theta\right)$ and $\left(x',y',\theta'\right)$. Therefore, formula \eqref{w_loc_m1_static} is intended to express a core of connectivity (local, since it is evaluated in a small cortical module) that is functional. 
We will see that the analogous for this area of formula \eqref{w_loc_m1_static} is provided by Bressloff-Cowan \cite{bressloff2003functional} and Sarti-Citti \cite{sarti2015constitution} models in the visual cortex.  
 
\subsubsection{Comparison of the static model with primary visual cortex V1}\label{comp_v1}

\begin{figure}[htbp]
\centering
\includegraphics[scale=0.24]
{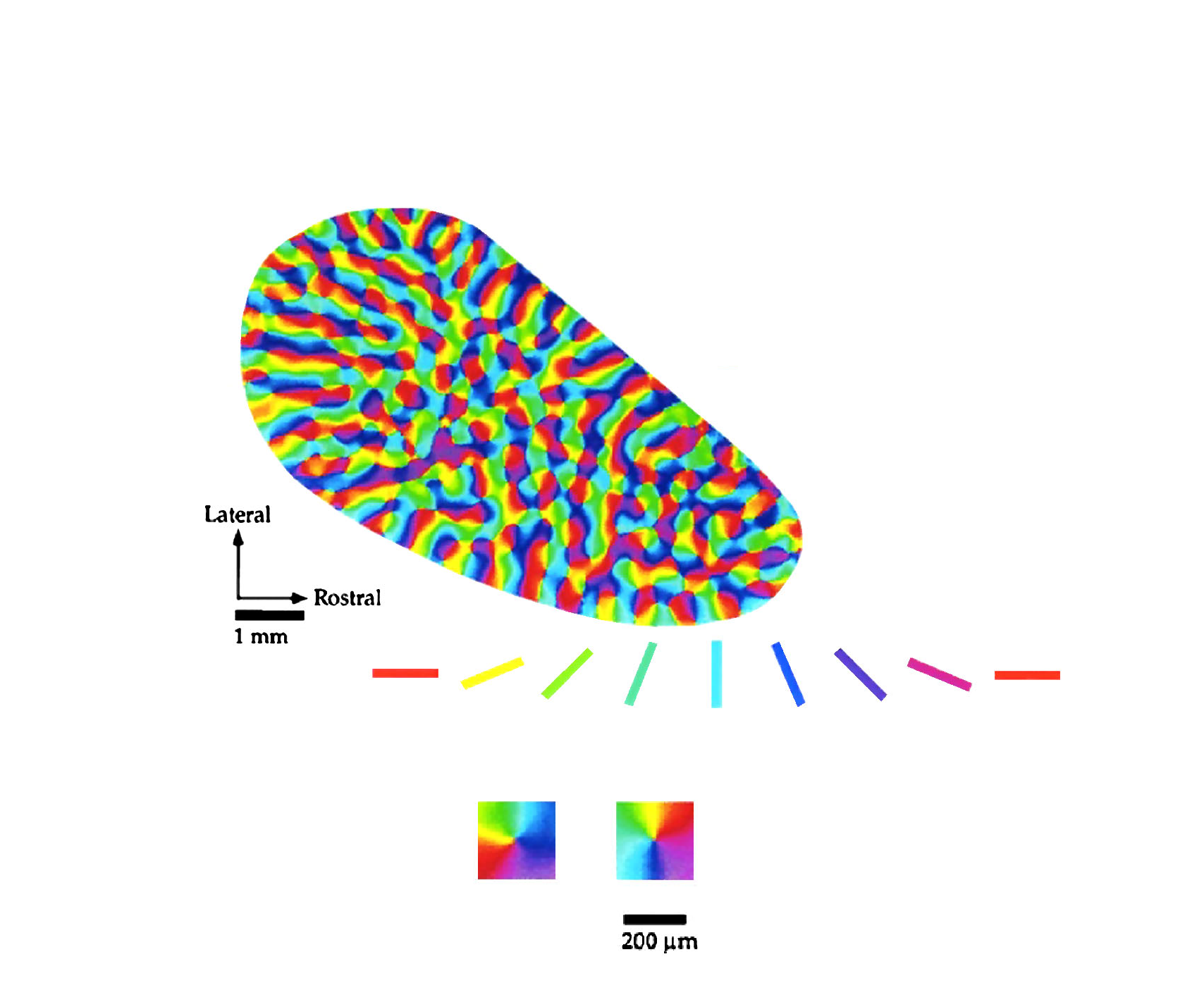}
\includegraphics[scale=0.4]{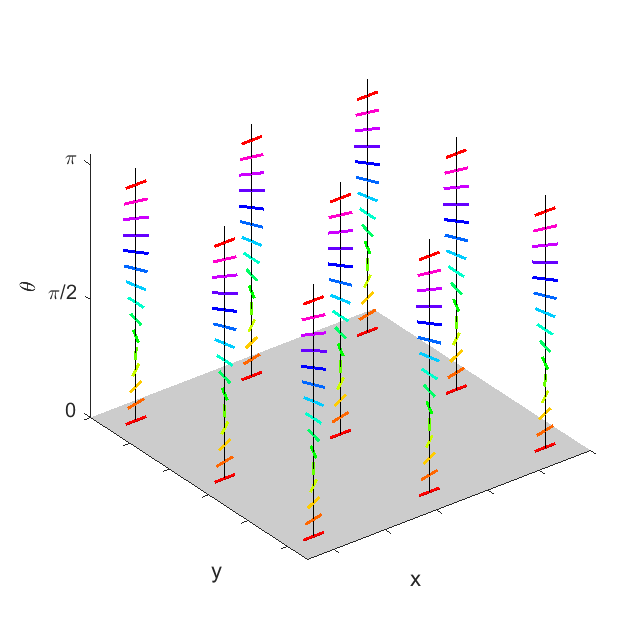}
\caption{(Left) Layout of orientation preferences in the visual cortex. At singular points (pinwheels), all orientations meet. Source: \cite{bosking1997orientation}. (Right) V1 modelled as a set of hypercolumns.}
\label{vision_picture}    
\end{figure}

An analogy on the selectivity behaviour of external features with neurons in the primary visual cortex area (V1) is evident. 
Hubel and Wiesel \cite{hubelandt1977functionalarchitectureofmacaquemonkeyvisual,hubel1995eye} discovered that to every retinal position is associated an hypercolumn of cells sensible to all possible orientations. A simplified representation is shown in the right side of Figure \ref{vision_picture}. 
Since the early `70s, a large number of
differential models were developed for visual cortex areas, starting with  Hoffmann \cite{hoffman1970higher}, Petitot and Tondut \cite{petitot1999vers}, Bressloff and Cowan \cite{bressloff2003functional}, Citti and Sarti \cite{citti2006cortical}, just to name a few of the main ones. Their models describe the functional architecture of V1 trough geometric frameworks such as contact bundles, jet bundles or Lie groups endowed with a sub-Riemannian metric.

In particular, the model expressed through the one-form \eqref{origine_di_tutto} 
matches
the one proposed by Citti-Sarti in 2006 \cite{citti2006cortical} for the description of image edge selectivity by V1 cells. Both for V1 and the arm area of M1, the total space of the fiber bundle is three-dimensional, whereas the cortical layers are of dimension two, so that a dimensional constraint has to be taken into account. 
In Figure \ref{vision_picture}a, the orientation preferences of simple cells in V1 are color coded and every hypercolumn is represented by a pinwheel. For the motor cortex, a ``directional map'' is suggested from Figure \ref{pinwheel_motori}, 
for which PDs are repeatedly arranged on the motor cortical layer in such a way that, within a given locale (hypercolumn), the full range of movement directions are represented. Moreover, as distance increases away from the center of the hypercolumn (black filled circle), up to the radius of the hypercolumn (120 $\mu$m), PDs diverge from that at the center of the circle (see Figures \ref{vision_picture} and \ref{model_M1} for a direct comparison).
We add 
that cells preferred directions in M1 are correlated across very small distances along the tangential dimension (see \cite{amirikian2003modular}) and this type of arrangement is consistent with the smooth variation of orientation preference observed in V1 (see \cite{hubel1963shape}). 
Moreover, the radii of the hypercolumns of arm area M1 and V1 are of the same order size and are respectively of 240 and 200$\mu$m.
 
One of the greatest difficulties and differences in modelling the functional architecture of M1 is the absence of an analogue of the simple cells receptive profiles, which we might call ``actuator profiles". Simple cells of visual areas are indeed identified by their receptive field (RF) which is the domain, subset of the retinal plane,  to which each cell is sensible in response to a visual stimulus. Activation of a cell's RF evokes the impulse response, which is called the receptive profile (RP) of the cell. A widely used model (\cite{jones1987evaluation}, \cite{daugman1985uncertainty}, \cite{lee1996image}) for the RP representation of a simple cell located at the retinal position $q$ and selective to the feature $p$, is in terms of Gabor filters $\psi_{\left(q,p\right)}: \mathbb{R}^2\rightarrow \mathbb{C}$, 
\begin{equation}\label{RP_2D}
\psi_{\left(q,p\right)}\left(x\right)= e^{ip\cdot x}e^{-\left(x-q\right)^2}.
\end{equation}
Although it is not well understood the presence or the definition of such functions for M1, we argue that the action of primary motor cortical cells occurs in a comparable way as in V1. This hypothesis is primarily supported by the tuning functions \eqref{prima_tuning_curve} and \eqref{positional_gradient_tc} expressed by Georgopoulos (see \ref{georg} and \cite{georgopoulos1982relations, schwartz1988primate}) and through the trajectory encoding model \eqref{encoding_model_esteso} provided by Hatsopoulos (see \ref{motor_coding} and \cite{Encoding}). Their models share the selective tuning of neurons by evaluating the alignment between an external input variable and the individual cell's preferred feature via a scalar product.
Analogously for V1, the linear term in \eqref{RP_2D} evaluates the aligning between cell's selective feature $p$ with respect to the input $x$.
Bressloff and Cowan \cite{bressloff2003functional} (see also \cite{bressloff2002geometric, wilson1973mathematical}) proposed to represent the stationary state of a population of simple cells through equation 
\begin{equation}\label{B-C}
a\left(\phi,r\right)= \frac{\mu}{\alpha\pi}\int_{0}^{\pi} \omega_{\text{LOC}}\left(\phi, \phi'\right)\sigma\left(a\left(\phi,r\right)\right)\diff \phi' + \frac{\mu\beta}{\alpha}\int_{\mathbb{R}}\omega_{\text{LAT}}\left(s\right)\sigma\left(a\left(\phi,r+ se_{\phi}\right)\right)\diff s,
\end{equation}
where $\omega_{\text{LOC}}$ and $\omega_{\text{LAT}}$ correspond to the strength of connections from the iso-orientation patch within and between the hypercolumns of V1, respectively. The function $a\left(\phi,r\right)$ is the activity in an iso-orientation patch at the point $r$ with orientation preference $\phi$, whereas $\sigma\left(a\right)$ is a smooth sigmoidal function of the activity $a$ and $\alpha$, $\mu$ are time and coupling constants. The weight $\omega_{\text{LOC}}$ represents the isotropic pattern of activity within any one hypercolumn,  
and in \cite{bressloff2003functional}, in the simplified case where the spatial frequencies are not taken into account, it is modelled by 
\begin{equation}\label{w_loc_cos_BC}
\omega_{\text{LOC}}^{\phi'}\left(\phi\right)= \frac{1}{2}\left(\cos\left(\phi- \phi'\right)+ \cos\left(\phi+\phi'\right)\right), \quad \phi,\phi'\in S^1.
\end{equation}
Equation \eqref{w_loc_cos_BC} 
is the analogous of the weighting function \eqref{w_g} assumed by Georgopoulos. In the work of Sarti and Citti \cite{sarti2015constitution} both contributions of $\omega_{LOC}$ and $\omega_{LAT}$ are modelled by means of a single connectivity kernel given by 
\begin{equation}\label{sarti_nucleo}
\omega\left(\left(\phi, r\right), \left(\phi', r'\right)\right)= e^{-d_c^2\left(\left(\phi, r\right), \left(\phi', r'\right)\right)},\quad \left(\phi, r\right), \left(\phi', r'\right)\in SE\left(2\right),
\end{equation}
where $d_c$ is a Carnot Carathéodory distance associated 
to the cortical feature space identified with the special Euclidean group $SE\left(2\right)\simeq\mathbb{R}^2\times S^1$. We will see in a higher dimensional model that a connectivity kernel of this type can be used to interpret the segmentation of neural activity in neural states (section \ref{data_analysis}).

\subsection{A 2D kinematic tuning model of movement directions}\label{cortical_features_structure}
Now we aim at realizing a unified neurogeometrical framework that generalizes the preceding model of movement fiber bundle structures.

We will describe a sub-Riemannian model such that motor cortical cells selective behaviour can be represented through integral curves of the cortical features space of time, position, direction of movement, speed and acceleration. The resulting model will present time dependent variables, which seems particularly natural for a model of movement.

\subsubsection{Integral curves and time dependent PD}\label{2.3.1}  

We represent motor cortical cell tuning variables by the triple $\left(t, x, y\right)\in\mathbb{R}^3$, which accounts for a specific hand's position in time. We also consider the variable $\theta\in S^1$ which encodes hand's movement direction, and the variables $v$ and $a$ which represent hand's speed and acceleration along $\theta$. 
The triple $\left(t, x, y\right)\in\mathbb{R}^3$ is assumed to belong to the base space of the new fiber bundle structure, whereas the variables 
$\left(\theta, v, a\right)\in S^1 \times \mathbb{R}^{2}$ form the selected features on the fiber over the point $\left(t, x, y\right)$. We therefore consider the 6D features set 
\begin{equation}\label{6D}
\mathcal{M}= \mathbb{R}^{3}_{\left(t,x,y\right)} \times S^1_{\theta} \times \mathbb{R}^{2}_{\left(v,a\right)},
\end{equation}
where this time the couple $\left(x,y\right)\in\mathbb{R}^2$ represents the cortical tuning for hand's position in a two dimensional space.

We refer to equation \eqref{encoding_model_esteso} to recall that the spike probability of a neuron is maximized in the direction of the movement fragment.  Therefore, as in section \ref{fiber_pos_dir}, the choice of the variables with their differential constraints induce the vanishing of the following 1-forms
\begin{align}\label{tre_1forme}
\omega_{1} = \cos\theta \diff x + \sin\theta \diff y - v\diff t= 0,\quad
\omega_{2} = -\sin\theta \diff x + \cos\theta \diff y= 0,\quad
\omega_{3} = \diff v -a\diff t= 0.
\end{align}
The one-form $\omega_{1}$ encodes the direction of velocity over time: the unitary vector $\left(\cos\theta, \sin\theta\right)$ is the vector in the direction of velocity, and its product with $\left(\dot{x}, \dot{y}\right)$ yields the speed.
As we already noted, conditions above are equivalent to find vector fields orthogonal to $\omega_i$. 
Consequently, the associated horizontal distribution $D^{\mathcal{M}}$ turns out to be spanned by the vector fields
\begin{align}\label{campi_2D}
X_{1}= v\cos\theta\frac{\partial}{\partial{x}} + v\sin\theta\frac{\partial}{\partial{y}}+ a\frac{\partial}{\partial{v}}+ \frac{\partial}{\partial{t}},\quad
X_{2}= \frac{\partial}{\partial{\theta}},\quad
X_{3}= \frac{\partial}{\partial{a}}.
\end{align}

Note that, if we prescribe time, movement direction and acceleration, we can deduce by integration first speed and then location. This is the reason why we  prescribe 
 only these 3 vector fields: $X_1$ prescribes the change in time, $X_2$ in the direction of movement and $X_3$ in the acceleration. In addition, 
not all curves are physically meaningful in this space: it is not possible for a curve to change its velocity $v=v(t)$ while the position $(x,y)$ remains constant. Hence we have to restrict the set of admissible curves and define horizontal ones.
Horizontal curves of the space are integral curves of the vector fields $X_1, X_2$ and $X_3$ and are of the form 

\begin{equation}\label{curvaa}
\gamma'\left(s\right)= \alpha_1\left(s\right)X_{1}\left(\gamma\left(s\right)\right)+ \alpha_2\left(s\right)X_{2}\left(\gamma\left(s\right)\right)+ \alpha_3\left(s\right)X_{3}\left(\gamma\left(s\right)\right),\\
\end{equation}
where the coefficients $\alpha_i$ are not necessarily constants.

We recalled in section \ref{motor_coding} that M1 cells are not selective to a single movement direction, but 
the preferred movement direction varies in time
\cite{churchland2007temporal, Encoding}. In particular, in Figure \ref{im_temporal_beh}b we reproduced data from \cite{churchland2007temporal}, where the PD of a single M1 neuron was represented as a curve dependent on time. 

We propose the curves expressed in \eqref{curvaa} as a model of the integrated selective behaviour of M1 neurons. 
Note in particular that the $t$ component of the horizontal curve $\gamma$ satisfies $t' = \alpha_1$. 
This means that the coefficient $\alpha_1$ is a modulation of the time, and can account for the difference between the external time and the perceived one. By simplicity, we will assume that the two times coincide, therefore we will assume $\alpha_1=1$ from now on. 
In the sequel we will see that it is possible to choose coefficients in equation \eqref{curvaa} which allow to recover the full fan of curves reproduced in Figure \ref{im_temporal_beh}. The expression of $\dot{\theta}$ described in \eqref{dotteta} is the coefficient of the vector field $X_2$ and the expression of $\dot{a}$ described in \eqref{a_punto_pari} is the coefficient of the vector field $X_3$:
\begin{equation}\label{curve_orizzontali_qui}
\dot{\gamma}\left(t\right)= X_{1}\left(\gamma\left(t\right)\right)+ \dot{\theta}\left(t\right) X_{2}\left(\gamma\left(t\right)\right)+ \dot{a}\left(t\right)X_{3}\left(\gamma\left(t\right)\right).
\end{equation}
The functions $t\mapsto \dot{\theta}\left(t\right)$ and $t\mapsto \dot{a}\left(t\right)$ represent, respectively, the rate of change of the selective tuning to movement direction and acceleration variables.
\subsubsection{Time-dependent neuronal population vector}\label{time_npv}
By analyzing the following commutation relations 

\begin{equation}\label{commutatori}
\begin{aligned}
\left[X_{1}, X_{2}\right]= v\sin\theta\frac{\partial}{\partial{x}}&- v\cos\theta\frac{\partial}{\partial{y}}=: X_{4},\quad\quad
\left[X_{3}, X_{1}\right]= \frac{\partial}{\partial{v}}=: X_{5},\\
&\left[X_{5}, X_{1}\right]= \cos\theta\frac{\partial}{\partial{x}}+ \sin\theta\frac{\partial}{\partial{y}}=: X_{6},
\end{aligned}
\end{equation}
we observe that $\left(X_i\right)_{i=1}^{6}$ are linearly independent. 
Therefore, all $\left(X_i\right)_{i=1}^{3}$ belonging to $D^{\mathcal{M}}$ together with their commutators span the whole tangent space at every point, meaning that H\"{o}rmander condition is fulfilled (see Appendix \ref{sub} for a brief review).
Thanks to H\"{o}rmander condition, it is possible to define a metric $d_{\mathcal{M}}$ in the cortical feature space $\mathcal{M}$. 
This allow to formally consider the analogous of the population vector \eqref{npv_esteso_r2_s1} defined in section \ref{npv_distance} (see as well Definition \ref{cc_distance_def}). Indeed, we will call as time dependent neural population vector an estimate of the collective behaviour of cells tuning around a cortical module centered at point $\eta_0\in\mathcal{M}$. We define it by means of the following 

\begin{equation}\label{npv_esteso_M}
P_{\mathcal{M}}\left(\eta_0\right):= \int_{E}h_{\eta_0}\left(\eta\right)\omega_{\mathcal{M}}\left(\eta_0,\eta\right) \diff\eta,
\end{equation}
where $E\subset\mathcal{M}$ is a neighbourhood of $\eta_0$ and the weighting function  
\begin{equation}\label{kernel_m}
\omega_{\mathcal{M}}\left(\eta_0,\eta\right)= e^{- d_{\mathcal{M}}\left(\eta_0,\eta\right)^2}
\end{equation}
encodes an estimate of the local connectivity between the cortical tuning points $\eta_0$ and $\eta$. The function $\eta\mapsto h_{\eta_0}\left(\eta\right)$ corresponds to the contribution provided by the variable $\eta$ in the population coding. As in Hatsopoulos model \eqref{encoding_model_esteso}, it is the spike probability of a neuron in response to the input variable $\eta_0$: 
\begin{equation}\label{mah}
h_{\eta_0}\left(\eta\right)= e^{\langle \eta_0, \eta \rangle}.
\end{equation} 
The definition of \eqref{kernel_m} embodies the same meaning as the weighting function \eqref{w_loc_m1_static} showed in the ``static" model and in the models for visual areas (see equations \eqref{w_loc_cos_BC}, \eqref{sarti_nucleo} in \ref{comp_v1} defined in \cite{bressloff2003functional} and  \cite{sarti2015constitution}). It represents the local interactions between cells within a cortical module by means of a distance of the cortical feature space. 

Due to the works of Nagel et al. \cite{nagel1985balls} and Montgomery \cite{Montgomery}, it is possible to provide a local approximation of distance $d_{\mathcal{M}}$ in terms of a homogeneous distance (see Definition \ref{def_distanza_hom}), as follows
\begin{equation}\label{estim_hom_npv}
d_{\mathcal{M}}\left(\eta_0,\eta\right)\simeq \left(\left|c_1 e_{1}\right|^{6} + \left|c_2e_{2}\right|^{6}+ \left|c_3e_{3}\right|^{6}+ \left|c_4e_{4}\right|^3+ \left|c_5e_{5}\right|^3+ \left|c_6e_{6}\right|^2 \right)^{\frac{1}{6}},
\end{equation}
where 
$\left(\eta_0,\eta\right)= \left(\left(t_0,x_0,y_0,\theta_0, v_0, a_0, \right),\left(t, x,y,\theta, v, a, \right)\right)\in\mathcal{M}^2$, $c_i$ are non negative constant coefficients, the number $6$ is the dimension of the space $\mathcal{M}$ and the increments $e_i$ are given by  
\begin{align}\label{e125}
e_1= t- t_0,\quad
e_2= \theta- \theta_0,\quad
e_5= \left(v- v_0\right)- \frac{t- t_0}{2}\left(a+ a_0\right)
\end{align}
and $e_4, e_6$ are solutions of system
\begin{align}
\begin{cases}
\dot{x}\left(s\right)&=  \left(e_{1}v\left(s\right)+ e_{6}\right)\cos\theta\left(s\right) + e_{4}v\sin\theta\left(s\right)\\
\dot{y}\left(s\right)&=  \left(e_{2}v\left(s\right)+ e_{6}\right)\sin\theta\left(s\right) - e_{4}v\cos\theta\left(s\right)\\
x\left(0\right)&= x_0,\quad y\left(0\right)= y_0\\
x\left(1\right)&= x_1,\quad y\left(1\right)= y_1,
\end{cases}
\end{align}
which in the linear case, for constant values of $\theta$, can be solved as
\begin{align}
e_4&= \frac{12\left(\left(x- x_0\right)\sin\theta- \left(y- y_0\right)\cos\theta\right)}{6\left(v_0+ v\right)- \left(t- t_0\right)\left(a- a_0\right) }, \\
e_6&= \left(x- x_0\right)\cos\theta + \left(y- y_0\right)\sin\theta- \frac{t- t_0}{12}\left(6\left(v_0+ v\right)- \left(t- t_0\right)\left(a- a_0\right)\right).
\end{align}
We report a proof of distance \eqref{estim_hom_npv} and of individual increments $e_i$ in Appendix \ref{app_B} (see in particular Remark \ref{contiremark}).

Finally we are led to the estimate of the increment $e_3$. We first note that 
\begin{equation*}
e_3\simeq a- a_0\quad\text{as}\quad a\rightarrow a_0. 
\end{equation*}
In addition, the neural states do not show a consistent selectivity to the acceleration amplitude, but only at its sign. For this reason we choose the increment $e_3$ as follows 
$$e_3= \arctan\left(\frac{v- v_0}{t- t_0}\right)\simeq \text{sgn}\left(\arctan\left(a- a_0\right)\right).$$ 

\section{Parameters fitting and numerical results}\label{result}

\subsection{Time dependent direction selectivity as local integral curves}\label{modellino?}

In this section, we discuss the structure of our model, identifying the coefficients of the integral curves proposed in \eqref{curve_orizzontali_qui} as suitable polynomials. This allows to describe the proposed set of curves as a space of finite dimension. Precisely it has the same dimension as the space of curves identified by \cite{churchland2007temporal}, and generate a fan, which is qualitatively  comparable with the experimentally discovered. After that, we perform a quantitative fitting between the modelled and measured curves, representing time dependent PDs.
The main analysis here is not to show that polynomial models can fit the average curves shown in \cite{churchland2007temporal}. Rather, the issue is to identify a set of parameters of the same dimension as the one found with a neural analysis, which accurately captures the same patterns of acceleration and movement direction in the data.

\subsubsection{The dimension of the space of parameters}
Our model is expressed as a family of curves defined in \eqref{curve_orizzontali_qui}. They only depend on the initial data, the interval where they are defined and the derivatives of the direction and acceleration.

In \cite{Encoding, churchland2007temporal}, given a temporal interval $\Delta T$, $m$ observations have been made at instants of time $t_1, \cdots, t_m$ and the measured preferred directions can be denoted by
\begin{equation}\label{tetatime}
\Big(\theta(t_1), \cdots, \theta(t_m)\Big). 
\end{equation}
We can represent \eqref{tetatime} through a continuous graph by representing the variable $\theta$ as a function of time 
(as it is shown in \cite{churchland2007temporal}, see Figure \ref{im_temporal_beh}b): 
\begin{equation}\label{theta_t}
\theta:[-T, T] \to S^1.
\end{equation}

This is equivalent of assuming that at every instant of time the preferred movement directions are respectively described by the unitary vectors
\begin{equation}\label{cos_theta_t}
\Big(\big(\cos(\theta(t_1)),  \sin(\theta(t_1)\big),\cdots,  \big(\cos(\theta(t_m)), \sin(\theta(t_m))\big)\Big),
\end{equation}
as it is visualized in Figure \ref{im_temporal_beh}a, from \cite{Encoding}. 

As for the temporal behaviour of the selective tuning of a single cell's PD, we observe that it can be linear, as it is shown for example by the black curve of Figure \ref{im_temporal_beh}b. 
We also see the presence of even curves, symmetric with respect to the interval where they are defined (as for example the dark green one). These will be described as polynomials of order two (or four).
In Figure \ref{im_temporal_beh}b we also see the presence of odd curves changing concavity, which will be represented as polynomials of order 3 or 5. In this way, polynomials of degree, up to order five, provide good models of the PD curves experimentally measured.

As a consequence, $\dot{\theta}\left(t\right)$ has to be a polynomial of degree up to 4 and there exist parameters $k_i$ such that 
\begin{equation}\label{dotteta}
\dot{\theta}\left(t\right) = k_0+ k_1 t+ k_2 t^2 + k_3 t^3+ k_4 t^4 ,
\end{equation}
and that all curves in the variable $\theta$ satisfying \eqref{curve_orizzontali_qui} will be characterized by these five parameters. 
The other coefficient which defines the model is the derivative of the acceleration. 
In the top of Figure \ref{im_temporal_beh}b the mean of the measured hand's speed profile is shown. It presents the typical bell-shaped trend already observed by  Morasso (\cite{morasso1981spatial}, \cite{Hatf}) and Flash and Hogan (\cite{FH}). 
Hence, we posit to characterize the function $t\mapsto v(t)$ by an even polynomial of fourth order. As a consequence, we will characterize $a$ as a polynomial of degree 3 and its  speed $\dot{a}$ as a polynomial of degree 2. It will be identified by 3 parameters $j_i$ which identify the map $t\mapsto \dot{a}\left(t\right)$ in this way
\begin{equation}\label{a_punto_pari}
\dot{a}\left(t\right)= j_0+ j_1 t + j_2t^2.
\end{equation}

In this model, only curves with polynomial coefficients are considered, so that each of these trajectories can also be identified as a point in a higher dimensional space. 
Since the initial instant of time $t_0=0$ and the initial position $(x_0, y_0)$ are fixed in all the experiment, the initial variable $\eta_0\in\mathcal{M}$ has only 3 free parameters. Then the problem depends on the time interval $T$, the 5 coefficients which define the trajectory of the preferred direction $\theta$, and the 3 coefficients for $\dot a$. In this way they form a space of dimension twelve. In particular, the dimension of the structure is of the same order as the one outlined by the principal components analysis performed in \cite{churchland2007temporal} and in \cite{Encoding}.

\subsubsection{A polynomial fitting}
Let us remark that this is not the only possible choice of parameters which allows to build a space of curves of dimension 12. We will here verify that the parameters we have chosen provide a good approximation for recovering both the profile of acceleration and of preferred direction depicted in Figure \ref{im_temporal_beh}b. Precisely, we will determine the optimal polynomial approximation of the coefficient of the curve proposed by our model through the least squares method \cite{bjorck1990least}. 

Each time dependent PDs of neurons has been visualized in \cite{churchland2007temporal} Figure 13 (here reproduced in Figure \ref{im_temporal_beh}b) as a curve surrounded by two curves of the same color. These lateral curves plot $95\%$ confidence interval and lines are suppressed when that interval was grater than $90$ degrees. Moreover, gray vertical lines mark an interval within which data are most reliable, defined between 70 ms before and after the time of strongest tuning. For this reason, we will focus on approximating the time-dependent PDs in this range. 

We propose in \eqref{dotteta} to estimate the expression of $\theta$ by polynomials up to order five. 
Table \ref{Tabel_1} summarizes the results obtained by applying the least-squares algorithm through polynomials of order two, three, four and five. In each case we compute the coefficient of determination (R$^2$) and the normalized root mean squared error (Nrmse) which provide an estimate of the standard deviation of the random component in the data. These coefficients are displayed for each single curves and on average.

\begin{table}[htbp]
\centering
\begin{tabular}{lllllllll}
\multicolumn{1}{c|}{}                & \multicolumn{4}{c|}{R$^2$}                                                                              & \multicolumn{4}{c}{Nrmse}                                                                          \\ \hline
\multicolumn{1}{l|}{Pol. Order}      & \multicolumn{1}{c}{II} & \multicolumn{1}{c}{III} & \multicolumn{1}{c}{IV} & \multicolumn{1}{c|}{V}      & \multicolumn{1}{c}{II} & \multicolumn{1}{c}{III} & \multicolumn{1}{c}{IV} & \multicolumn{1}{c}{V} \\ \hline
\multicolumn{1}{l|}{(a) Black}       & 0.9981                 & 0.9993                  & 0.9994                 & \multicolumn{1}{l|}{0.9999} & 0.01314 & 0.00764                 & 0.007344               & 0.003575              \\
\multicolumn{1}{l|}{(b) Blue}        & 0.9758                 & 0.9976                  & 0.9979                 & \multicolumn{1}{l|}{0.9996} & 0.04452                & 0.01405                 & 0.01311                & 0.005884              \\
\multicolumn{1}{l|}{(c) Magenta}     & 0.9751                 & 0.9949                  & 0.9989                 & \multicolumn{1}{l|}{0.9995} & 0.03812                & 0.01724                 & 0.00811                & 0.005603              \\
\multicolumn{1}{l|}{(d) Olive green} & 0.9992                 & 0.9995                  & 0.9997                 & \multicolumn{1}{l|}{0.9997} & 0.008106               & 0.006355                & 0.005332               & 0.004825              \\
\multicolumn{1}{l|}{(e) Rose}        & 0.9668                 & 0.9803                  & 0.9921                 & \multicolumn{1}{l|}{0.9940} & 0.05274                & 0.04064                 & 0.02574                & 0.02237               \\
\multicolumn{1}{l|}{(f) Red}         & 0.9431                 & 0.9970                  & 0.9976                 & \multicolumn{1}{l|}{0.9995} & 0.05065                & 0.0117                  & 0.01037                & 0.004712              \\
\multicolumn{1}{l|}{(g) Bordeaux}    & 0.7973                 & 0.9967                  & 0.9986                 & \multicolumn{1}{l|}{0.9992} & 0.0813                 & 0.01035                 & 0.006722               & 0.005274              \\
\multicolumn{1}{l|}{(h) Green}       & 0.9847                 & 0.9993                  & 0.9997                 & \multicolumn{1}{l|}{0.9998} & 0.03014                & 0.006389                & 0.003872               & 0.003854              \\
\multicolumn{1}{l|}{(i) Dark Green}  & 0.8496                 & 0.8631                  & 0.9936                 & \multicolumn{1}{l|}{0.9936} & 0.1332                 & 0.1272                  & 0.02764                & 0.02767               \\
\multicolumn{1}{l|}{(j) Violet}      & 0.9840                 & 0.9904                  & 0.9945                 & \multicolumn{1}{l|}{0.9955} & 0.04008                & 0.03097                 & 0.02346                & 0.02141               \\
\multicolumn{1}{l|}{(k) Yellow}      & 0.9930                 & 0.9954                  & 0.9981                 & \multicolumn{1}{l|}{0.9994} & 0.0247                 & 0.01998                 & 0.01305                & 0.007396              \\ \hline
\multicolumn{1}{l|}{Means}           & 0.9515                 & 0.9831                  & 0.9973                 & \multicolumn{1}{c|}{0.9981} & 0.04698                & 0.0266                  & 0.01316                & 0.01023               \\ \hline
                                     &                        &                         &                        &                             &                        &                         &                        &                       \\
                                     &                        &                         &                        &                             &                        &                         &                        &                       \\
                                     &                        &                         &                        &                             &                        &                         &                        &                      
\end{tabular}
\caption{Results of the polynomial approximations through the least squares method. For each curve shown in Figure \ref{oddcurves}, the values of the R$^2$ coefficient and the normalized root mean squared error for polynomials of order two, three, four, five are given. In the last row, the averages of the values of R$^2$ and nrmse are provided.}
\label{Tabel_1}
\end{table}

\begin{figure}[htbp]
\centering
\subfloat[Black]{\includegraphics[height=4cm]{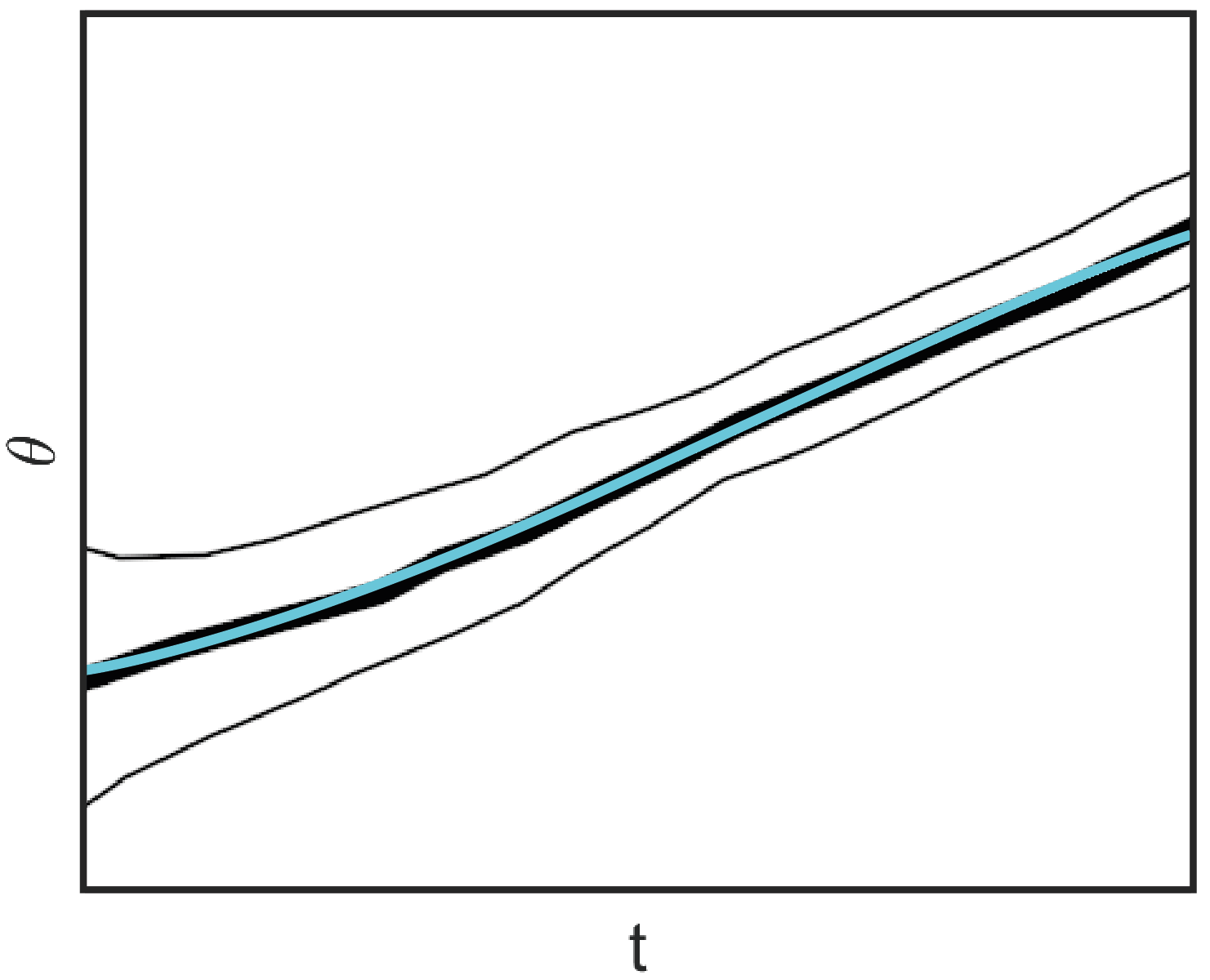}}\quad
\subfloat[Blue]{\includegraphics[height=4cm]{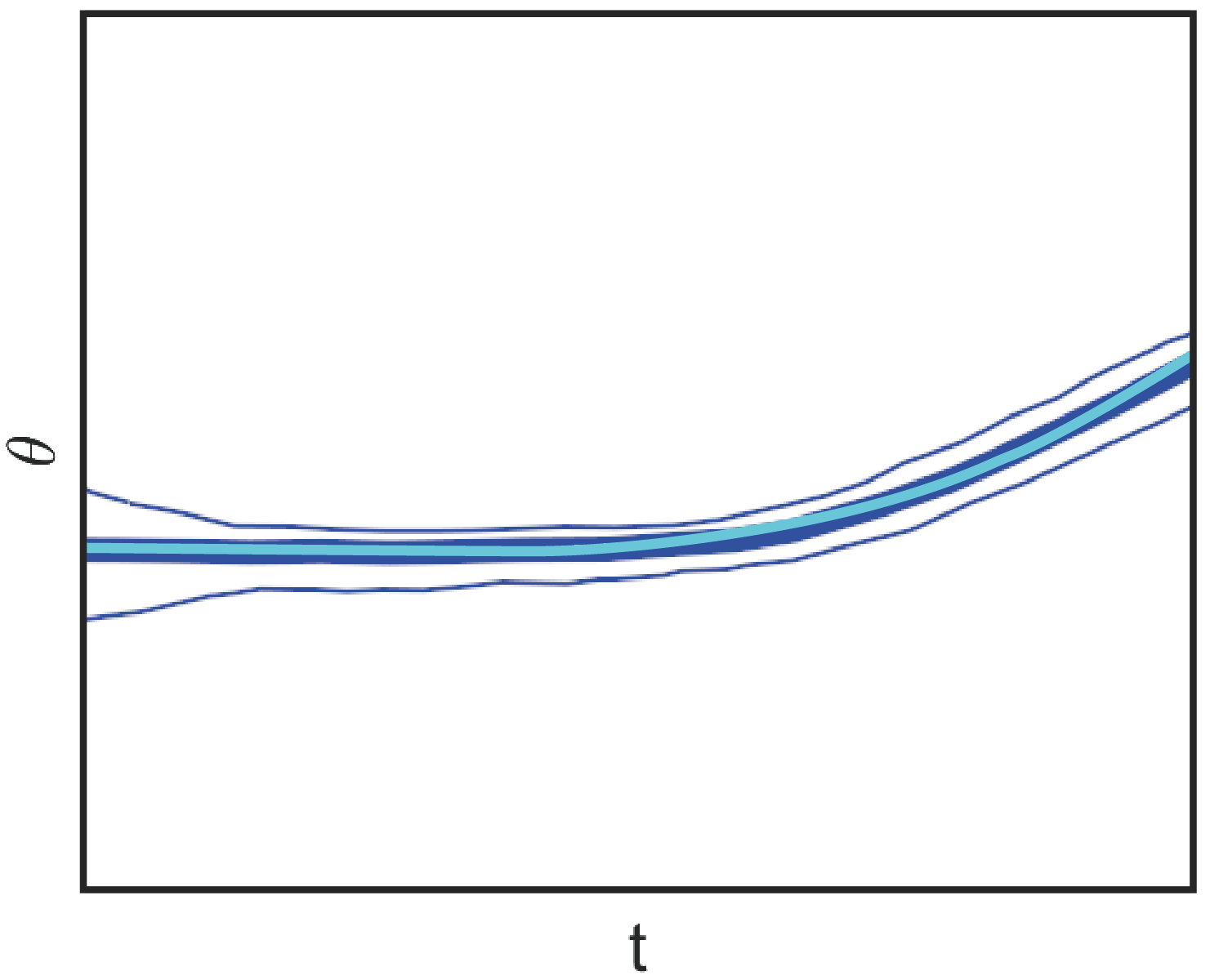}}\quad
\subfloat[Magenta]{\includegraphics[height=4cm]{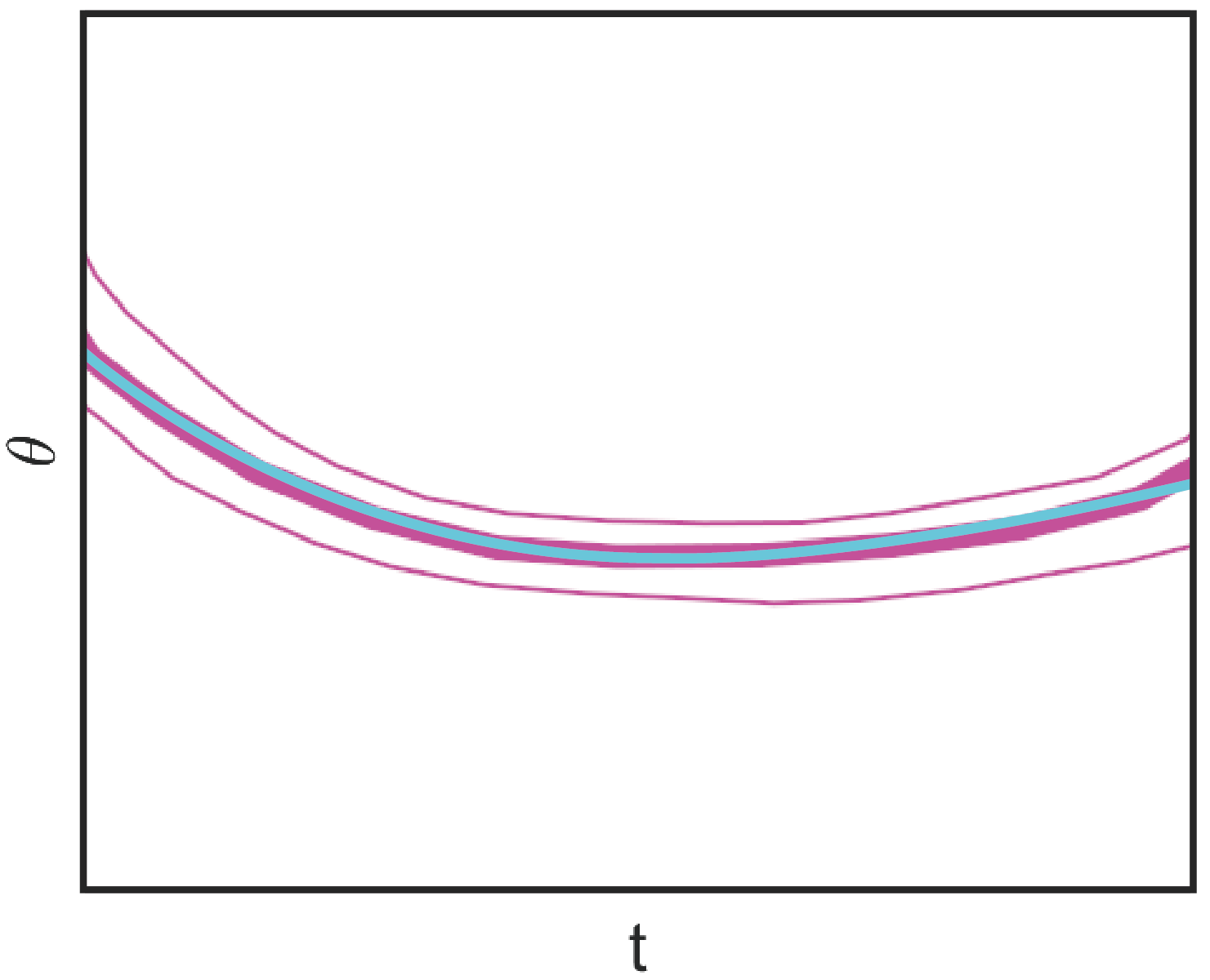}}\\
\subfloat[Olive green]{\includegraphics[height=4cm]{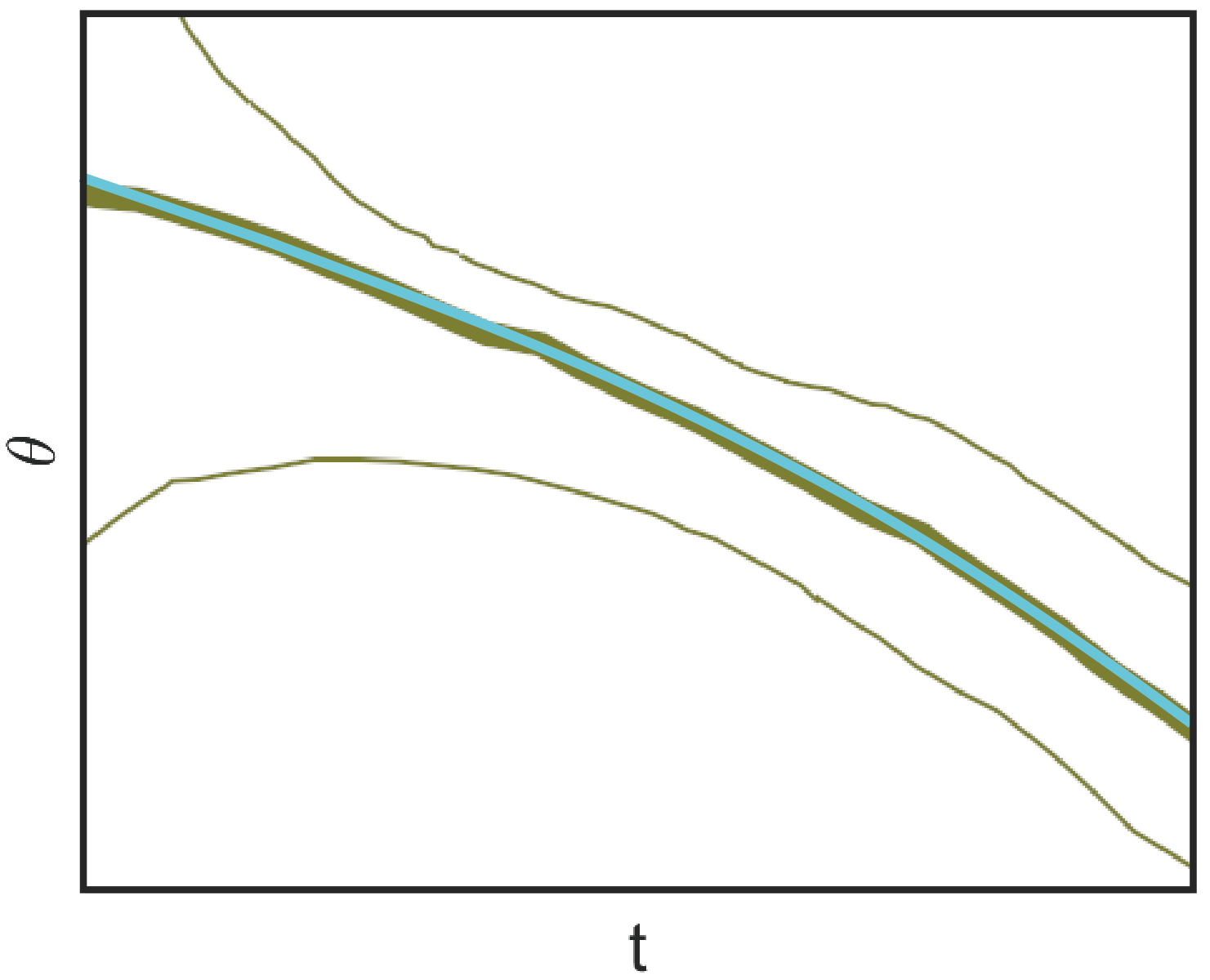}}\quad
\subfloat[Rose]{\includegraphics[height=4cm]{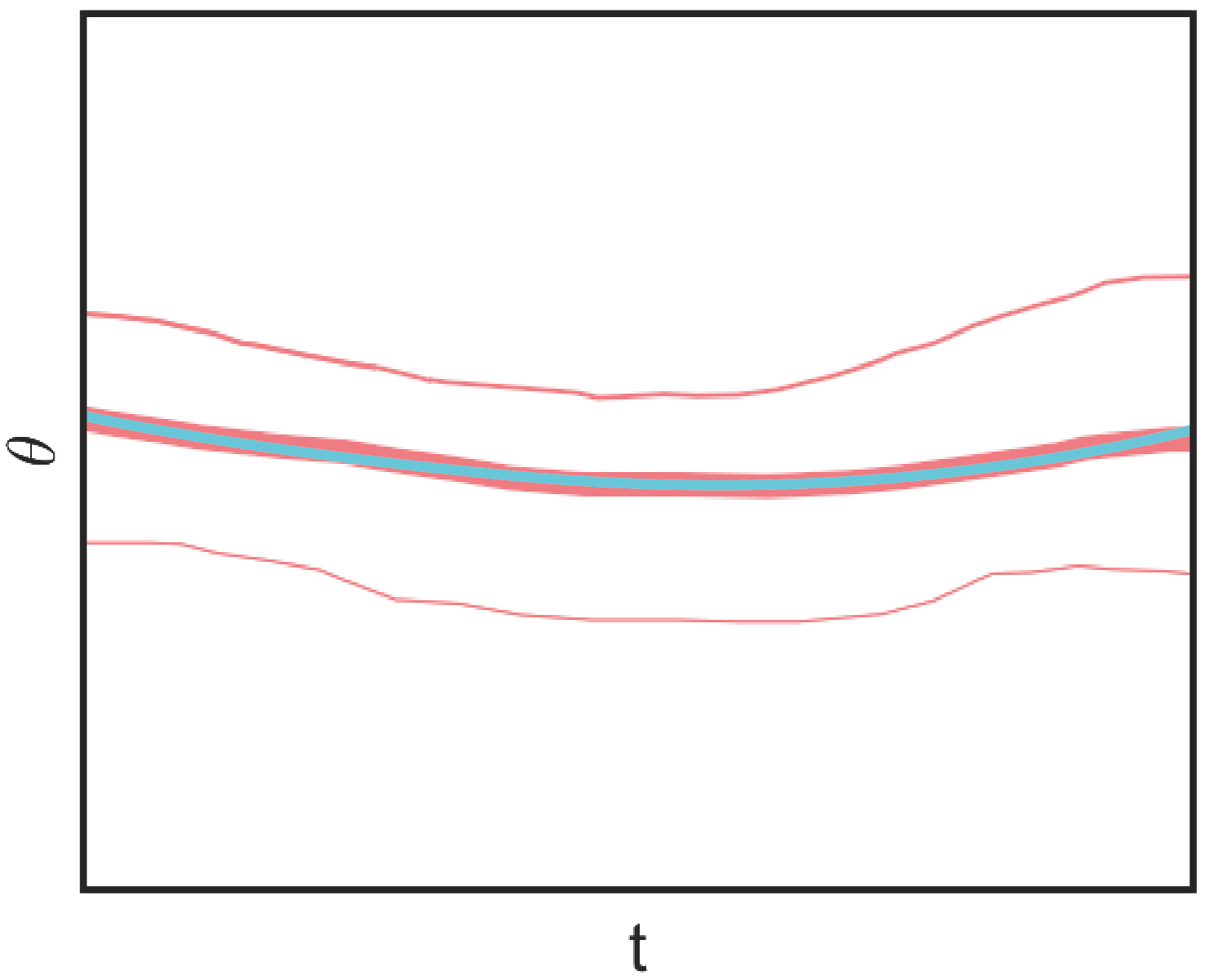}}\quad
\subfloat[Red]{\includegraphics[height=4cm]{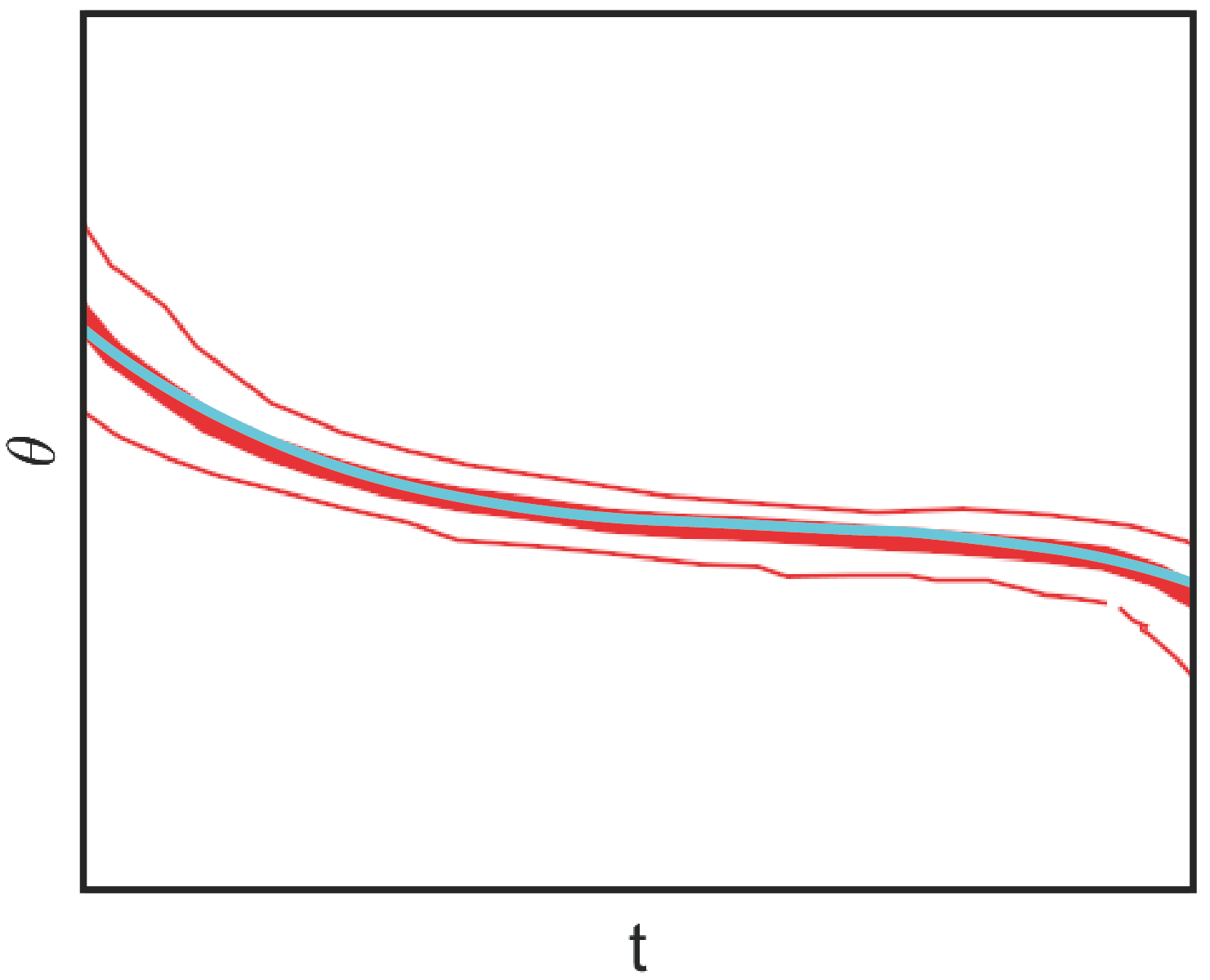}}\\
\subfloat[Bordeaux]{\includegraphics[height=4cm]{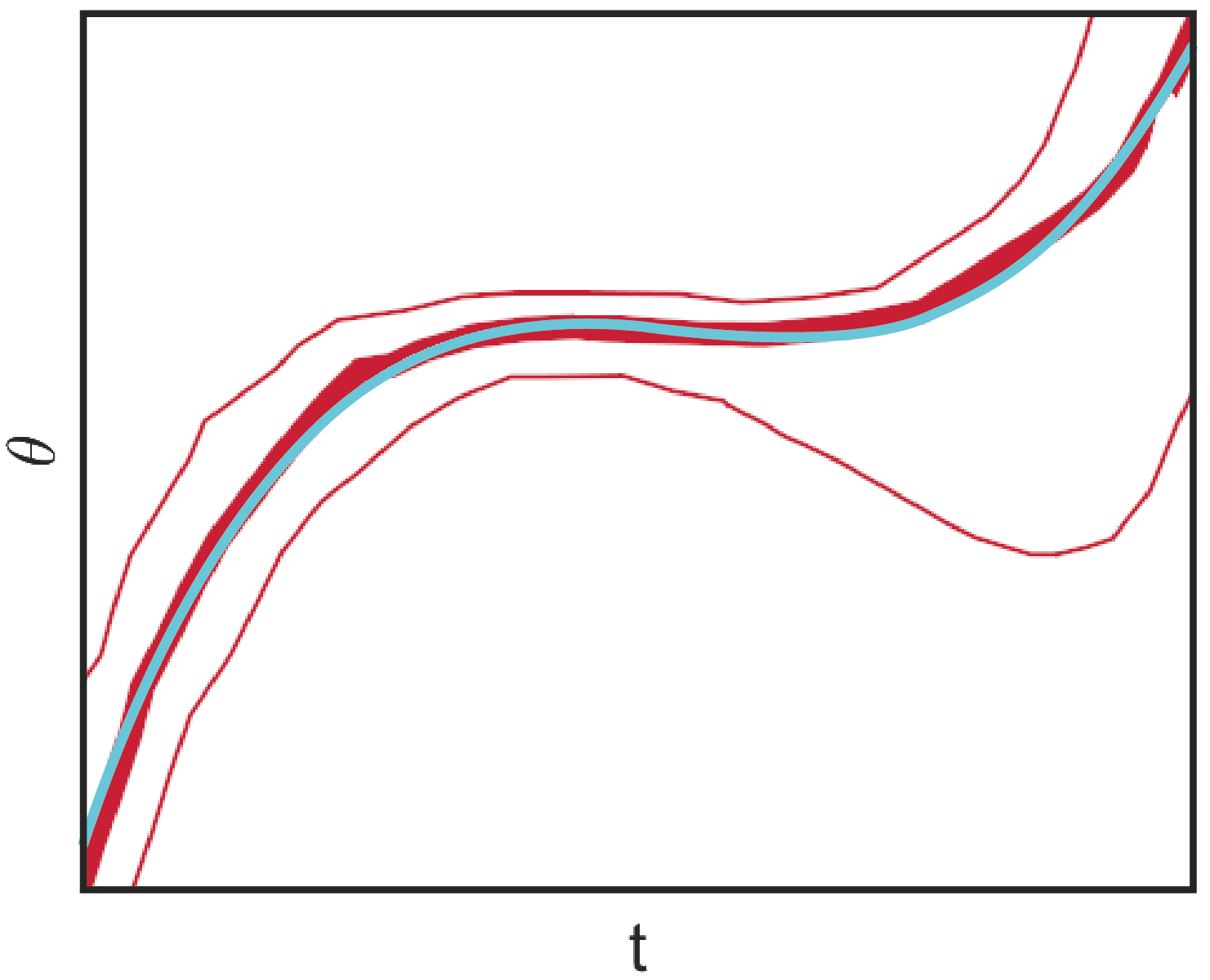}}\quad
\subfloat[Green]{\includegraphics[height=4cm]{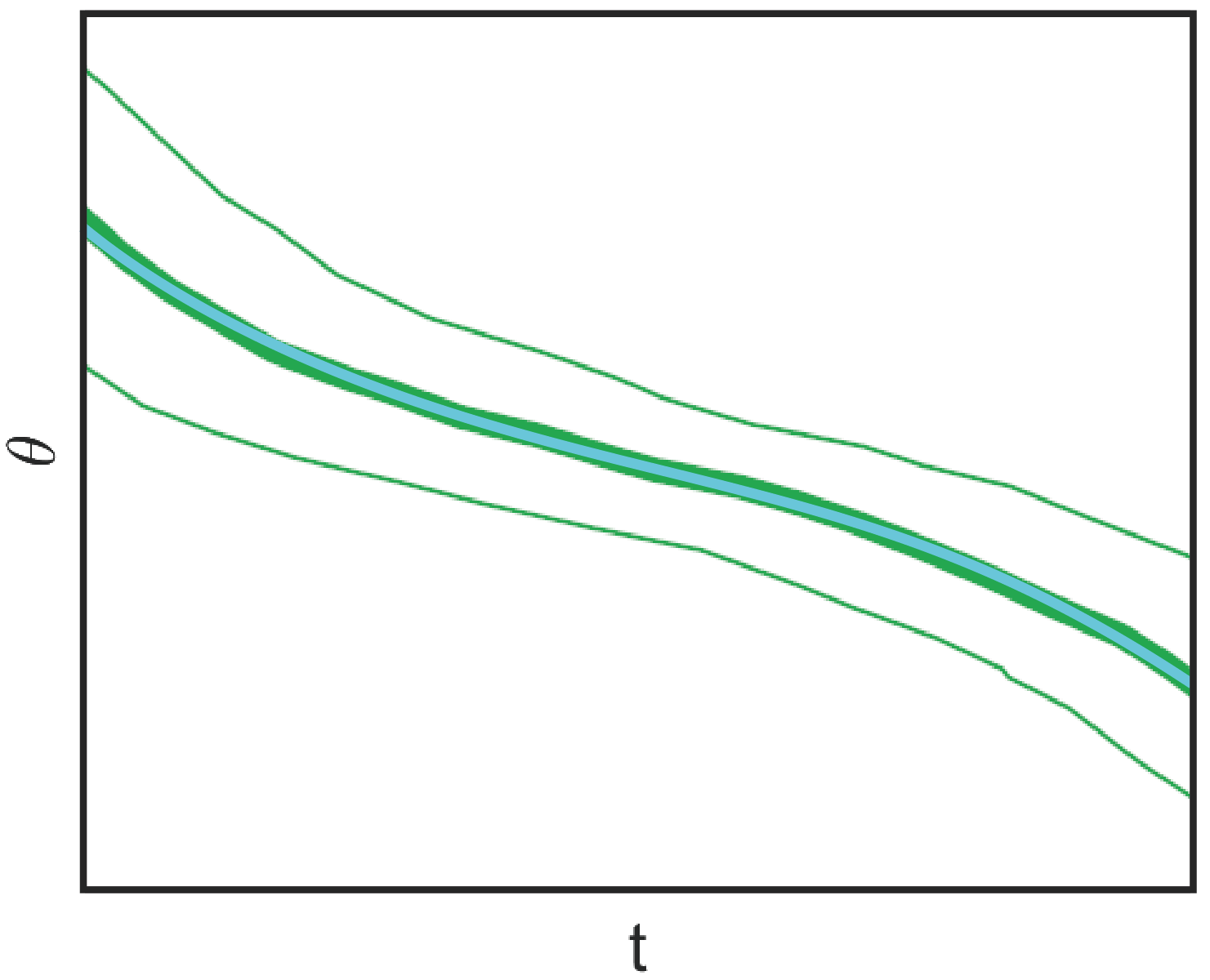}}\quad
\subfloat[Dark green]{\includegraphics[height=4cm]{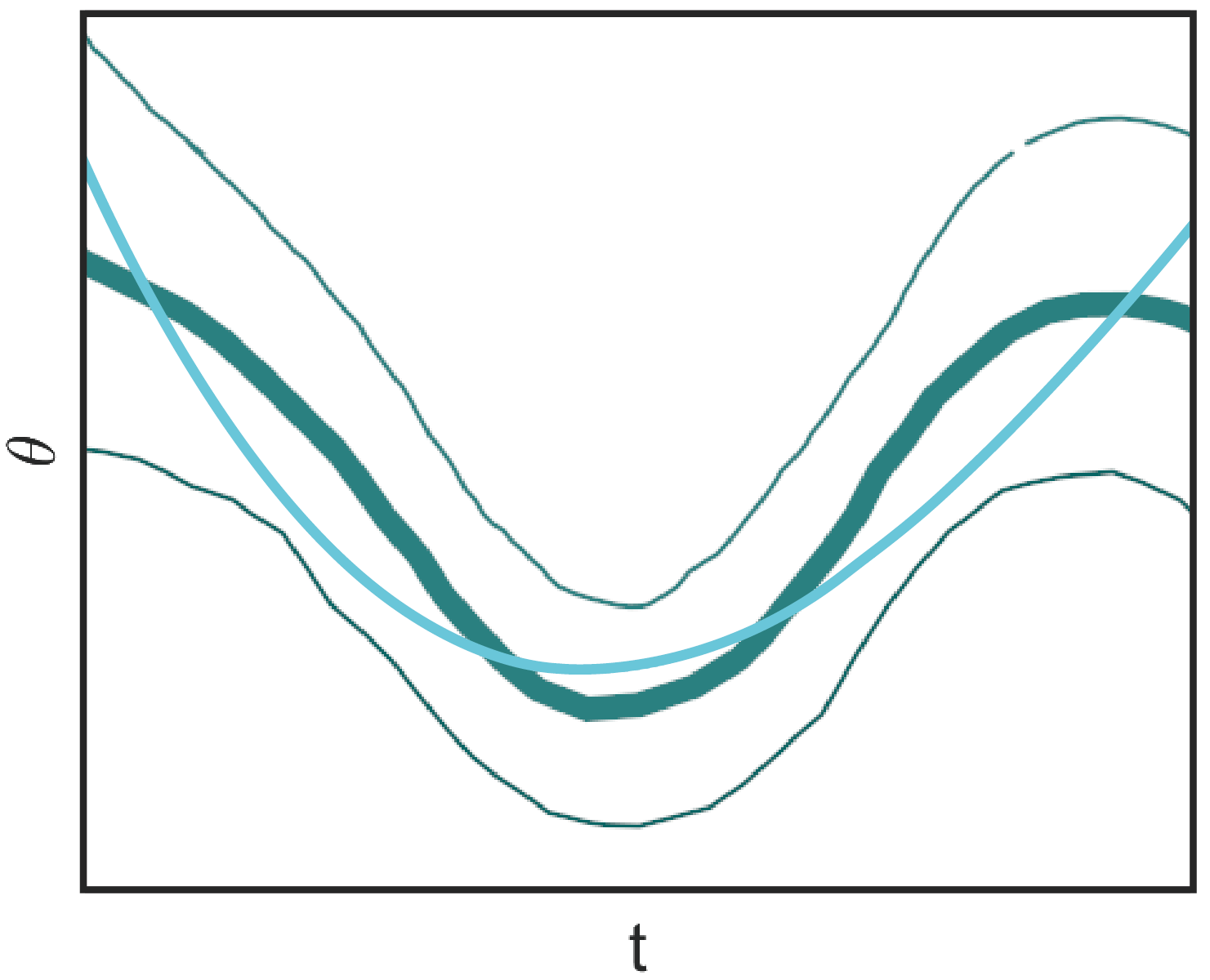}}\\
\subfloat[Violet]{\includegraphics[height=4cm]{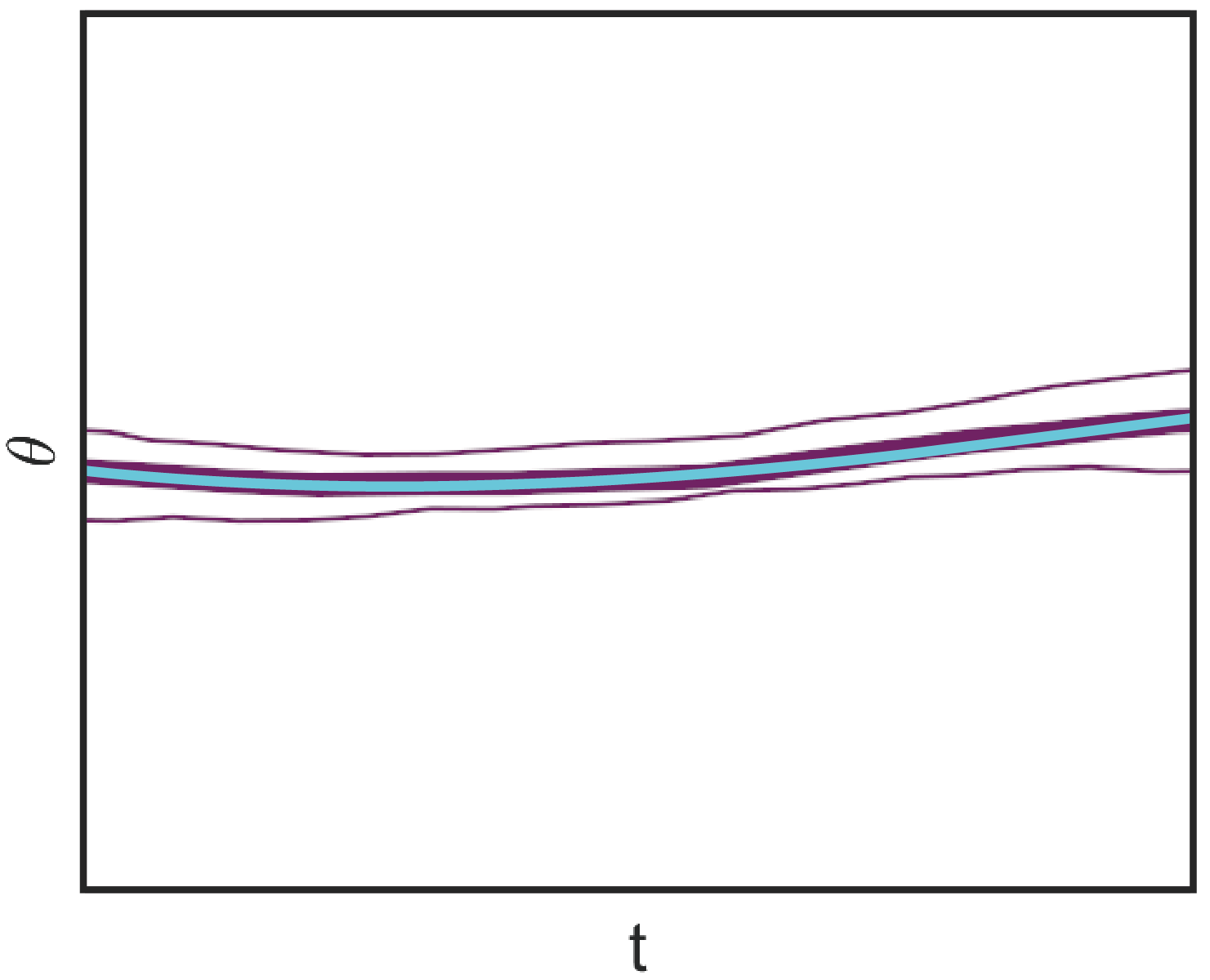}}\quad
\subfloat[Yellow]{\includegraphics[height=4cm]{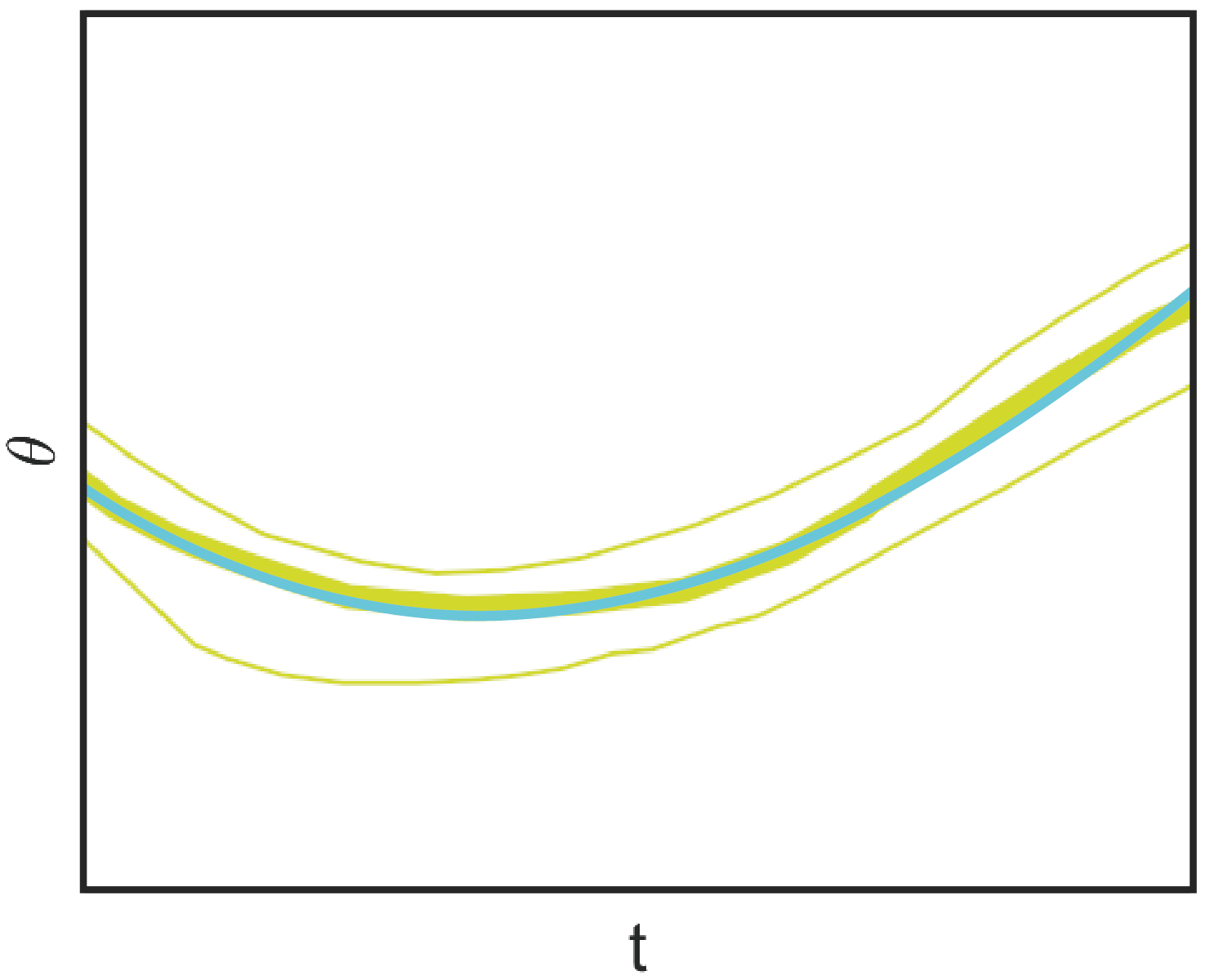}}
\caption{Time dependent PDs and polynomial fitting. For each box, a third-order polynomial fitting obtained by the least-squares method (cyan curve) is represented.}
\label{oddcurves}
\end{figure}

With polynomials of order two we get an average R$^2$ of 0.9515 and a good approximation over 0.99 only for the curves (a) Black, (d) Olive green and (k) Yellow.  
We can also observe from Figure \ref{oddcurves} that these curves are convex (up or down). 
 
\begin{figure}[htbp]
\centering
\includegraphics[height=4cm]{9_v2}\qquad\qquad
\includegraphics[height=4cm]{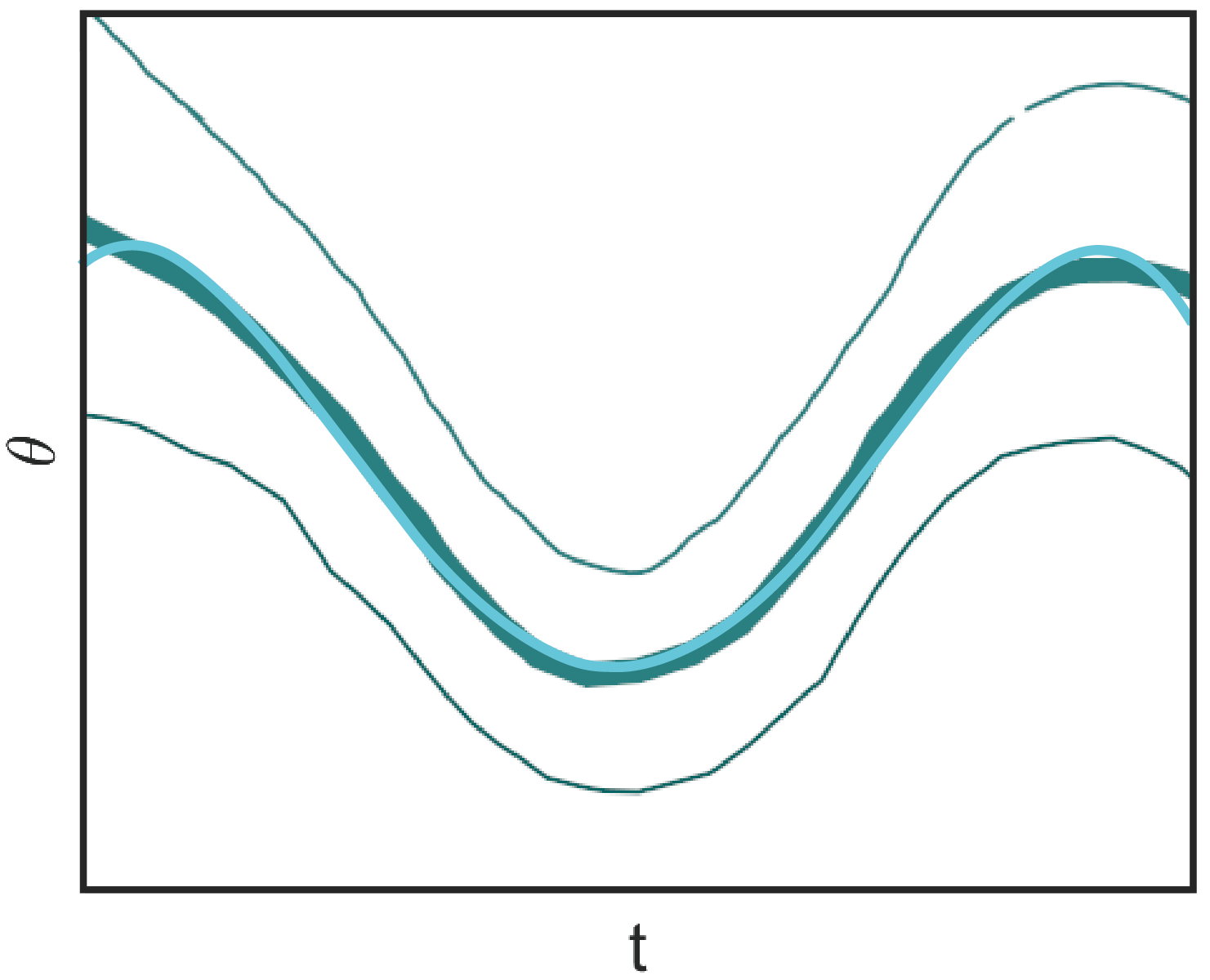}
\caption{Comparison of third (left) and fourth-order polynomial fitting (right) for time-dependent PD, colored in dark green. Fourth-order polynomial fit for the velocity profile extracted from the top panel of the figure \ref{im_temporal_beh}b.}
\label{curva_verde}
\end{figure}

Approximation of third orders are shown in the second column of Table \ref{Tabel_1}. The fitting of all curves is also visualized in Figure \ref{oddcurves}. The colored-trace curves are the curves extracted from the experimental data of Figure \ref{im_temporal_beh}b (Figure 13, from \cite{churchland2007temporal}), whereas the cyan ones denote the polynomial approximation. 
In Table \ref{Tabel_1}, the coefficient of determination (R$^2$) on average is 0.9831, and always greater than 0.99, except for the (e) Rose and the (i) Dark green curves. 
In particular, the dark green curve presents a very high error (Nrme$= 0.1272$): R$^2$ is 0.8630 and, as can be seen from the representation in Figure \ref{oddcurves}, the fitting is not accurate. Nonetheless, the approximated curve belongs to the confidence strip outlined by the thinner dark green curves (see also Figure \ref{curva_verde}, left). 

Better approximation is obtained for polynomials of order four, whose coefficient R$^2$ indeed reaches 0.9936 (see Figure \ref{curva_verde}, right as an example). Indeed, with polynomials of order four we obtain an optimal approximation with R$^2$ always over 0.99. Clearly, the approximation is even better for approximation of order five. 

\begin{figure}[htbp]
\centering
\includegraphics[height=4.5cm]{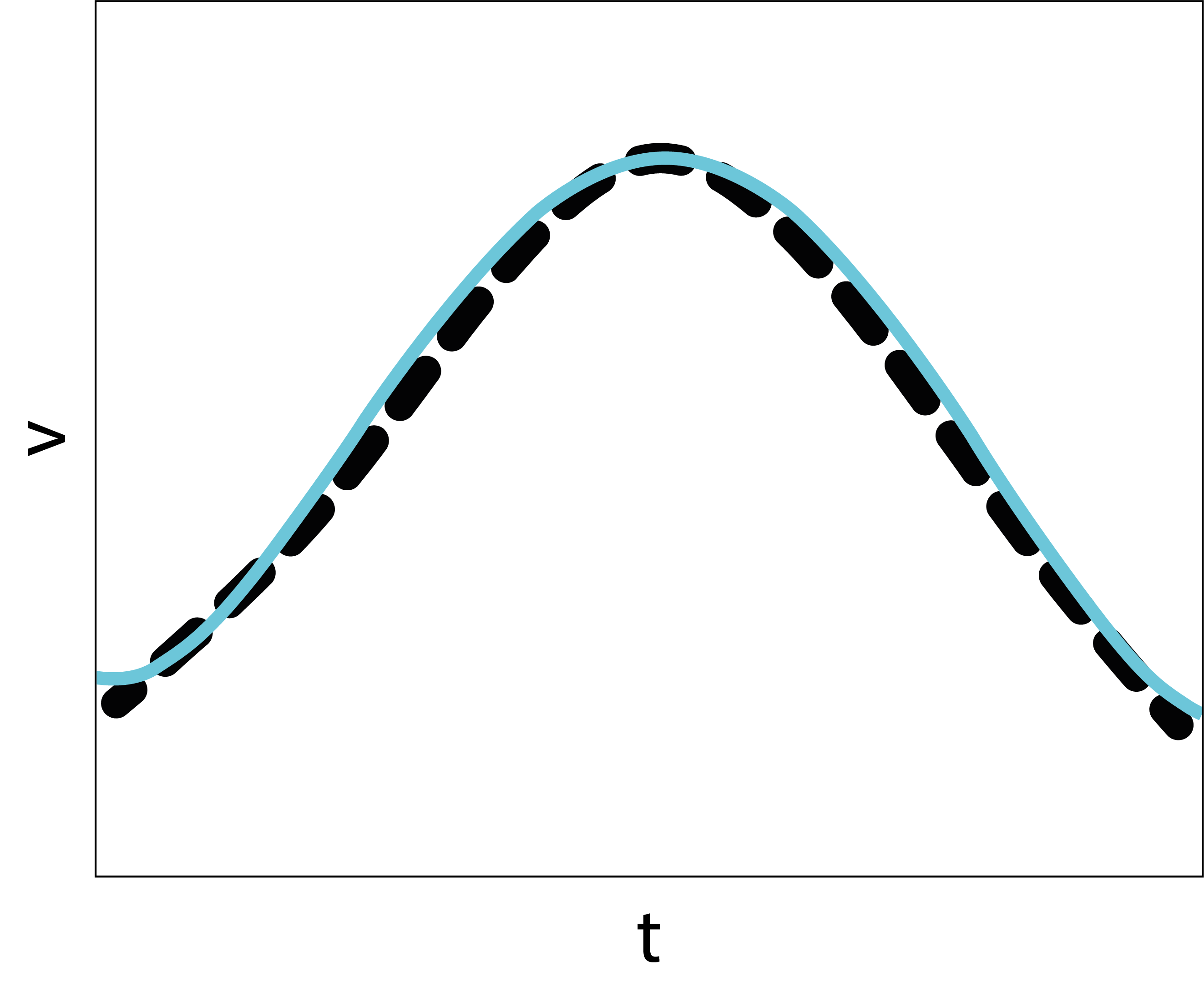}
\caption{Fourth-order polynomial fitting for the velocity profile extracted from the top panel of Figure \ref{im_temporal_beh}b. R$^2= 0.9968$, while nrmse$= 0.01786$.}
\label{fitting_acc}
\end{figure}

Finally, we show in Figure \ref{fitting_acc} the approximation obtained with a fourth-order polynomial for the velocity profile represented in the top panel of Figure \ref{im_temporal_beh}b. The factor R$^2$ in this case is 0.9968 and normalized root mean-squared error 0.01786.   
Further, we note that equation \eqref{a_punto_pari} allows to recover the bell shape velocity profile proposed by Flash and Hogan model \cite{FH}, whose expression is  
\begin{equation}\label{speed_profile}
v\left(t\right)= v_0\left(\left(\frac{t}{T}\right)^4- 2\left(\frac{t}{T}\right)^2 +1\right). 
\end{equation}
This polynomial fits the experimental velocity for $v_0= 140$ (cm$/$s) and $T= 0.07$s. This can be considered an ideal perfectly symmetric approximation of the velocity. We find a similar polynomial in Figure \ref{fitting_acc} with a small coefficient in the first order term, which takes into account the lack of symmetry in the experimental data.

\subsubsection{A qualitative comparison}

\begin{figure}[htbp]
\centering
\includegraphics[scale= 0.8]{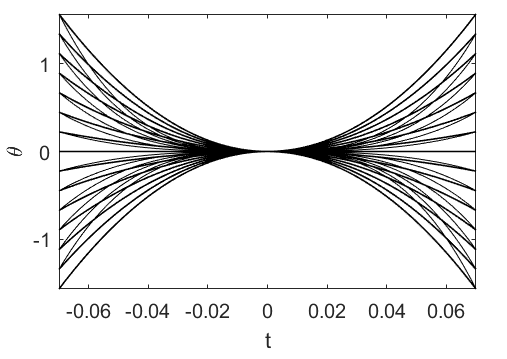}\qquad\qquad
\caption{Model example of M1 cells temporal selective tuning as a family of 30 integral curves solutions of system \eqref{sistema_curve_integrali_polinomi}. Movement direction components solutions over the temporal window $\left[-0.07, 0.07\right]$s. Different curves correspond to different choices of second and third order polynomials in equation \eqref{sistema_curve_integrali_polinomi}.}
\label{curve_integrali_disegnino}
\end{figure}

To recover the fan in \cite{churchland2007temporal} (see Figure \ref{im_temporal_beh}b), we note that the curves have been re-ordered in such a way to have the same direction and 
velocity at the point $t=0$. 
The whole set of curves, with the prescribed initial condition $\eta_0\in\mathcal{M}$ at time $t=0$, represents more precisely the cortical cells selectivity with respect to position, direction of movement, speed and acceleration:
\begin{equation}\label{sistema_curve_integrali_polinomi}
\begin{cases}
\dot{\gamma}\left(t\right)= X_{1}\left(\gamma\left(t\right)\right)+ p\left(t\right)X_{2}\left(\gamma\left(t\right)\right)+ q\left(t\right)X_{3}\left(\gamma\left(t\right)\right)\\
\gamma\left(0\right)=\eta_0.
\end{cases}
\end{equation}
Figure \ref{curve_integrali_disegnino} shows a family of integral curves the functions $t\mapsto p\left(t\right)$ and $t\mapsto q\left(t\right)$ are polynomials up to fourth and second order (see equations \eqref{dotteta} and \eqref{a_punto_pari}). All curves are characterized by a speed component having a bell-shaped profile, according to equation \eqref{speed_profile}, whose maximum value occurs at time $t=0$. This time is conceived to be the instant of a cell's strongest tuning with respect to the speed and movement direction variables, and, in correspondence of it, all curves meet at point $\left(x,y,\theta\right)= \left(0,0,0\right)$ 
 (see red and black dots of the plots in Figure \ref{curve_integrali_disegnino}).
 
\begin{figure}[htbp]
\centering
\includegraphics[scale=0.45]{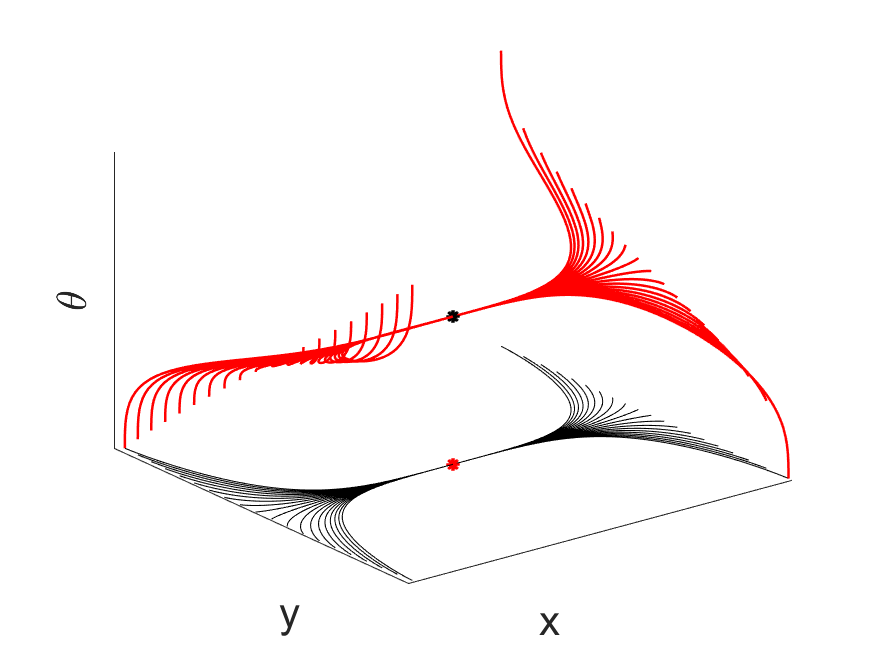}\qquad\quad
\includegraphics[scale=0.45]{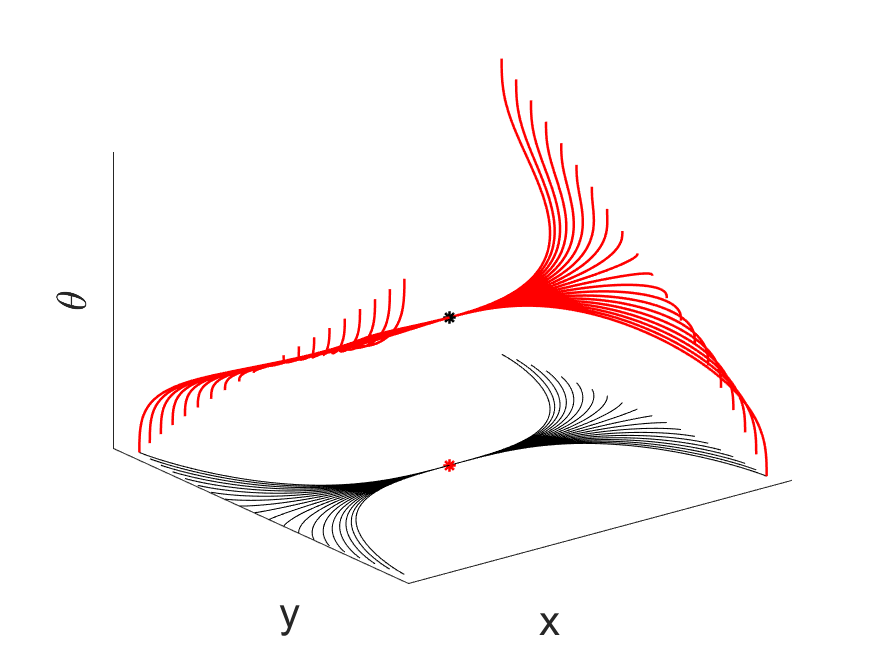}\\
\includegraphics[scale=0.45]{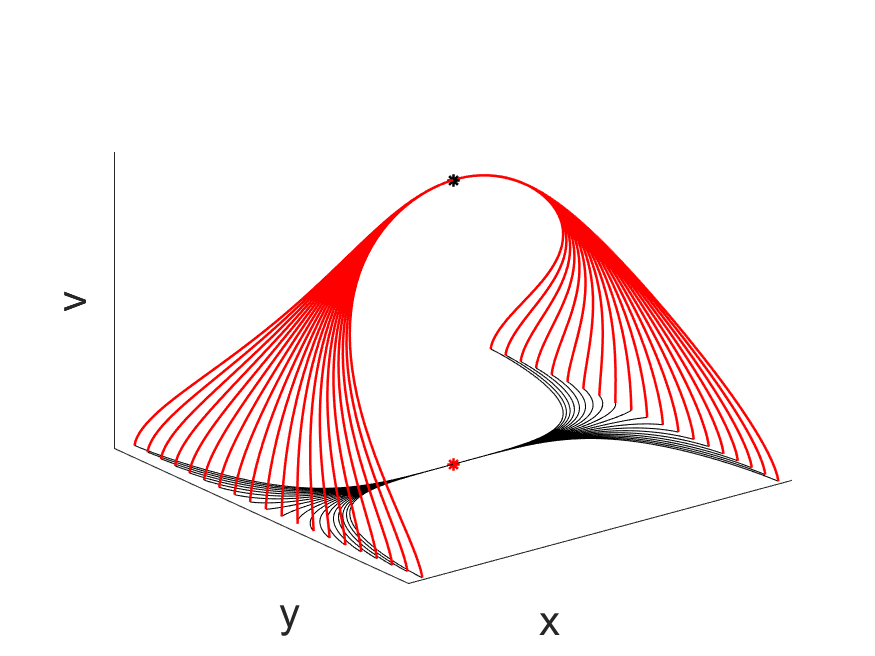}\qquad\quad
\includegraphics[scale=0.45]{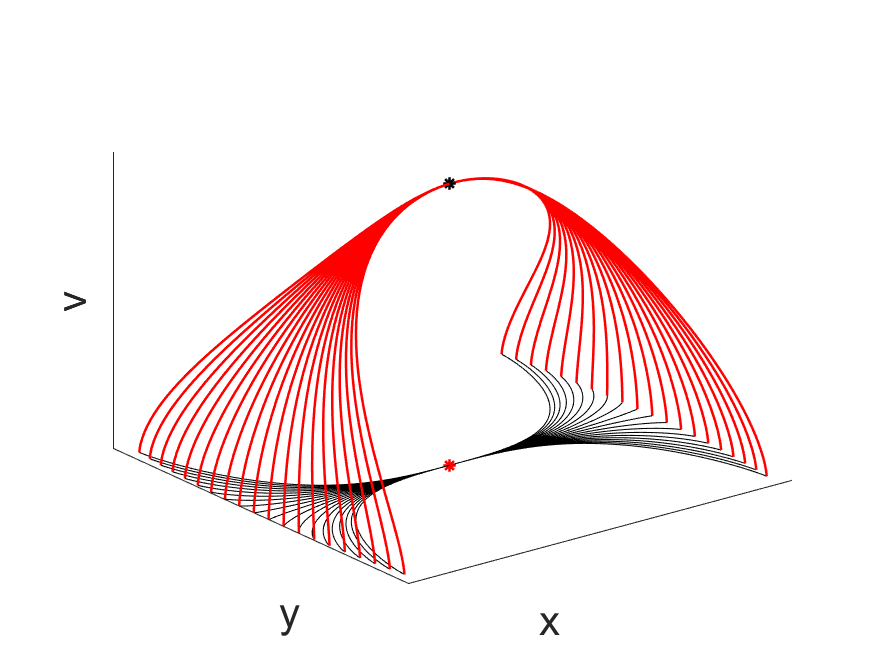}\\
\includegraphics[scale=0.45]{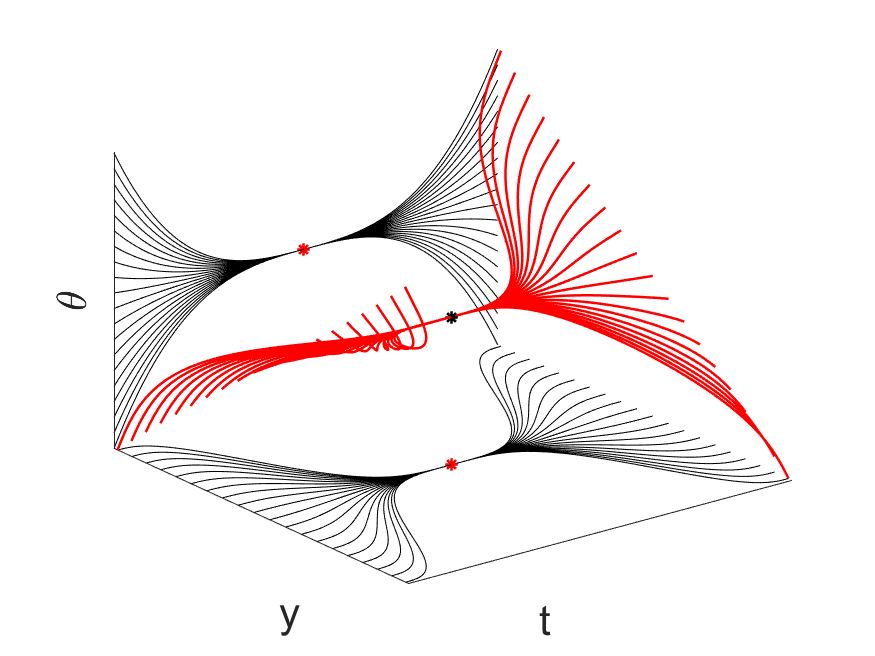}\qquad\quad
\includegraphics[scale=0.45]{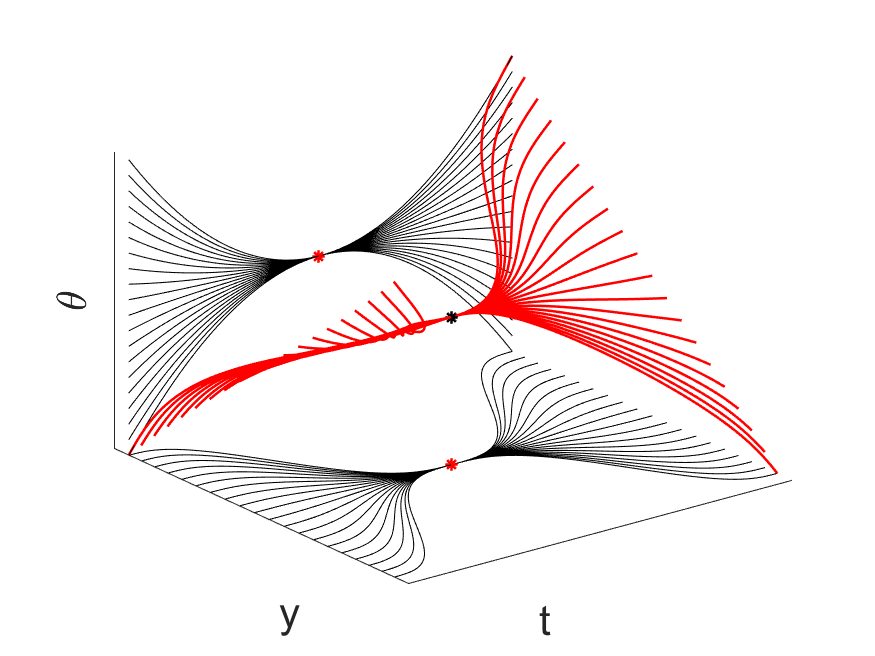}
\caption{Other model examples of M1 cells tuning patterns. Fan of curves projections over the variables $\left(x,y,\theta\right)$, $\left(x,y,v\right)$ and $\left(t,y,\theta\right)$. Different curves correspond to different choices of third (left column) and second order (right column) polynomials in equation \eqref{sistema_curve_integrali_polinomi}.} 
\end{figure}

\begin{figure}[htbp]
\centering
\includegraphics[scale=0.2]
{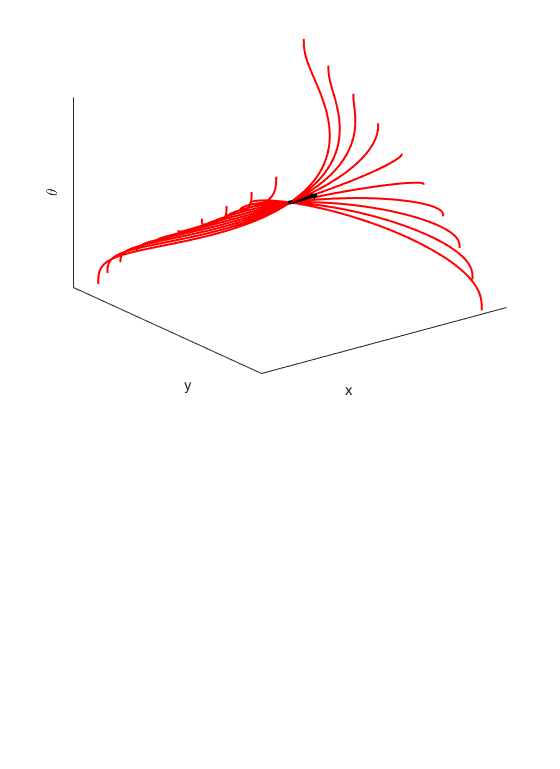}
\quad
\includegraphics[scale=0.75]{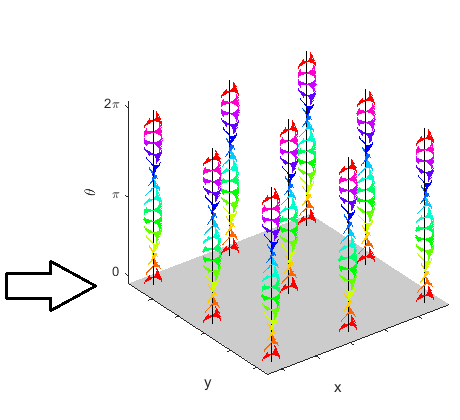}
\caption{Kinematic bundle model for time-dependent directionally tuned M1 cells. We only represent the $\left(x,y,\theta\right)$ components of system \eqref{sistema_curve_integrali_polinomi} applied for movement direction tuning which linearly changes over time. Left, fixed a PD at the central point, a solutions pattern which spreads over a temporal window is represented. Right, over each cell's selective tuning point $\left(x,y\right)$ of the base bundle at time $t=0$, there is a fiber of different patterns having strongest tuning for a preferred movement direction which spans the interval $\left[0,2\pi\right]$.}
\label{im_kinematic_bundle}
\end{figure}

We underline that in the ``static" simplified model presented in section \ref{fiber_pos_dir}, a point of the space is assumed to be a neuron characterized by its positions and directions of movement. 
In this context, a point of the previous structure corresponds to an ``instantaneous" cell selective movement parameter. 
Here, in the temporal bundle, the selective behaviour of a single neuron is represented by a whole trajectory expressed as a solution of \eqref{sistema_curve_integrali_polinomi}.
We show a simplified representation of the new fiber bundle in Figure \ref{im_kinematic_bundle}, where we have depicted only the $\left(x,y,\theta\right)$ components to facilitate a comparison with Figure \ref{model_M1} and to directly observe the extension of the temporal model with respect to the static one.
The central arrow in the left graph of Figure \ref{im_kinematic_bundle} has the same meaning as those depicted in the previous ``static'' fiber bundle, but now, as seen in the right part of the image, the fiber has a higher dimension representing the spread of time-selective behaviour. 

\subsection{Comparison with the time dependent receptive profiles in V1}\label{comp_cocci}
We will now compare the model of movement in the arm area of cortex M1 with models of movement coded in the visual cortex V1. We stress that the analysis on the comparison between visual and motor cortical cells is not based on their functionality. Visual cells are indeed characterized by their receptive profiles which detect features of the visual stimulus; on the other hand, cells in M1 are characterized by ``actuator profiles" and whose properties have been synthesized in \ref{georg} and \ref{motor_coding}.
The analogy is based on the coding of their related features, and on the structure of the functional geometry of the two areas.

\begin{figure}[htbp]
\centering
\includegraphics[scale=0.4]{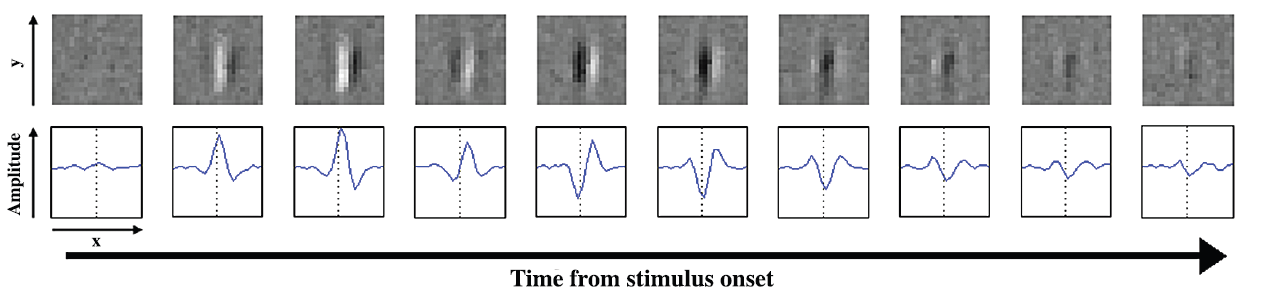}
\caption{Time course of a simple cell's RP. A simple cell's RF, represented in the second row as a curve of 1D RP vaying in time. The representation is analogous to the one used for the analysis on PD vectors which vary over time in M1 (see Figure \ref{im_temporal_beh}a). Source: \cite{cocci2014spatio}}.
\label{gabor_3d_1}
\end{figure}

\begin{figure}[H]
\centering
\includegraphics[scale=0.45]{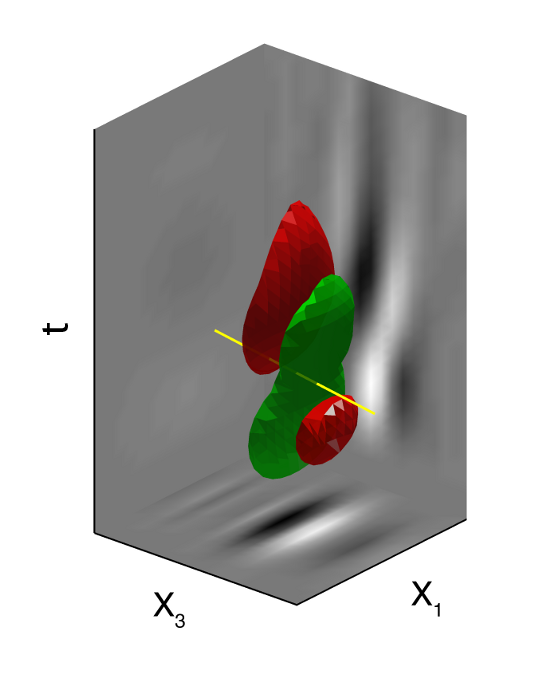}
\caption{Representation of a 3D-continuum receptive profile. It is the analogous of directionally tuned cells behaviour depicted in Figure \ref{im_temporal_beh}b. Source: \cite{cocci2014spatio}.}
\label{gabor_3d_2}
\end{figure}

We briefly recalled in the Introduction and in \ref{comp_v1} that simple cells in V1 detect positions, local orientations, however complex cells also encode parameters of movement via their receptive profiles (\cite{deangelis1993spatiotemporal, deangelis1995receptive}). 
For a given fixed position and orientation, cells receptive profiles sensible to movement are represented as a family of RPs varying in time (see \cite{jones1987evaluation}, \cite{deangelis1995receptive}):  
\begin{equation}\label{RP}
\Big( RP(t_1), \cdots, RP(t_m)\Big). 
\end{equation}
This complex receptive profile can be modelled as a curve in the space of 2D profiles. The graph of a continuous curve of receptive profiles 
\begin{equation}
RP: [-T,T]\to\mathbb{R}^2
\end{equation}
has been represented by G. Cocci \cite{cocci2014spatio} as a 3D volume defining a higher-dimensional profile selective of time frequency and velocity parameters (see Figures \ref{gabor_3d_1}, \ref{gabor_3d_2}). 

The variation of the RP from one frame to the next encodes the velocity of movement. Hence, we argue that the model of M1 (\ref{modellino?}) is analogous, but more general to the model of movement in V1, since we coded the variation of a cell's preferred movement direction not only via the first derivative, but via higher order derivatives.

\begin{figure}[H]
\centering
\includegraphics[scale=0.85]{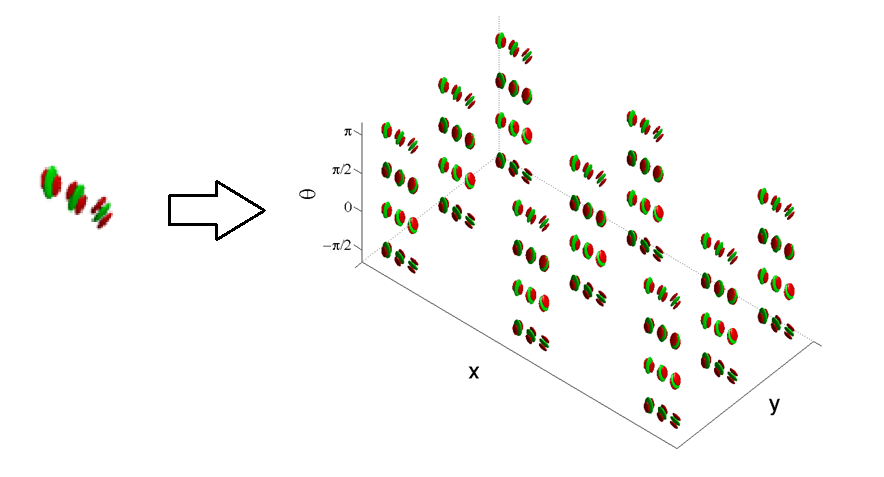}
\caption{(Left) Representation of a cell's RP over time. (Right) Schematization of the spatio-temporal fiber bundle of V1. For each spatio-temporal point $\left(x,y,t\right)$ there is a two dimensional fiber of possible local detected orientations $\theta$ and local velocity $v$. See Figure \ref{im_kinematic_bundle} for a direct comparison with M1. Images from \cite{cocci2014spatio}.}
\label{gabor_3d_cocci_bundle}
\end{figure}

Accordingly, 
the movement-receptive profile family was represented 
in Cocci's model as a fiber bundle, 
with a
base formed by position and time selective behaviour  $\left(q_1,q_2, t\right)$ and with 
engrafted variables $\left(\theta, v\right)$,  accounting for the orientation and velocity tuning 
over the point $\left(q_1,q_2, t\right)$. 
For $\theta$ fixed, there is therefore a 
one-parameter family of RPs
depending on the velocity variable 
(as depicted in Figure \ref{gabor_3d_cocci_bundle} left). This is the analogous of the fan of curves with varying PDs we represented in Figure \ref{im_kinematic_bundle} left, for a given fixed central PD. Then, in the model of movement for V1, the entire fan is obtained by varying orientation and position variables (see Figure \ref{gabor_3d_cocci_bundle} right), from which a total space of dimension five arises.

\section{Spatio-temporal grouping model for M1}\label{data_analysis}
In this section, we will proceed with the sub-Riemannian model for M1 cells by focusing on the coding of directional trajectories. 
As we recalled in section \ref{neural_states_movement_output}, neural activity shows coherent behaviours represented in terms of movement trajectories pointing to a specific pattern of movement decomposition (for further details see \cite{kadmon2019movement} and Figures \ref{fig_4}, \ref{fig_ret} as references).
The authors of \cite{kadmon2019movement} underlined that they tried to obtain the neural decomposition in fragments by using various distances proposed in literature (see the list below).
Precisely, they tested six models that included tuning for: movement direction; movement
direction gain modulated by speed; direction of the acceleration vector; direction of the acceleration vector gain modulated by the magnitude of the acceleration vector; both movement direction and direction of the acceleration vector; or both movement direction and direction of the acceleration vector, each gain modulated by the magnitude of the corresponding vector:
\begin{enumerate}
\small
\item $fr_i\left(t- \tau\right)= B_{0_{i}}+ B_{1_{i}}\cos\left(\theta\left(t\right)- \theta_{\text{PD}_{i}}\right)$;
\item $fr_i\left(t- \tau\right)= B_{0_{i}}+ B_{1_{i}}\lVert \vec{v}\left(t\right)\rVert\cos\left(\theta\left(t\right)- \theta_{\text{PD}_{i}}\right)$;
\item $fr_i\left(t- \tau\right)= B_{0_{i}}+ B_{1_{i}}\cos\left(\theta_a\left(t\right)- \theta_{a_{\text{PD}_{i}}}\right)$;
\item $fr_i\left(t- \tau\right)= B_{0_{i}}+ B_{1_{i}}\lVert \vec{a}\left(t\right)\rVert\cos\left(\theta_a\left(t\right)- \theta_{a_{\text{PD}_{i}}}\right)$;
\item $fr_i\left(t- \tau\right)= B_{0_{i}}+ B_{1_{i}}\cos\left(\theta\left(t\right)- \theta_{\text{PD}_{i}}\right)+ B_{2_{i}}\cos\left(\theta_a\left(t\right)- \theta_{a_{\text{PD}_{i}}}\right)$;
\item $fr_i\left(t- \tau\right)= B_{0_{i}}+ B_{1_{i}}\lVert \vec{v}\left(t\right)\rVert\cos\left(\theta\left(t\right)- \theta_{\text{PD}_{i}}\right)+ B_{2_{i}}\lVert \vec{a}\left(t\right)\rVert\cos\left(\theta_a\left(t\right)- \theta_{a_{\text{PD}_{i}}}\right).$
\end{enumerate}

In the numbered list above (taken from \cite{kadmon2019movement}), $fr_i$ denotes the instantaneous firing rate of neuron $i$, $\tau$ the
time lag between neural activity and kinematic output (taken
to be 100ms), $B_{0_{i}}$ the baseline firing rate, $B_{1_{i}}$ and $B_{1_{i}}$ modulation depths, whereas $\lVert \vec{v}\left(t\right)\rVert$ and $\lVert \vec{a}\left(t\right)\rVert$ represent the magnitude of the
velocity and acceleration vectors, respectively, $\theta$ and $\theta_a$ are the directions of the velocity and acceleration vectors, and  $\theta_{\text{PD}}$ and $\theta_{a_{\text{PD}}}$ are the preferred velocity and acceleration angles of neuron $i$.

However, none of these distances were successful in yielding the desired neural decomposition. This failure can have two explanations: either the considered kinematic parameters are not sufficient to recover the decomposition, and more parameters are coded in the brain, or a more complex distance is needed.

Here, we will show that with our distance $d_{\mathcal{M}}$ defined in \ref{time_npv} (see Equation \eqref{estim_hom_npv}) which takes into account the differential relations between the variables exactly provides the same decomposition. The algorithm we will apply is a variant of $k$-means which considers the presence of this distance: first we perform a change of variables induced by distance $d_{\mathcal{M}}$ and then we apply the $k$-means in the new variables. This provides an answer to the problem we posed above, and clarifies that the set of kinematic variables considered up to now is sufficient to recover the cortical decomposition, and it is most probably the set of parameters responsible for the considered task in this area.

Specifically, we will test the pattern of movement decomposition through a local estimate of distance $d_{\mathcal{M}}$ \eqref{estim_hom_npv} and of the associated kernel    
\begin{equation}\label{kernel_m_capitolo_grouping}
\omega_{\mathcal{M}}\left(\eta_0,\eta\right)= e^{- d_{\mathcal{M}}\left(\eta_0,\eta\right)^2}, \quad \eta_0, \eta\in \mathcal{M},
\end{equation}
where the cortical feature space $\mathcal{M}= \mathbb{R}^{3}_{\left(t,x,y\right)} \times S^1_{\theta} \times \mathbb{R}^{2}_{\left(v,a\right)}$ is expressed in terms of kinematic variables. 
We also show that the same decomposition can not be recovered with a simpler algorithm based on the Euclidean distance, or a weighted Euclidean distance.

\subsection{Spectral analysis}\label{spectral_lett}
The most classical model describing the cortical activity is the mean field equation of Ermentrout and Cowan \cite{ermentrout1980large} and Bressloff and Cowan \cite{bressloff2002visual},  \cite{bressloff2003functional}. This equation describes the evolution of the cortical activity depending on a connectivity kernel. Sarti and Citti \cite{sarti2015constitution} proved a relation between the stable states of Bressloff and Cowan equation and perceptual units of the visual input in V1. We also refer to the works of Faugeras et al. \cite{faugeras2009persistent}, Cocci et al. \cite{cocci2015cortical}, Favali et al. \cite{favali2017local}. 
In perfect analogy, we assume that the neural states corresponding to movement trajectories can be interpreted as a form of clustering. 
A vast literature is available on the theoretical and practical aspects of several clustering algorithms (see e.g. \cite{von2007tutorial} as a summary of the most known techniques). 
A classical method is the $k$-means algorithm, which clusters the data according to the Euclidean distance. As a result, clouds of points are correctly grouped. 
However, when dealing with aligned data, accurate clustering can not be achieved without first performing a change of variable which strongly correlates aligned points. This step is called spectral cluster algorithm, since the change of variable is induced by eigenvectors of a distance function.

To this end we define an affinity matrix $A$,
whose elements $\left(a_{ij}\right)$ have values proportional to the similarity of point $\eta_i$ to point $\eta_j$, for which $A$ can be defined by
\begin{equation}\label{affinity_matrix_general}
A = a_{ij}= e^{-d^2\left(\eta_i,\eta_j\right)},
\end{equation}
where $d$ is a suitable distance over the considered space. Principally, there exist two classes of spectral clustering techniques \cite{lafon2006diffusion}: methods for locality-preserving embeddings of large data sets, that project the data points onto the eigenspaces of the affinity matrices (\cite{coifman2006diffusion}, \cite{belkin2003laplacian}, \cite{roweis2000nonlinear}), and methods for data segregation and partitioning, that basically perform an
additional clustering step taking as input the projected data set (\cite{perona1998factorization}, \cite{weiss1999segmentation}, \cite{shi2000normalized}, \cite{meilua2001random}, \cite{ng2001spectral}).
It has been originally shown by Perona \cite{perona1998factorization} that the first eigenvector of $A$ can represent a background/foreground contour separation.
To reduce error due to noise, the affinity matrix can be suitably normalized.
Many normalizations have been proposed (e.g. \cite{butler2006spectral}, \cite{ng2001spectral}, \cite{shi2000normalized}): one of the most widely applied is the one presented by Meila and Shi \cite{meilua2001random} since 
it reveals properties of the underlying weighted graph by ways of the Markov chain, providing a probabilistic foundation of the clustering algorithm. Indeed, the authors defined a Markov-type matrix $P$ as follows
\begin{equation}\label{markov}
P = D^{-1}A, \quad D\; \text{diagonal matrix, }\quad d_i = \sum_{j=1}^n a_{ij}.
\end{equation}
In general the matrix $P$ will not be symmetric, but its eigenvalues are real, positive and smaller than one, while the eigenvectors have real components (\cite{lafon2006diffusion}, \cite{coifman2006diffusion}). 
The clustering properties of the eigenvectors of $P$ can
be clearly understood in the ideal case, when $P$ is a block diagonal matrix. 
In this case a natural grouping is obtained via  the projection on the eigenvectors.
If the affinity matrix $A$ is already a block matrix, the multiplication by $D$ can be avoided. 

In the general case, if the matrix $P$ is not diagonal, a $k$-means algorithm in the coordinates induced by eigenvectors will provide the classification for our problem.

\begin{enumerate}
\item Calculate the normalized affinity matrix $P = D^{-1}A$ (skip this step if $A$ is a block diagonal matrix).
\item Call  $\lbrace \left(\lambda_i, u_i\right)\rbrace_{i=1}^k$  
 the first $k$ eigenvalues and the corresponding eigenvectors of the matrix $P$. 
\item If $P$ is a block matrix, projection on the eigenvectors forms the clusters for the decomposition. If $P$ is not diagonal, perform a $k$-means algorithm in the coordinates induced by the eigenvectors
\end{enumerate}


\subsection{The spectral clustering method in the feature space $\mathcal{M}$}\label{sub_section_aff_m}
In this section we will apply the connectivity kernel \eqref{kernel_m_capitolo_grouping} for the definition of an affinity matrix.  We will show that the salient groups obtained due to spectral clustering are in agreement with the neural results present in \cite{kadmon2019movement}.

More precisely, given a set of reaching paths, we discretize \eqref{kernel_m_capitolo_grouping} by means of the real symmetric affinity matrix $A_{ij}$:
\begin{equation}\label{affinity_matrix_tesi}
A_{i,j}= \omega_{\mathcal{M}}\left(\left(x_i, y_i, \theta_i, v_i, a_i, t_i\right), \left(x_j, y_j, \theta_j, v_j, a_j, t_j\right)\right),
\end{equation}
that contains the connectivity information between all the
kinematic variables. We then apply to \eqref{affinity_matrix_tesi} the algorithm provided in section \ref{spectral_lett}.
The eigenvectors associated with the largest eigenvalues of the affinity matrix will be short movement trajectories compatible with the movement fragments of Hatsopoulos et al. \cite{Encoding,reimer2009problem} and with the trajectories found in Kadmon Harpaz et al. \cite{kadmon2019movement} for the pattern of movement decomposition (see Figures \ref{fig_ret} and \ref{fig_4} as references).

\subsubsection{Results}
We will provide two test cases in which the connectivity kernel \eqref{kernel_m_capitolo_grouping} is applied to a set of movement trajectories. 

\medskip

\noindent \textbf{Test 1: Simulation of a center-out task}\\
As a first example, we will analyze a trajectory of movement performing a center-out task, as it is represented in Figure \ref{fig_ret} (from \cite{kadmon2019movement}). Below we show an approximation of the image.

\begin{figure}[H]
\centering
\includegraphics[scale= 0.4]{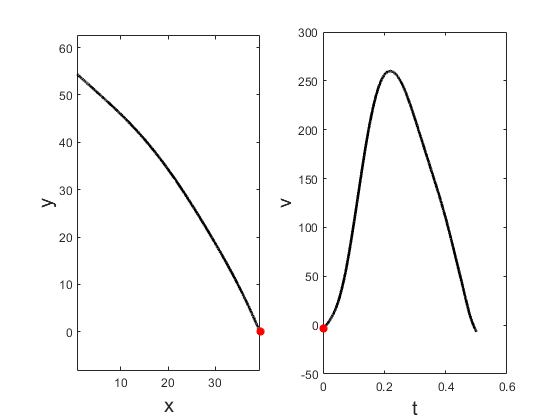}
\caption{Reaching path and speed profile of a center-out task: approximation of Figure \ref{fig_ret}. (Left) Reaching path over the $\left(x,y\right)$ plane. (Right) Speed profile over the  $\left(t,v\right)$ plane. The red dot represents the movement starting position.}
\label{approx_fig_ret}
\end{figure}

The motion trajectory, as in the paper \cite{kadmon2019movement}, is characterized by two graphs, one on the $\left(x,y\right)$ plane, the reaching path, and one on the $\left(t,v\right)$ plane corresponding to the velocity profile. The red dot in Figure \ref{approx_fig_ret} identifies the starting point of the movement. 

\begin{figure}[htbp]
\centering
\subfloat[]{\includegraphics[scale=0.35]{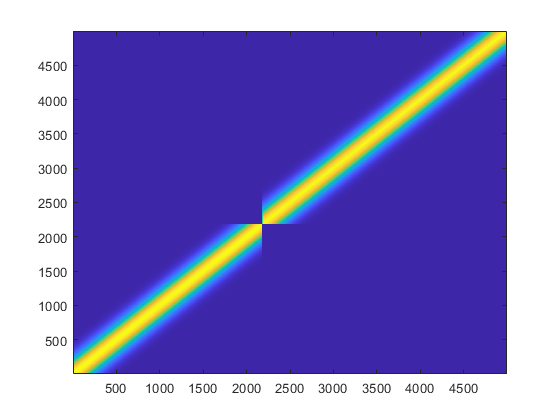}}
\subfloat[]{\includegraphics[scale= 0.33]{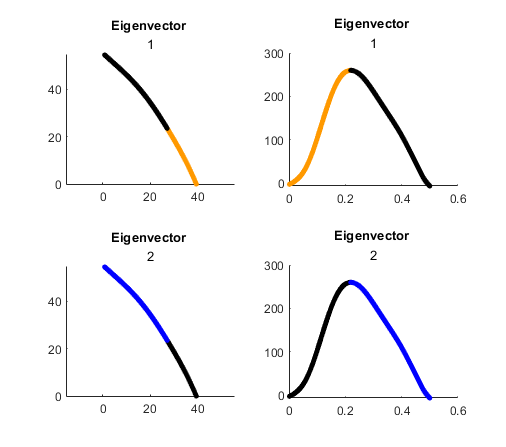}}
\subfloat[]{\includegraphics[scale= 0.33]{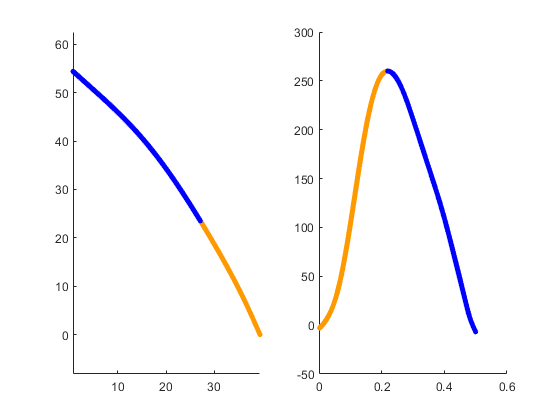}}
\caption{ Spectral clustering. (a) The Affinity matrix \eqref{affinity_matrix_tesi}. Yellow-colored areas represent points of higher affinity.  (b) Eigenvectors projections over the reaching trajectory. (c) Results of movement decomposition. Eigenvectors identify acceleration and deceleration trajectories according with the experimental results of \cite{kadmon2019movement}. See Figure \ref{fig_ret} for a direct comparison}
\label{aff_m_ret}
\end{figure}

In this very simple case, movement direction is almost constant with only one target point to be reached and just one maximum point of the speed profile.

As we clarified  \eqref{estim_hom_npv}, distance $d_{\mathcal{M}}$ depends on suitable coefficients 
$c_1, \cdots, c_6$. 
In particular, we normalized the temporal component $e_1$ with respect to the time window for which a neuron is found to be selective, which averages out to be equal to 0.4 sec \cite{Encoding}.  This choice is also consistent with the typical duration of increasing and decreasing velocity profiles present in \cite{kadmon2019movement} (see, for example, Figure \ref{fig_4}). 
 Moreover, in all the tests we made, we obtained the best results with a weight $c_1= 10$ and all the other constants identically 1, so as to give more importance to the temporal aspect of the trajectory. 

The resulting affinity matrix is clearly divided into blocks (see Figure \ref{aff_m_ret}). These blocks are the eigenvectors associated to the two major eigenvalues of the affinity matrix and represent the clusters of the pattern of movement decomposition. By projecting the eigenvectors over the reaching trajectory, it turns out that these correspond precisely to the acceleration and deceleration phases of the movement task. Acceleration-phase is orange-colored and deceleration-phase is blue-colored as the neural states represented in Figure \ref{fig_ret}.

In this simple case, the same results could have been obtained using a weighted Euclidean $k$-means clustering. Precisely if 
$\left(\left(t_0, x_0,y_0,\theta_0, v_0, a_0, \right),\left(t_1,x_1,y_1,\theta_1, v_1, a_1\right)\right)$, and  $c_1, \cdots, c_6$ are positive weights, a general distance can be defined as  
\begin{equation}\label{eucl_ad}
||\eta_0-\eta_1|| = \left(\sum_{i=1}^6  \left|c_i(\eta_0(i)-\eta_1(i))\right|^{2}\right)^{\frac{1}{2}}
\end{equation}

If all coefficients are identically 1, we recover the standard Euclidean distance, 
but a better  clustering result is obtained with a 
 $c_1= 10$, while all the other coefficients are chosen to be $1$. Note that the coefficient $c_1=10$ is exactly the same used in the sub-Riemannian case.

\begin{figure}[H]
\centering
\subfloat[]{\includegraphics[scale=0.3]{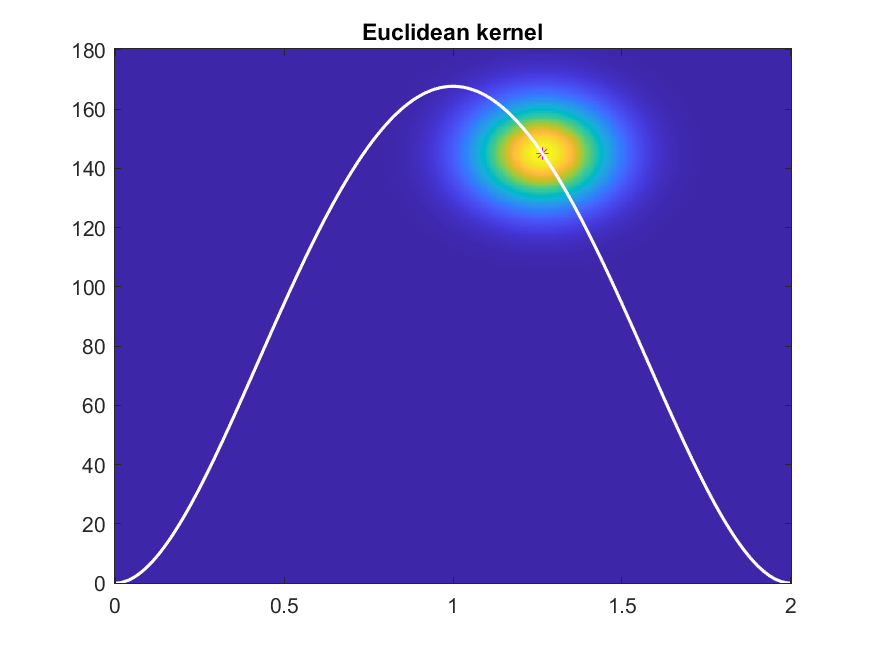}}\quad\quad
\subfloat[]{\includegraphics[scale=0.3]{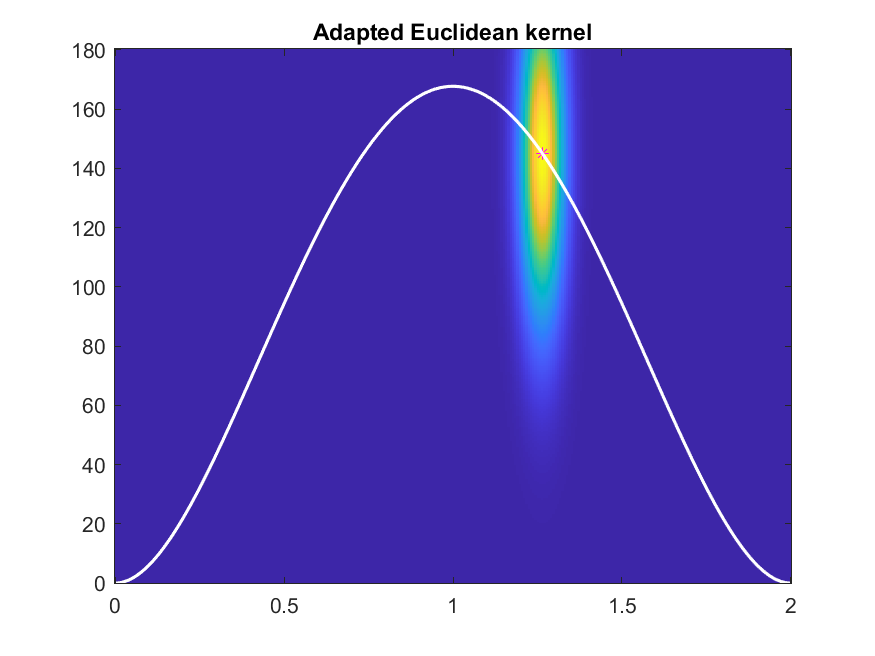}}\quad\quad
\subfloat[]{\includegraphics[scale=0.3]{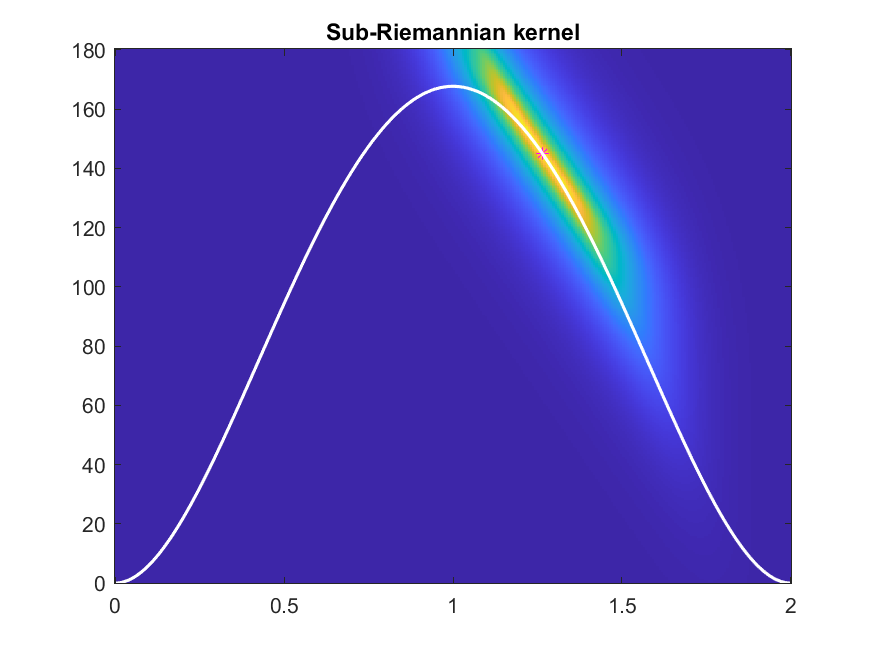}}\\
\caption{Different kernel representations over a smooth velocity profile. (a) Euclidean kernel. (b) Euclidean kernel, with a relevant weight on the time variable. (c) Sub-riemannian kernel.}
\label{plot_kernels}
\end{figure}

Below we show a representation of three kernels: the  Euclidean one, the Euclidean one with the considered weight, and  the sub-Riemannian one. The representation shown in Figure \ref{plot_kernels} is the application of each of the three kernels to a bell-shaped velocity profile in the $(t,v)$ plane (the corresponding path in the $(x,y)$ plane is a segment).

\medskip

As can be observed from Figure \ref{plot_kernels}, the subriemannian kernel is tangent to the curve at the violet-colored point $(t_0, v_0)$, since the coefficient $e_5$ defined in \eqref{e125} which expresses the increment in the velocity variable $v$ also depends on the accelleration $a$, which clearly is the derivative of $v$. On the contrary, the Euclidean one, is always aligned with the axes, and  does not use any information regarding the derivative of the curve. For this reason only the sub-Riemannian one leads to a principle of good continuation in velocity.



The advantage of the subriemannian kernel is that it can capture more complex relationships between points, as it takes into account the intrinsic geometry of the data space. 
Significant differences between the kernels can be observed when applied to movements that are not smooth, with sudden changes in velocity and curvature, as we will see in the next test on a more complicated path.


\medskip

\textbf{Test 2: Simulation of a random target pursuit task}

\begin{figure}[H]
\centering
\includegraphics[scale=0.4]{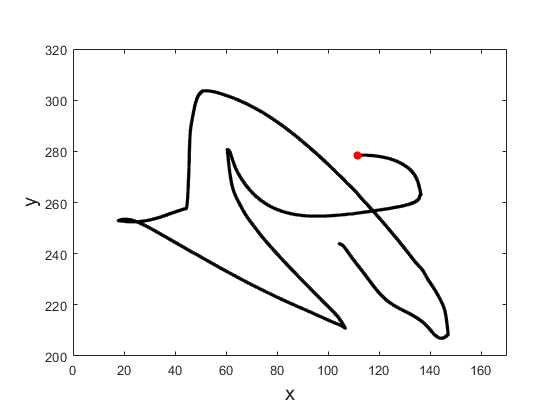}
\includegraphics[scale=0.4]{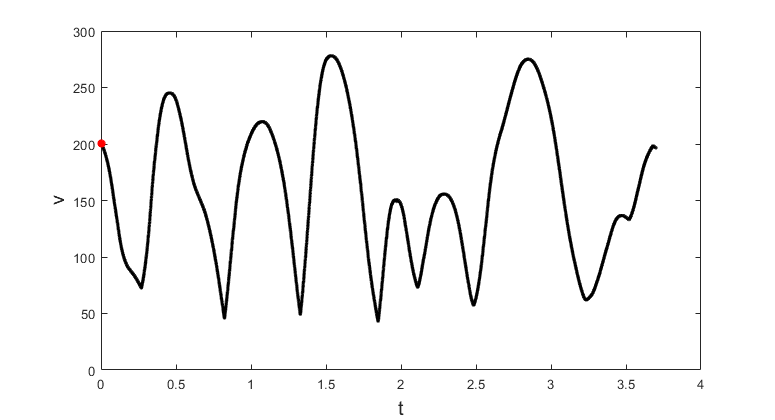}
\caption{Reaching path and speed profile of a random target pursuit task: approximation of Figure \ref{fig_4}. The red dot represents the movement starting position, while the blue one represents the first target to be reached.}
\label{approx_fig_2}
\end{figure}

In this experiment, we consider the random target pursuit task depicted in Figure \ref{approx_fig_2} (adapted from \cite{kadmon2019movement}), and compare the grouping performed with a standard Euclidean $k-$means algorithm with the spectral clustering with the sub-Riemannian distance $d_{\mathcal{M}}$ introduced in the previous section. We consider a longer trajectory in order to have many different slopes of the trajectory in different points, and compare the algorithms in different situations.

We first apply $k$-means clustering to the trajectory data with a generalized Euclidean distance, defined in \eqref{eucl_ad} with the same coefficient $c_1=10$.

\begin{figure}[H]
\centering
\includegraphics[scale=0.35]{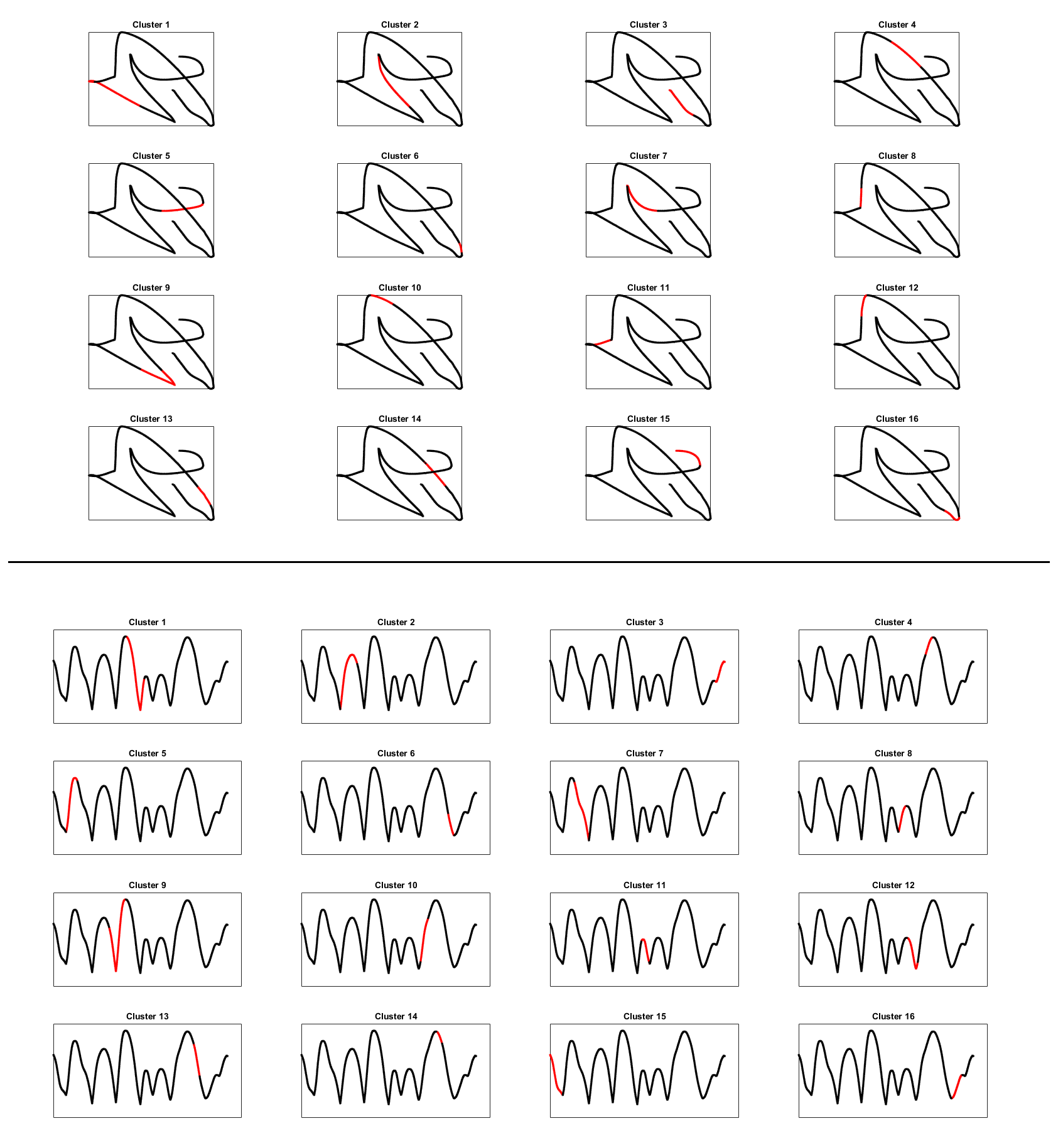}
\caption{Euclidean $k-$means. Eigenvectors projections over the reaching trajectory on the $\left(x,y\right)$ and $\left(t,v\right)$ planes.} 

\label{kmeans}
\end{figure}

\begin{figure}[H]
\centering
\subfloat[Euclidean]{\includegraphics[scale=0.3]{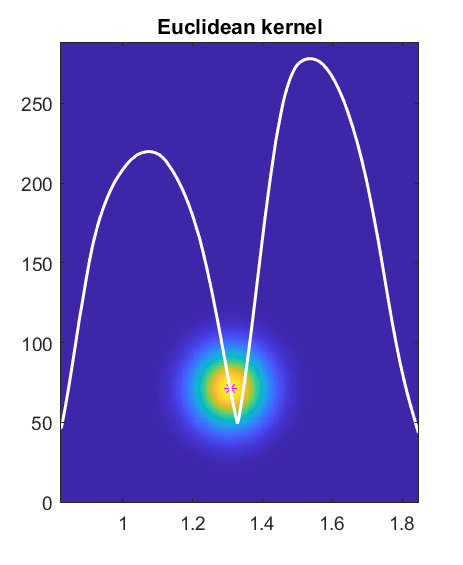}}\qquad\qquad
\subfloat[Riemannian]{\includegraphics[scale=0.3]{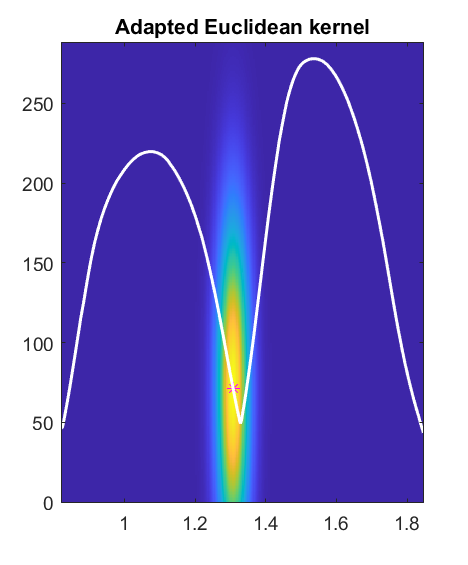}}\qquad\qquad
\subfloat[Sub-Riemannian]{\includegraphics[scale=0.3]{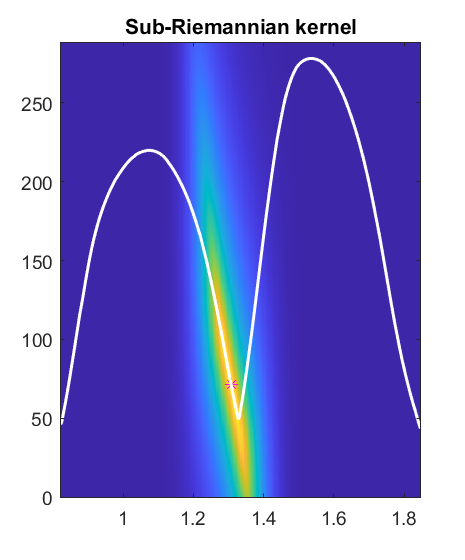}}
\caption{Kernel representations on a non-smooth segment on the $(t,v)$ plane of Figure \ref{approx_fig_2} .}
\label{plot_kernel_zoom}
\end{figure}

The result is displayed in Figure \ref{kmeans}. However, not even with this coefficient, the  segmentation is totally coherent with the expected results. Some sub-trajectories have abrupt changes in direction and speed, while others appear to be overly fragmented (Clusters $3$, $4$, $6$, $10$, $13$, $14$, $16$).  In addition, clusters $1$, $2$ and $9$ fail to capture the phases of acceleration and deceleration despite the changes in acceleration and direction of movement. This is because Euclidean $k$-means relates each point with  its neighbors in the direction of the axes, without considering the slope of the trajectory. This becomes clear while comparing Figures \ref{plot_kernels} and \ref{plot_kernel_zoom}. The Euclidean kernel is the same in the two cases, and clusters correctly when the change of slope is slow (Figure \ref{plot_kernels}),
while covers points in ascending and descending phase in presence of an abrupt change of the speed.



\begin{figure}[H]
\centering
\includegraphics[scale=0.3]{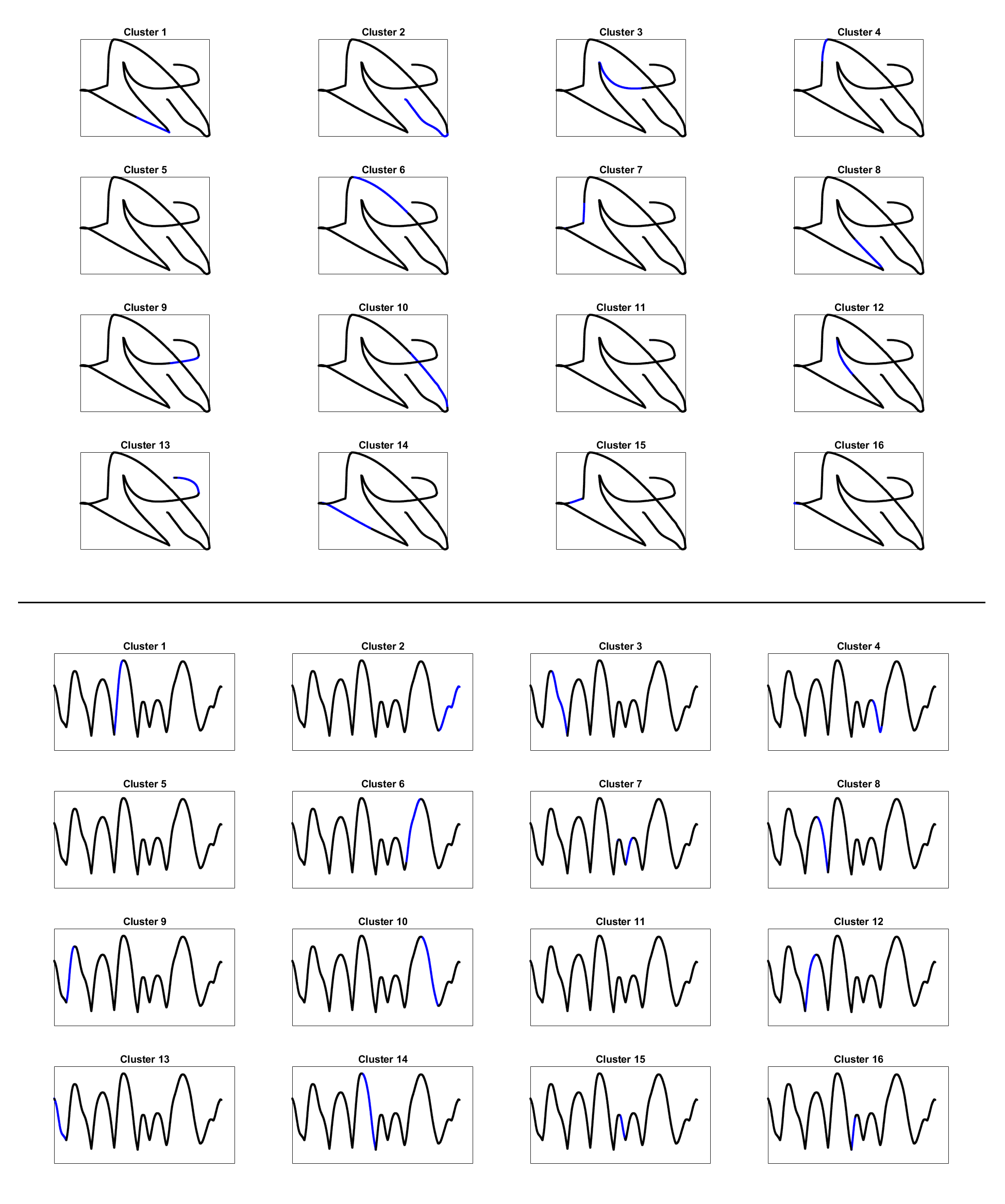}
\caption{Sub-Riemannian clustering. Eigenvectors projections over the reaching trajectory on the $\left(x,y\right)$ and $\left(t,v\right)$ planes. 
}
\label{subR_dec}
\end{figure}

Then we apply the sub-Riemannian clustering algorithm, always with the coefficient $c_1= 10$. As we see in in Figure \ref{subR_dec}, this algorithm allows us to correctly identify movement fragments as the acceleration and deceleration phases of the trajectory. The main reason is that 
sub-Riemannian distance correctly codes the differential constraints between the variables, strongly relating time, speed and acceleration in such a way that we can not change the velocity of a point, without changing its position. 
In Figure \ref{plot_kernel_zoom}, we zoomed on a segment of the trajectory in the $(t,v)$ plane identified by Euclidean cluster 9. For this segment of the trajectory, we applied the three kernels to a point on the curve to highlight the role of the tangency of the sub-Riemannian kernel. Due to the sudden change of speed, the Euclidean kernels tend to group points belonging
to different acceleration/deceleration phases, while the sub-Riemannian one tends to aligne with the shape of the curve, and performs the desired grouping.

Hence, we claim that the distance $d_{\mathcal{M}}$ \eqref{estim_hom_npv} is adequate, not only because of the properties of the kinematic space, but also because of the classification given by the clustering algorithm. We emphasize how instead the distances tested in the paper \cite{kadmon2019movement} (see the list recalled in \ref{spectral_lett}) or a Euclidean $k-$means algorithm did not justify the classification results into movement trajectories.
 We therefore introduced a distance that allows to perform a kernel component analysis which is the phenomenological counterpart of the neural PCAs provided by Kadmon Harpaz et al \cite{kadmon2019movement}. 

\section{Conclusions}\label{concl}
Our model proposes a geometric setting to explain the neural behaviour of the motor arm area M1. By getting inspiration from Georgopoulos neural models \cite{georgopoulos1982relations, georgopoulos2015columnar}, we provided a fiber bundle structure which is able to describe the hypercolumnar organization of the cortical area. On this structure, we considered the selective tuning of M1 neurons of kinematic variables by especially focusing on their temporal behaviour (\cite{Encoding}, \cite{churchland2007temporal}). 
We then extended the previous structure by considering that the cortex can code the time dependent direction of movement expressed as movement fragments as attested by experimental data of Hatsopoulos et al. \cite{Encoding} and Churchland et al. \cite{churchland2007temporal}. This led to consider a higher dimensional fiber bundle which codes movement fragments in the fiber: these were described as integral curves of the geometric structure with subriemannian metrics. Finally, in this space we defined a distance which models the weighting functions measured in the cortex and allows to recover an estimate of the neuronal population vector. 
The problem of identifying cortical activity patterns and their associated phenomenological primitives has been extensively studied in the visual cortex, specifically to identify perceptual units. Here we have applied the approach of \cite{sarti2015constitution} who related the emergent states of cortical activity studied in \cite{koffka2013principles} with visual perceptual units obtained with grouping algorithms.
Similarly, we linked the neural states found in \cite{kadmon2019movement} to elementary trajectories obtained through a clustering algorithm based on a Sub-Riemannian distance. 
We decompose movement trajectories into curves of acceleration or deceleration with a specific plane direction. These trajectories are well in accordance with the motor patterns measured in Kadmon Harpaz et al. \cite{kadmon2019movement} and Hatsopoulos et al. \cite{Encoding}. 
We emphasize that by working only on kinematic variables we recovered the same neural classification acquired by electrode array. 
This also proves that the 
kinematic parameters we identified are sufficient to completely explain the process of movement decomposition into trajectory fragments observed in \cite{kadmon2019movement}.

In the future, we will expand the model to a space of movement trajectories, where admissible distances and variations can be defined. This could be of neural support, as it could better explain the data presented in Hatsopoulos et al. \cite{Encoding}, Churchland et al. \cite{churchland2007temporal}, and Kadmon Harpaz et al. \cite{kadmon2019movement}, which strongly support the role of movement trajectories in motor encoding.

\appendix
\section{Appendices}
\subsection{Notions on fiber bundles and subriemannian distance}\label{sub}	
In this section we give the necessary mathematical definitions and theorems which are required in the framework of our study (we refer to \cite{jost2008riemannian}  and \cite{agrachev2019comprehensive, Montgomery} for the details of explanations provided). We are particularly interested in manifolds which have the structure of a fiber bundle. We therefore recall here the basic definitions

\begin{definition}\label{fiber bundle}
A fiber bundle $\left(E, M, F, \pi\right)$, on a manifold $M$ is a manifold $E$
together with a smooth surjective map $\pi  : E \to M$ such that, for all $ p \in M$, the following properties hold:
\begin{itemize}
\item 
The fiber $E_{p} := \pi^{-1}(p)$ is isomorphic to the set $F$ (typically a vector space or a group), called fiber.
\item 
There is a neighbourhood $U$ of $p$ in $M$ and a diffeomorphism $\chi : \pi^{-1}(U) \to U \times F$ such that
$\forall q \in U$, $\chi{|_{E_q}}
: E_q \to {q} \times F$.
\end{itemize}
The space $E$ is called total space, the manifold $M$ is the base, the vector space $E_p$ is the fiber over $p$ and the map $\chi$ is called a local trivialization. \end{definition}

\medskip

In the model we provide, the space is endowed with a structure of sub-Riemannian manifold, which can be summarized as follows.

A notion necessary to define a subriemannian manifold is horizontal distribution.
\begin{definition}\label{def_distribuzione}
Given an $n$-dimensional manifold $M$ and $k$ linearly independent vector fields $X_1, X_2, \cdots, X_k$ defined
on $M$, we denote $D_q = \spn \lbrace X_1, X_2,\cdots, X_k \rbrace \left(q\right)$, for all $q\in M$. 
The collection of all $D_q$ is the horizontal distribution $D\subset TM$. 
\end{definition}
Let us explicitly note that in general the dimension of $D_q$, denoted with $k$, is strictly smaller than the dimension $n$ of the manifold $M$. 
\begin{definition}
We call Lie Algebra generated by $X_{1}, \dots, X_{k}$ and denoted as 
$\mathcal{L}\left(X_{1}, \dots, X_{k}\right)$
the linear span of the operators $X_{1}, \dots, X_{k}$ and their commutators of any order. 
\end{definition}
\begin{definition}\label{grado}
According to the notations
\begin{align*}
D^{0}&= \lbrace 0\rbrace\\
D\,&= \spn\left(X_{1}, \dots , X_{k}\right)\\
D^2&= \spn\left(X_i, \left[X_{j}, X_{s}\right], \: X_i, X_j\in D, \: X_s\in D\right)\\
D^m&=  \spn\left(X_i, \left[X_{j}, X_{s}\right], \: X_i, X_j\in D^{m-1}, \: X_s\in D\right)
\end{align*}
We will say that a vector field $X$ has degree $m$ if $X\in D^{m}\setminus D^{m-1}$.  In this case we write $\text{deg}\left(X\right)= m$. 
\end{definition}

\begin{remark}\label{campifamosi}
The horizontal distribution defined in \ref{2.3.1} is spanned by the vector fields $X_1, X_2, X_3$ expressed in \eqref{campi_2D}, and the commutators have been computed in \eqref{commutatori}. According to Definition \ref{grado}, the degree of the commutators is the following one $$X_4 =[X_1, X_2],\; X_5 =[X_3, X_1], \text{ so that } \text{deg}(X_4) = \text{deg}(X_5) = 2$$ 
$$X_6 =[X_5, X_1] = [[X_3, X_1], X_1], \text{ so that } \text{deg}(X_4) = 3.$$ 
\end{remark}

\begin{definition}
Let $M$ be a manifold of dimension $n$ and let $\left(X_{j}\right)_{j=1}^k$ be a family of smooth vector fields defined on $M$. 
If the condition
$$\mathcal{L}\left(X_{1}, \dots, X_{k}\right)_{|_{p}}= T_{p}M\quad,\;\forall p\in M$$
is satisfied, we say that the vector fields $\left(X_{j}\right)_{j=1}^k$ are H\"ormander vector fields or satisfy the H\"ormander condition.
\end{definition}
A distribution $D$ generated by H\"{o}rmander vector fields is called bracket-generating. From this point, we can give the following definition

\begin{definition}\label{subRmanifold}
A subriemannian manifold is a triple $(M, D, \langle\cdot,\cdot \rangle_{g})$, where $M$ is a differentiable manifold, $D$ is a
bracket generating distribution and $\langle\cdot,\cdot \rangle_{g}$ is a scalar product on $D$. 
\end{definition}

The geometry of the space is described through objects whose tangent vectors belong to the fixed distribution $D$. In particular 
\begin{definition}
 A curve $\gamma:[a,b]\to M$ 
is called horizontal if it is absolutely continuous and $\dot \gamma(t) \in D_ { \gamma (t)}$ for every $t$.  
\end{definition}

Under the H\"ormander condition, a connectivity property holds true (Chow's Theorem \cite{chow2002systeme}) and a distance can be defined in two steps, as follows.

First, we introduce a notion of length 
\begin{definition}
The length of a horizontal curve $\gamma:[a,b]\to M$ is 
\begin{equation}\label{length_subr}
l\left(\gamma\right)= \int_{a}^{b} \left\|\dot{\gamma}\left(t\right)\right\|_{g} dt,
\end{equation}
where $\left\|\dot{\gamma}\left(t\right)\right\|_{g}= \sqrt{\left\langle \dot{\gamma}\left(t\right), \dot{\gamma}\left(t\right)\right\rangle_{g}}$ is computed using the inner product on the horizontal space $D_{\gamma\left(t\right)}$. The norm $\left\| \cdot \right\|_{g}$ is usually called horizontal norm.
\end{definition}
Then we can give a definition of distance in terms of minimal length path joining two arbitrary points of the manifold.
\begin{definition}\label{cc_distance_def}
The Carnot-Carathéodory distance between two arbitrary points $p$ and $q$ of a sub-Riemannian manifold is given by
\begin{equation}\label{carnot-car_distance}
d\left(p, q\right)= \inf\left\{l\left(\gamma\right): \gamma\; \text{is a horizontal curve connecting}\: p\;\text{and}\; q\right\}.
\end{equation}
\end{definition}

Montgomery \cite{Montgomery} proved the existence of length-minimizers, so that the $\inf$ in \eqref{carnot-car_distance} can be replaced by a minimum.

\subsection{{Homogeneous distance on $\mathcal{M}$}}\label{app_B}
The horizontal distribution is naturally endowed with a structure of Lie algebra through the bracket. Moreover, the commutator is a first order operator obtained as a difference of second order derivatives, so that it has degree 2. Analogously a higher order commutator will have higher degree. 
We already computed in Remark \ref{campifamosi} the degree of the vector fields $\left(X_i\right)_{i=1}^6$ defined in \eqref{campi_2D} and \eqref{commutatori}.

Since the distance has to be considered as a function of degree 1, then in the definition, the coefficients of the vector fields of degree 1 will be be raised to a power 1, the coefficients of vector fields of degree 2 will be raised to a power $\frac{1}{2}$ and the coefficients of degree 3 to a power $\frac{1}{3}$. As a consequence, the distance will be defined as follows:

\begin{definition}\label{def_distanza_hom}
Given the space $\mathcal{M}$ defined in \eqref{6D} and a family of constant positive coefficients $\lbrace c_i\rbrace_{i=1}^{6}$, we call homogeneous distance between two points $\eta_0$, $\eta_1\in \mathcal{M}$
\begin{equation}\label{distanza_exp_6D}
d_{\mathcal{H}}\left(\eta_0,\eta_1\right)= \left(\sum_{i=1}^3 \left|c_i e_{i}\right|^{6} + \sum_{i=4}^5 \left|c_i e_{i}\right|^{3}+ \left|c_6 e_{6}\right|^2 \right)^{\frac{1}{6}}.
\end{equation}
\end{definition}
The result provided in \cite{nagel1985balls} ensures that the Carnot-Carathéodory distance $d_{\mathcal{M}}$ expressed in terms of equation \eqref{carnot-car_distance} is locally equivalent to \eqref{distanza_exp_6D}. 

\begin{remark}\label{contiremark}

The vector fields $\left(X_i\right)_{i=1}^6$ form a basis for $T\mathcal{M}$ (see section \ref{time_npv}), hence the velocity of a curve $\gamma$ in the tangent space can be written as a linear combination of $\left(X_i\right)_{i=1}^6$:
\begin{align*}
\dot{\gamma}\left(s\right)=& e_{1}X_{1} + e_2 X_{2}+ e_{3}X_{3}+ e_{4}X_{4}+ e_{5}X_{5}+ e_{6}X_{6}=\\
=& e_{1}\begin{pmatrix} 
   v\cos\theta \\
   v\sin\theta \\
   0\\
   a\\
   0\\
   1 
\end{pmatrix} + e_{2}\begin{pmatrix} 
   0 \\
   0 \\
   1  \\
   0\\
   0\\
   0
\end{pmatrix}+ e_{3}\begin{pmatrix} 
   0 \\
   0 \\
   0 \\
   0\\
   1\\
   0
\end{pmatrix}+ e_{4}\begin{pmatrix}
   v\sin\theta \\
   -v\cos\theta \\
   0 \\
   0\\
   0\\
   0
\end{pmatrix}+ e_{5}\begin{pmatrix}
   0 \\
   0 \\
   0 \\
   1\\
   0\\
   0
\end{pmatrix}+ e_{6}\begin{pmatrix}
   \cos\theta \\
   \sin\theta \\
   0 \\
   0\\
   0\\
   0
\end{pmatrix}.
\end{align*} 
In order to define a homogeneous distance in $\mathcal{M}$, we search for an explicit expression of the increments $e_i$. We study the associated system
\begin{align*}
\begin{cases}
\dot{\gamma}\left(s\right)=& e_{1}X_{1} + e_{2} X_{2}+ e_{3}X_{3}+ e_{4}X_{4}+ e_{5}X_{5}+ e_{6}X_{6}\\
\gamma\left(0\right)=& \left(x_{0},y_{0},\theta_{0},v_{0},a_{0},t_{0}\right)\\
\gamma\left(1\right)=& \left(x_{1},y_{1},\theta_{1},v_{1},a_{1},t_{1}\right),\\
\end{cases}
\end{align*}
and we get 
\begin{align*}
\dot{\gamma}_{1}\left(s\right)=& \dot{x}=  e_{1}v\cos\theta + e_{4}v\sin\theta + e_{6}\cos\theta\\
\dot{\gamma}_{2}\left(s\right)=& \dot{y}=  e_{1}v\sin\theta - e_{4}v\cos\theta + e_{6}\sin\theta\\
\dot{\gamma}_{3}\left(s\right)=& \dot{\theta}=  e_{2}\\
\dot{\gamma}_{4}\left(s\right)=& \dot{v}=  e_{1}a + e_{5}\\
\dot{\gamma}_{5}\left(s\right)=& \dot{a}=  e_{3}\\
\dot{\gamma}_{6}\left(s\right)=& \dot{t}=  e_{1}.
\end{align*}
We immediately obtain $e_{1}= t_{1}- t_{0}$, $e_{2}= \theta_{1}- \theta_{0}$ and $e_{3}= a_{1}- a_{0}$.
In this way we also get $v\left(s\right)= e_{1}e_{3}\frac{s^{2}}{2} + e_{1}a_{0}s + e_{5}s+ v_{0}$ and consequently $ e_{5}= \left(v_{1}-v_{0}\right) -\frac{e_{1}}{2}\left(a_{0}+a_{1}\right)$.
Hence, by calling $\tilde x = x \cos (\theta) + y \sin(\theta)$ and  $\tilde x = x \sin (\theta) - y \cos(\theta)$ we get
\begin{align*}
\dot{\tilde\gamma}_{1}(s)=&
e_{1}v  + e_{6} - \tilde y e_2 \\
\dot{\tilde \gamma}_{2} (s)=&    e_{4}v + \tilde x e_2.
\end{align*}

Hence integrating $\dot{\tilde\gamma}_{1}$ and $\dot{\tilde\gamma}_{2}$ between $0$ and $1$ we obtain the expressions of $\tilde x_{1}- \tilde x_{0}$ and $\tilde y_{1}- \tilde y_0$, and subsequently $e_4$ and $e_6$. 

In particular, if $e_2= 0$ the two equation decouple. Since $v$ is a polynomial, $e_4$ and $e_6$ turn out to be 
\begin{align*}
e_4&= - \frac{12\left(\tilde y_1- \tilde y_0\right)}{6\left(v_0+ v_1\right)- e_1 e_3} = \frac{12\left(\left(x_1- x_0\right)\sin\theta_0- \left(y_1- y_0\right)\cos\theta_0\right)}{6\left(v_0+ v_1\right)- e_1 e_3}, \\
\\
e_6&= (\tilde x_1- \tilde x_0)- \frac{e_1}{12}\left(6\left(v_0+ v_1\right)- e_1 e_3\right)\\& = \left(x_1- x_0\right)\cos\theta_0 + \left(y_1-y_0\right)\sin\theta_0- \frac{e_1}{12}\left(6\left(v_0+ v_1\right)- e_1 e_3\right).
\end{align*}

If $e_2 \not=0$, the system is a standard oscillator, with polynomial forcing term. Hence it can be directly integrated to find $e_4$ and $e_6$.  
\end{remark}

\section*{Declarations}
\textbf{Ethical Approval}
Not applicable.\\
\newline
\textbf{Authors' contributions}
All authors reviewed the manuscript.\\
\newline
\textbf{Competing interests}
The authors declare no conflict of interest.\\
\newline
\textbf{Funding}
This project has received funding from the European Union’s Horizon 2020 research and innovation programme under the Marie Skłodowska-Curie grant agreement No 777822.\\
\newline
\textbf{Availability of data and materials}
Graphical representations of the model are available upon request.

\bibliographystyle{splncs04}

\begin{thebibliography}{}
\providecommand{\url}[1]{\texttt{#1}}
\providecommand{\urlprefix}{URL }
\providecommand{\doi}[1]{https://doi.org/#1}

\end{thebibliography}


\addcontentsline{toc}{section}{References}
\begin{thebibliography}{10}

\bibitem{Hatf}
William Abend, Emilio Bizzi, and Pietro Morasso.
\newblock Human arm trajectory formation.
\newblock {\em Brain: a journal of neurology}, 105(Pt 2):331--348, 1982.

\bibitem{aflalo2006partial}
Tyson~N Aflalo and Michael~SA Graziano.
\newblock Partial tuning of motor cortex neurons to final posture in a
  free-moving paradigm.
\newblock {\em Proceedings of the National Academy of Sciences},
  103(8):2909--2914, 2006.

\bibitem{aflalo2006possible}
Tyson~N Aflalo and Michael~SA Graziano.
\newblock Possible origins of the complex topographic organization of motor
  cortex: reduction of a multidimensional space onto a two-dimensional array.
\newblock {\em Journal of Neuroscience}, 26(23):6288--6297, 2006.

\bibitem{agrachev2019comprehensive}
Andrei Agrachev, Davide Barilari, and Ugo Boscain.
\newblock {\em A comprehensive introduction to sub-Riemannian geometry}, volume
  181.
\newblock Cambridge University Press, 2019.

\bibitem{ajemian2001model}
Robert Ajemian, Daniel Bullock, and Stephen Grossberg.
\newblock A model of movement coordinates in the motor cortex:
  posture-dependent changes in the gain and direction of single cell tuning
  curves.
\newblock {\em Cerebral Cortex}, 11(12):1124--1135, 2001.

\bibitem{amirikian2003modular}
Bagrat Amirikian and Apostolos~P Georgopoulos.
\newblock Modular organization of directionally tuned cells in the motor
  cortex: is there a short-range order?
\newblock {\em Proceedings of the National Academy of Sciences},
  100(21):12474--12479, 2003.

\bibitem{ashe1994movement}
James Ashe and Apostolos~P Georgopoulos.
\newblock Movement parameters and neural activity in motor cortex and area 5.
\newblock {\em Cerebral cortex}, 4(6):590--600, 1994.

\bibitem{belkin2003laplacian}
Mikhail Belkin and Partha Niyogi.
\newblock Laplacian eigenmaps for dimensionality reduction and data
  representation.
\newblock {\em Neural computation}, 15(6):1373--1396, 2003.

\bibitem{bjorck1990least}
{\AA}ke Bj{\"o}rck.
\newblock Least squares methods.
\newblock {\em Handbook of numerical analysis}, 1:465--652, 1990.

\bibitem{bosking1997orientation}
William~H Bosking, Ying Zhang, Brett Schofield, and David Fitzpatrick.
\newblock Orientation selectivity and the arrangement of horizontal connections
  in tree shrew striate cortex.
\newblock {\em Journal of neuroscience}, 17(6):2112--2127, 1997.

\bibitem{bressloff2002visual}
Paul~C Bressloff and Jack~D Cowan.
\newblock The visual cortex as a crystal.
\newblock {\em Physica D: Nonlinear Phenomena}, 173(3-4):226--258, 2002.

\bibitem{bressloff2003functional}
Paul~C Bressloff and Jack~D Cowan.
\newblock The functional geometry of local and horizontal connections in a
  model of v1.
\newblock {\em Journal of Physiology-Paris}, 97(2-3):221--236, 2003.

\bibitem{bressloff2002geometric}
Paul~C Bressloff, Jack~D Cowan, Martin Golubitsky, Peter~J Thomas, and
  Matthew~C Wiener.
\newblock What geometric visual hallucinations tell us about the visual cortex.
\newblock {\em Neural computation}, 14(3):473--491, 2002.

\bibitem{butler2006spectral}
Steve Butler, Fan Chung, et~al.
\newblock Spectral graph theory.
\newblock {\em Handbook of linear algebra}, page~47, 2006.

\bibitem{caminiti1990making}
Roberto Caminiti, Paul~B Johnson, and Antonio Urbano.
\newblock Making arm movements within different parts of space: dynamic aspects
  in the primate motor cortex.
\newblock {\em Journal of Neuroscience}, 10(7):2039--2058, 1990.

\bibitem{carpenter2012neurophysiology}
Roger Carpenter and Benjamin Reddi.
\newblock {\em Neurophysiology: a conceptual approach}.
\newblock CRC Press, 2012.

\bibitem{chow2002systeme}
Wei-Liang Chow.
\newblock {\"U}ber systeme von linearen partiellen differential-gleichungen
  erster ordnung.
\newblock In {\em The Collected Papers Of Wei-Liang Chow}, pages 47--54. World
  Scientific, 2002.

\bibitem{churchland2007temporal}
Mark~M Churchland and Krishna~V Shenoy.
\newblock Temporal complexity and heterogeneity of single-neuron activity in
  premotor and motor cortex.
\newblock {\em Journal of neurophysiology}, 97(6):4235--4257, 2007.

\bibitem{citti2006cortical}
Giovanna Citti and Alessandro Sarti.
\newblock A cortical based model of perceptual completion in the
  roto-translation space.
\newblock {\em Journal of Mathematical Imaging and Vision}, 24(3):307--326,
  2006.

\bibitem{cocci2014spatio}
Giacomo Cocci.
\newblock Spatio-temporal models of the functional architecture of the visual
  cortex.
\newblock 2014.

\bibitem{cocci2015cortical}
Giacomo Cocci, Davide Barbieri, Giovanna Citti, and Alessandro Sarti.
\newblock Cortical spatiotemporal dimensionality reduction for visual grouping.
\newblock {\em Neural computation}, 27(6):1252--1293, 2015.

\bibitem{coifman2006diffusion}
Ronald~R Coifman and St{\'e}phane Lafon.
\newblock Diffusion maps.
\newblock {\em Applied and computational harmonic analysis}, 21(1):5--30, 2006.

\bibitem{daugman1985uncertainty}
John~G Daugman.
\newblock Uncertainty relation for resolution in space, spatial frequency, and
  orientation optimized by two-dimensional visual cortical filters.
\newblock {\em JOSA A}, 2(7):1160--1169, 1985.

\bibitem{deangelis1995receptive}
Gregory~C DeAngelis, Izumi Ohzawa, and Ralph~D Freeman.
\newblock Receptive-field dynamics in the central visual pathways.
\newblock {\em Trends in neurosciences}, 18(10):451--458, 1995.

\bibitem{deangelis1993spatiotemporal}
Gregory~C DeAngelis, Izumi Ohzawa, and RD~Freeman.
\newblock Spatiotemporal organization of simple-cell receptive fields in the
  cat's striate cortex. i. general characteristics and postnatal development.
\newblock {\em Journal of neurophysiology}, 69(4):1091--1117, 1993.

\bibitem{economo2018distinct}
Michael~N Economo, Sarada Viswanathan, Bosiljka Tasic, Erhan Bas, Johan
  Winnubst, Vilas Menon, Lucas~T Graybuck, Thuc~Nghi Nguyen, Kimberly~A Smith,
  Zizhen Yao, et~al.
\newblock Distinct descending motor cortex pathways and their roles in
  movement.
\newblock {\em Nature}, 563(7729):79--84, 2018.

\bibitem{ermentrout1980large}
G~Bard Ermentrout and JD~Cowan.
\newblock Large scale spatially organized activity in neural nets.
\newblock {\em SIAM Journal on Applied Mathematics}, 38(1):1--21, 1980.

\bibitem{evarts1967representation}
Eo~Vo Evarts.
\newblock Representation of movements and muscles by pyramidal tract neurons of
  the precentral motor cortex, 1967.

\bibitem{faugeras2009persistent}
Olivier Faugeras, Romain Veltz, and Fran{\c{c}}ois Grimbert.
\newblock Persistent neural states: stationary localized activity patterns in
  nonlinear continuous n-population, q-dimensional neural networks.
\newblock {\em Neural computation}, 21(1):147--187, 2009.

\bibitem{favali2017local}
Marta Favali, Giovanna Citti, and Alessandro Sarti.
\newblock Local and global gestalt laws: A neurally based spectral approach.
\newblock {\em Neural computation}, 29(2):394--422, 2017.

\bibitem{flash1992timing}
Tamar Flash, Ealan Henis, Rivka Inzelberg, and AD~Korczyn.
\newblock Timing and sequencing of human arm trajectories: normal and abnormal
  motor behaviour.
\newblock {\em Human movement science}, 11(1-2):83--100, 1992.

\bibitem{flash2005motor}
Tamar Flash and Binyamin Hochner.
\newblock Motor primitives in vertebrates and invertebrates.
\newblock {\em Current opinion in neurobiology}, 15(6):660--666, 2005.

\bibitem{FH}
Tamar Flash and Neville Hogan.
\newblock The coordination of arm movements: an experimentally confirmed
  mathematical model.
\newblock {\em Journal of neuroscience}, 5(7):1688--1703, 1985.

\bibitem{georgopoulous1984representation}
AP~Georgopoulos.
\newblock The representation of movement direction in the motor cotex: Single
  cell and population studies.
\newblock {\em Dynamic aspects of neocortical function}, pages 501--524, 1984.

\bibitem{georgopoulos1984static}
AP~Georgopoulos, R~Caminiti, and JF~Kalaska.
\newblock Static spatial effects in motor cortex and area 5: quantitative
  relations in a two-dimensional space.
\newblock {\em Experimental Brain Research}, 54(3):446--454, 1984.

\bibitem{georgopoulos1988neural}
Apostolos~P Georgopoulos.
\newblock Neural integration of movement: role of motor cortex in reaching.
\newblock {\em The FASEB Journal}, 2(13):2849--2857, 1988.

\bibitem{georgopoulos1988spatial}
Apostolos~P Georgopoulos.
\newblock Spatial coding of visually guided arm movements in primate motor
  cortex.
\newblock {\em Canadian journal of physiology and pharmacology},
  66(4):518--526, 1988.

\bibitem{georgopoulos2015columnar}
Apostolos~P Georgopoulos.
\newblock Columnar organization of the motor cortex: direction of movement.
\newblock In {\em Recent Advances on the Modular Organization of the Cortex},
  pages 123--141. Springer, 2015.

\bibitem{georgopoulos1992motor}
Apostolos~P Georgopoulos, James Ashe, Nikolaos Smyrnis, and Masato Taira.
\newblock The motor cortex and the coding of force.
\newblock {\em Science}, 256(5064):1692--1695, 1992.

\bibitem{georgopoulos1982relations}
Apostolos~P Georgopoulos, John~F Kalaska, Roberto Caminiti, and Joe~T Massey.
\newblock On the relations between the direction of two-dimensional arm
  movements and cell discharge in primate motor cortex.
\newblock {\em Journal of Neuroscience}, 2(11):1527--1537, 1982.

\bibitem{georgopoulos1988primate}
Apostolos~P Georgopoulos, Ronald~E Kettner, and Andrew~B Schwartz.
\newblock Primate motor cortex and free arm movements to visual targets in
  three-dimensional space. ii. coding of the direction of movement by a
  neuronal population.
\newblock {\em Journal of Neuroscience}, 8(8):2928--2937, 1988.

\bibitem{georgopoulos2007mapping}
Apostolos~P Georgopoulos, Hugo Merchant, Thomas Naselaris, and Bagrat
  Amirikian.
\newblock Mapping of the preferred direction in the motor cortex.
\newblock {\em Proceedings of the National Academy of Sciences},
  104(26):11068--11072, 2007.

\bibitem{georgopoulos1986neuronal}
Apostolos~P Georgopoulos, Andrew~B Schwartz, and Ronald~E Kettner.
\newblock Neuronal population coding of movement direction.
\newblock {\em Science}, 233(4771):1416--1419, 1986.

\bibitem{graziano2008intelligent}
Michael Graziano.
\newblock {\em The intelligent movement machine: An ethological perspective on
  the primate motor system}.
\newblock Oxford University Press, 2008.

\bibitem{graziano2007mapping}
Michael~SA Graziano and Tyson~N Aflalo.
\newblock Mapping behavioral repertoire onto the cortex.
\newblock {\em Neuron}, 56(2):239--251, 2007.

\bibitem{graziano2002cortical}
Michael~SA Graziano, Charlotte~SR Taylor, Tirin Moore, and Dylan~F Cooke.
\newblock The cortical control of movement revisited.
\newblock {\em Neuron}, 36(3):349--362, 2002.

\bibitem{Encoding}
Nicholas~G Hatsopoulos, Qingqing Xu, and Yali Amit.
\newblock Encoding of movement fragments in the motor cortex.
\newblock {\em Journal of Neuroscience}, 27(19):5105--5114, 2007.

\bibitem{hoffman1970higher}
William~C Hoffman.
\newblock Higher visual perception as prolongation of the basic lie
  transformation group.
\newblock {\em Mathematical Biosciences}, 6:437--471, 1970.

\bibitem{hogan1984organizing}
Neville Hogan.
\newblock An organizing principle for a class of voluntary movements.
\newblock {\em Journal of neuroscience}, 4(11):2745--2754, 1984.

\bibitem{holdefer2002primary}
RN~Holdefer and LE~Miller.
\newblock Primary motor cortical neurons encode functional muscle synergies.
\newblock {\em Experimental Brain Research}, 146(2):233--243, 2002.

\bibitem{hubel1995eye}
David~H Hubel.
\newblock {\em Eye, brain, and vision.}
\newblock Scientific American Library/Scientific American Books, 1995.

\bibitem{hubel1963shape}
David~H Hubel and TN~Wiesel.
\newblock Shape and arrangement of columns in cat's striate cortex.
\newblock {\em The Journal of physiology}, 165(3):559--568, 1963.

\bibitem{hubel1962receptive}
David~H Hubel and Torsten~N Wiesel.
\newblock Receptive fields, binocular interaction and functional architecture
  in the cat's visual cortex.
\newblock {\em The Journal of physiology}, 160(1):106--154, 1962.

\bibitem{hubelandt1977functionalarchitectureofmacaquemonkeyvisual}
DH~HubelandT and N~Wiesel.
\newblock Functionalarchitectureofmacaquemonkeyvisual cortex.
\newblock {\em Proc. Roy. Soc. Lond. B}, 198(1-59), 1977.

\bibitem{jones1987evaluation}
Robert Jones.
\newblock The evaluation and grading of placement performance.
\newblock {\em Teaching Public Administration}, 7(1):31--43, 1987.

\bibitem{jost2008riemannian}
J{\"u}rgen Jost and Jeurgen Jost.
\newblock {\em Riemannian geometry and geometric analysis}, volume 42005.
\newblock Springer, 2008.

\bibitem{kadmon2019movement}
Naama Kadmon~Harpaz, David Ungarish, Nicholas~G Hatsopoulos, and Tamar Flash.
\newblock Movement decomposition in the primary motor cortex.
\newblock {\em Cerebral cortex}, 29(4):1619--1633, 2019.

\bibitem{kalaska2009intention}
John~F Kalaska.
\newblock From intention to action: motor cortex and the control of reaching
  movements.
\newblock {\em Progress in Motor Control}, pages 139--178, 2009.

\bibitem{kannan2004clusterings}
Ravi Kannan, Santosh Vempala, and Adrian Vetta.
\newblock On clusterings: Good, bad and spectral.
\newblock {\em Journal of the ACM (JACM)}, 51(3):497--515, 2004.

\bibitem{kettner1988primate}
Ronald~E Kettner, Andrew~B Schwartz, and Apostolos~P Georgopoulos.
\newblock Primate motor cortex and free arm movements to visual targets in
  three-dimensional space. iii. positional gradients and population coding of
  movement direction from various movement origins.
\newblock {\em Journal of Neuroscience}, 8(8):2938--2947, 1988.

\bibitem{koffka2013principles}
Kurt Koffka.
\newblock {\em Principles of Gestalt psychology}.
\newblock Routledge, 2013.

\bibitem{krishna1999genetic}
K~Krishna and M~Narasimha Murty.
\newblock Genetic k-means algorithm.
\newblock {\em IEEE Transactions on Systems, Man, and Cybernetics, Part B
  (Cybernetics)}, 29(3):433--439, 1999.

\bibitem{lafon2006diffusion}
Stephane Lafon and Ann~B Lee.
\newblock Diffusion maps and coarse-graining: A unified framework for
  dimensionality reduction, graph partitioning, and data set parameterization.
\newblock {\em IEEE transactions on pattern analysis and machine intelligence},
  28(9):1393--1403, 2006.

\bibitem{lee1996image}
Tai~Sing Lee.
\newblock Image representation using 2d gabor wavelets.
\newblock {\em IEEE Transactions on pattern analysis and machine intelligence},
  18(10):959--971, 1996.

\bibitem{meilua2001random}
Marina Meil{\u{a}} and Jianbo Shi.
\newblock A random walks view of spectral segmentation.
\newblock In {\em International Workshop on Artificial Intelligence and
  Statistics}, pages 203--208. PMLR, 2001.

\bibitem{Montgomery}
Richard Montgomery.
\newblock {\em A Tour of Subriemannian Geometries, Their Geodesics and
  Applications}.
\newblock Mathematical Surveys and Monographs 91. American Mathematical
  Society, 2006.

\bibitem{moran1999motor}
Daniel~W Moran and Andrew~B Schwartz.
\newblock Motor cortical representation of speed and direction during reaching.
\newblock {\em Journal of neurophysiology}, 82(5):2676--2692, 1999.

\bibitem{morasso1981spatial}
Pietro Morasso.
\newblock Spatial control of arm movements.
\newblock {\em Experimental brain research}, 42(2):223--227, 1981.

\bibitem{mountcastle1997columnar}
Vernon~B Mountcastle.
\newblock The columnar organization of the neocortex.
\newblock {\em Brain: a journal of neurology}, 120(4):701--722, 1997.

\bibitem{mussa2004neural}
Ferdinando~A Mussa-Ivaldi and Sara~A Solla.
\newblock Neural primitives for motion control.
\newblock {\em IEEE Journal of Oceanic Engineering}, 29(3):640--650, 2004.

\bibitem{nagel1985balls}
Alexander Nagel, Elias~M Stein, and Stephen Wainger.
\newblock Balls and metrics defined by vector fields i: Basic properties.
\newblock {\em Acta Mathematica}, 155:103--147, 1985.

\bibitem{naselaris2006large}
Thomas Naselaris, Hugo Merchant, Bagrat Amirikian, and Apostolos~P
  Georgopoulos.
\newblock Large-scale organization of preferred directions in the motor cortex.
  ii. analysis of local distributions.
\newblock {\em Journal of neurophysiology}, 96(6):3237--3247, 2006.

\bibitem{ng2001spectral}
Andrew Ng, Michael Jordan, and Yair Weiss.
\newblock On spectral clustering: Analysis and an algorithm.
\newblock {\em Advances in neural information processing systems}, 14, 2001.

\bibitem{omrani2017perspectives}
Mohsen Omrani, Matthew~T Kaufman, Nicholas~G Hatsopoulos, and Paul~D Cheney.
\newblock Perspectives on classical controversies about the motor cortex.
\newblock {\em Journal of neurophysiology}, 118(3):1828--1848, 2017.

\bibitem{paninski2004spatiotemporal}
Liam Paninski, Matthew~R Fellows, Nicholas~G Hatsopoulos, and John~P Donoghue.
\newblock Spatiotemporal tuning of motor cortical neurons for hand position and
  velocity.
\newblock {\em Journal of neurophysiology}, 91(1):515--532, 2004.

\bibitem{perona1998factorization}
Pietro Perona and William Freeman.
\newblock A factorization approach to grouping.
\newblock In {\em European Conference on Computer Vision}, pages 655--670.
  Springer, 1998.

\bibitem{petitot2003neurogeometry}
Jean Petitot.
\newblock The neurogeometry of pinwheels as a sub-riemannian contact structure.
\newblock {\em Journal of Physiology-Paris}, 97(2-3):265--309, 2003.

\bibitem{petitot2017neurogeometrie}
Jean Petitot.
\newblock {\em Neurog{\'e}om{\'e}trie de la vision}.
\newblock Springer, 2017.

\bibitem{petitot1999vers}
Jean Petitot and Yannick Tondut.
\newblock Vers une neurog{\'e}om{\'e}trie. fibrations corticales, structures de
  contact et contours subjectifs modaux.
\newblock {\em Math{\'e}matiques et sciences humaines}, 145:5--101, 1999.

\bibitem{reimer2009problem}
Jacob Reimer and Nicholas~G Hatsopoulos.
\newblock The problem of parametric neural coding in the motor system.
\newblock In {\em Progress in motor control}, pages 243--259. Springer, 2009.

\bibitem{roweis2000nonlinear}
Sam~T Roweis and Lawrence~K Saul.
\newblock Nonlinear dimensionality reduction by locally linear embedding.
\newblock {\em science}, 290(5500):2323--2326, 2000.

\bibitem{sarti2015constitution}
Alessandro Sarti and Giovanna Citti.
\newblock The constitution of visual perceptual units in the functional
  architecture of v1.
\newblock {\em Journal of computational neuroscience}, 38(2):285--300, 2015.

\bibitem{schwartz2007useful}
Andrew~B Schwartz.
\newblock Useful signals from motor cortex.
\newblock {\em The Journal of physiology}, 579(3):581--601, 2007.

\bibitem{schwartz1988primate}
Andrew~B Schwartz, Ronald~E Kettner, and Apostolos~P Georgopoulos.
\newblock Primate motor cortex and free arm movements to visual targets in
  three-dimensional space. i. relations between single cell discharge and
  direction of movement.
\newblock {\em Journal of Neuroscience}, 8(8):2913--2927, 1988.

\bibitem{scott20056}
Stephen~H Scott.
\newblock 6 conceptual frameworks for interpreting motor cortical function: New
  insights from multiple-joint paradigm.
\newblock 2005.

\bibitem{shi2000normalized}
Jianbo Shi and Jitendra Malik.
\newblock Normalized cuts and image segmentation.
\newblock {\em IEEE Transactions on pattern analysis and machine intelligence},
  22(8):888--905, 2000.

\bibitem{stark2009motor}
Eran Stark, Rotem Drori, and Moshe Abeles.
\newblock Motor cortical activity related to movement kinematics exhibits local
  spatial organization.
\newblock {\em Cortex}, 45(3):418--431, 2009.

\bibitem{teka2017motor}
Wondimu~W Teka, Khaldoun~C Hamade, William~H Barnett, Taegyo Kim, Sergey~N
  Markin, Ilya~A Rybak, and Yaroslav~I Molkov.
\newblock From the motor cortex to the movement and back again.
\newblock {\em PloS one}, 12(6):e0179288, 2017.

\bibitem{todorov2000direct}
Emanuel Todorov.
\newblock Direct cortical control of muscle activation in voluntary arm
  movements: a model.
\newblock {\em Nature neuroscience}, 3(4):391--398, 2000.

\bibitem{von2007tutorial}
Ulrike Von~Luxburg.
\newblock A tutorial on spectral clustering.
\newblock {\em Statistics and computing}, 17(4):395--416, 2007.

\bibitem{weiss1999segmentation}
Yair Weiss.
\newblock Segmentation using eigenvectors: a unifying view.
\newblock In {\em Proceedings of the seventh IEEE international conference on
  computer vision}, volume~2, pages 975--982. IEEE, 1999.

\bibitem{wilson1973mathematical}
Hugh~R Wilson and Jack~D Cowan.
\newblock A mathematical theory of the functional dynamics of cortical and
  thalamic nervous tissue.
\newblock {\em Kybernetik}, 13(2):55--80, 1973.

\bibitem{WISE200110137}
S.P. Wise.
\newblock Motor cortex.
\newblock In Neil~J. Smelser and Paul~B. Baltes, editors, {\em International
  Encyclopedia of the Social \& Behavioral Sciences}, pages 10137--10140.
  Pergamon, Oxford, 2001.

\end{thebibliography}

\end{document}